\documentclass{aa}
\pdfoutput=1

\usepackage{graphicx}
\usepackage{natbib}

\usepackage{txfonts}
\usepackage{units}

\begin{document}

   \title{A large sample of Kohonen selected E+A (post-starburst) galaxies from the Sloan Digital Sky Survey}

   \author{H. Meusinger\inst{1,2}
          \and
          J. Br\"unecke\inst{1,2}
          \and 
          P. Schalldach\inst{1}
          \and
          A. in der Au\inst{3}
          }

   \institute{T\"uringer Landessternwarte, Sternwarte 5, 07778 Tautenburg, Germany, 
              \email{meus@tls-tautenburg.de}
         \and
              Universit\"at Leipzig, Fakult\"at f\"ur Physik und Geowissenschaften, Linnestra{\ss}e 5, 04103 Leipzig, Germany 
         \and
              Texture-Editor GbR, Dornr\"oschenstra{\ss}e 48, 81739 Munich, Germany
             }

   \date{Received XXXXXX; accepted XXXXXX}

 
  \abstract
   {
   The galaxy population in the contemporary Universe is characterised by a clear bimodality, 
   blue galaxies with significant ongoing star formation and red galaxies with only a little. 
   The migration between the blue and the red cloud of galaxies is an issue of active research.
   Post starburst (PSB) galaxies are thought to be observed in the short-lived transition phase. 
   }
   {
   We aim to create a large sample of local PSB galaxies from the Sloan Digital Sky Survey (SDSS).
   Another aim is to present a tool set for an efficient search in a large database of SDSS spectra based on Kohonen 
   self-organising maps (SOMs).
   }
   {
   We computed a huge Kohonen SOM for $\sim 10^6$ spectra from SDSS Data Release 7. 
   The SOM is made fully available, in combination with an interactive user interface, for the astronomical community.
   We selected a large sample of PSB galaxies taking advantage of the clustering behaviour of the SOM. 
   The morphologies were inspected on deep co-added SDSS images.
   We used the Portsmouth galaxy property computations to study the evolutionary stage of the PSB galaxies and
   archival multi-wavelength data to search for hidden AGNs.    
   }
   {
   We compiled a catalogue of 2\,665 PSB galaxies with redshifts $z < 0.4$.
   In the colour-mass diagram, the PSB sample is clearly concentrated towards the region 
   between the red and the blue cloud. The relative frequency of distorted PSB galaxies is at 
   least 57\% for EW(H$\delta$)$ > 5$\AA\,, significantly higher than in the comparison sample. 
   The search for AGNs based on conventional selection criteria in the radio and MIR results in 
   a low AGN fraction of $\sim 2 - 3$\%. We confirm an MIR excess in the mean SED of the E+A sample that may indicate hidden AGNs, though other sources are also possible.
   }
   {}

  \titlerunning{Kohonen-selected E+A galaxies}
  \authorrunning{H. Meusinger et al.}

   \maketitle
%
%
%
\section{Introduction}
\label{sec:Introduction}

The bimodality of the galaxy distribution in the colour-luminosity (or colour-stellar mass) plane and the 
migration of galaxies between the red and the blue cloud is an important issue in galaxy evolution 
research
\citep[e.g.][]{Strateva_2001,
2003MNRAS.341...33K, 
Blanton_2003,
Baldry_2004,
Gabor_2011,
Rodriguez_2014,
Knobel_2015}.
E+A galaxies are thought to be best candidates for systems in that transformation stage towards early-type 
galaxies in the red sequence \citep{Yang_2008,Wong_2012}.
The rare type of E+A galaxies is defined by optical spectra that indicate a combination of
characteristics from old stellar populations typical of elliptical galaxies on the one hand and strong 
Balmer absorption lines, mostly from A stars indicating a recent episode of substantial star formation, on the 
other hand \citep{Dressler_Gunn_1983,Dressler_Gunn_1992,Couch_1987,Kaviraj_2007,Bergvall_2016}.
Alternatively the term K+A galaxies is used \citep[e.g.][]{Melnick_2013,Melnick_2015}
refering to spectra of an old stellar population dominated by K giants superimposed
by a strong population of A stars, without restriction on morphology.
In this paper, we will use the term E+A galaxies.

The absence of strong [\ion{O}{ii}] or H$\alpha$ emission lines in the spectra of E+A galaxies 
indicates that there is currently no substantial visible star formation \citep[e.g.][]{Couch_1987,Quintero_2004,Goto_2007a,Wu_2014}. 
The strong impact of A-type stars in the spectrum is a sign of a substantial stellar population 
with an age corresponding to or less than the main-sequence lifetime of A stars from a 
starburst less than about one Gyr ago. A significant part of about 10\% to 60\% 
of the stellar mass of the galaxy was created in that starburst
\citep{Kaviraj_2007,Choi_2009,Swinbank_2012,Melnick_2013}. 
An alternative interpretation of the optical spectra could be a still on-going starburst that is obscured by dust
\citep{Poggianti_2000}.
Based on 20 cm radio continuum observations, \citet{Goto_2004} has shown that this dusty starburst scenario
can be excluded for the majority of his sample of 34 E+A galaxies. 
A previous starburst, rather than just a truncation of the star formation, is also required by the optical
and near-infrared colours \citep{Balogh_2005}. 
E+A galaxies are therefore considered as prototypical post-starburst (PSB) galaxies,
observed in a short-lived transition phase from the blue cloud towards the red sequence. 
For luminous E+A galaxies ($M(z) < -22$) the star formation rates (SFR) in the starburst seem to be high enough 
to qualify them as successors of luminous and ultra-luminous infrared galaxies (LIRGs and ULIRGs) \citep{Kaviraj_2007,Liu_2007}.  At least some ULIRGs may evolve to quasars and the intermediate stage may be represented by PSB quasars that show the spectral signatures of both a luminous active galactic nucleus (AGN) and a PSB stellar population 
\citep{Brotherton_1999,Cales_2015,Melnick_2015}.

It has long been recognised that tidal interactions and merging of gas-rich galaxies can act as triggers
for star formation \citep{Toomre_1972,Larson_1978,Hopkins_2008} and can be a major driver of starbursts 
\citep[e.g.][]{Barton_2000,Snyder_2011}. In hierarchical models, galaxy mergers are a key mechanism of structure 
formation and evolution. Major mergers, minor mergers, and tidal interactions with close companions can all 
perturb the structure of the involved galaxies on time scales of the order of a Gyr 
\citep[e.g.][]{Mihos_1996,DiMatteo_2007,Duc_2013} where tidally induced star formation seems to be triggered 
very soon after the closest approach \citep{Barton_2000,Holincheck_2016}.
Strong correlations have been found between the lopsidedness in the outer parts and the youth of the stellar 
population in the central regions \citep{Reichard_2009} and also between the proximity of a nearby neighbour 
and the average specific SFR \citep{Li_2008}.   

Gravitational interactions are expected to also play a role as trigger for E+A galaxies.
The time scale of the spectral signature of A stars from a PSB stellar population coincides with the time scale
of the appearance of strong tidal structures. 
The majority of the local PSB galaxy population have neither early- nor late-type galaxy morphologies
\citep{Wong_2012} where tidal features have been found in many cases
\citep[e.g.][]{Zabludoff_1996,Goto_2005,Yang_2008,Yamauchi_2008,Sell_2014}. 
Using the largest sample studied till then, \citet{Goto_2005} has investigated the environment of
266 E+A galaxies from scales of a typical distance of satellite galaxies to the scale of large-scale structure. 
He found that E+A galaxies have an excess of local galaxy density only at scales of dynamical interaction
with closely accompanying galaxies, but not at scales of galaxy clusters. 
About 30\% of the galaxies in this sample were found to have dynamically disturbed signatures of interactions or mergers.
In a sample of 21 E+A galaxies observed with the {\it HST} studied by \citet{Yang_2008}, at least 55\% were found to show dramatic tidal features indicative of mergers. Most of the galaxies from this sample lie in the field, well outside of rich clusters.
An even higher percentage (75\%) of galaxies with distorted morphology was found by \citet{Sell_2014} for a small sample of young PSB galaxies at $z \sim 0.6$.

While there is strong evidence that gravitational interactions trigger starbursts, the processes that lead to the quenching of star formation are poorly understood. The energetic output from an AGN triggered by a major merger is thought to be an effective quenching mechanism 
\citep[e.g.][]{Springel_2005,Hopkins_2006,Booth_2013}.  
The luminosity function of the PSB galaxy population seems to closely follow that of AGNs indicating a link
between starburst and AGN activity, where AGNs are difficult to detect either because of dust obscuration or AGN domination \citep{Bergvall_2016}. 
However, direct evidence for AGN-induced quenching is still sparse \citep{Heckman_2014}.
AGNs reside almost exclusively in massive galaxies, the fraction of galaxies with AGN strongly decreases at stellar masses below $10^{11} M_\odot$ \citep{2003MNRAS.346.1055K}. 
Also the quenching efficiency seems to show a strong trend with stellar mass and luminosity consistent with the energetic feedback from supernovae for log\,$M_\ast/M_\odot \la 10$, while the major effect may come from AGNs at higher masses \citep{Kaviraj_2007}.
Empirical evidence for AGNs in PSB galaxies is sparse. Based on the analysis of line ratios \citet{Yan_2006} suggested that most PSB galaxies may harbour AGN. Direct indications of AGNs in individual PSB galaxies was reported for a few cases only \citep{Liu_2007,Georgkakis_2008}. 
Other studies of samples of E+A galaxies did not confirm a substantial number of luminous AGNs. \citet{Swinbank_2012} found 20 - 40\% of their sample of 11 E+A galaxies to have 1.4 GHz radio emission suggestive of low-luminosity AGNs, but they concluded that AGNs do not play a dramatic role for the host galaxies, or the time scale of AGN feedback is short.   
\citet{De_Propris_2014} found that no E+A galaxy in their local sample hosts an AGN with substantial luminosity. 
\citet{Sell_2014} concluded that their sample of 12 young PSB galaxies at $z \sim 0.6$, selected from a larger parent sample as the most likely to host AGNs, represent massive merger remnants with high-velocity gaseous outflows primarily driven by compact starbursts rather than AGNs.
On the other hand, these studies do not conclusively rule out that AGNs may play a role in some point of the 
evolution and that the quenching of star formation and AGN activity rapidly follow each other 
\citep{Melnick_2015}.

The creation of sufficiently large samples is crucial for statistical studies of E+A galaxies but is difficult 
because of the rarity of this object type. The situation has become strongly improved with the availability of large spectroscopic surveys, particularly the Sloan Digital Sky Survey \citep[SDSS;][]{York_2000}. 
Large and homogeneous samples of E+A galaxies from the SDSS were derived particularly by 
\citet{Goto_2003},\citet{Goto_2005}, and \citet{Goto_2007a}.  
The tremendous amount of data produced by the SDSS requires efficient methods for browsing the huge archive. 
The search for narrowly defined spectral types in such a data set makes sophisticated tools desirable.  
Artificial neural network algorithms provide an efficient tool. We developed the software tool ASPECT to compute large Kohonen self-organising maps \citep[SOMs;][]{Kohonen_2001} for up to one million SDSS spectra \citep{inderAu_2012}. 
In previous studies, we computed SOMs for the quasar spectra of the SDSS data release seven \citep[DR7;][]{Abazajian_2009} 
to select unusual quasars \citep{Meusinger_2012,Meusinger_2014}. 
Thereafter we applied this technique to galaxies, stars, and unknowns in the SDSS DR7, 
as well as to the quasar spectra in the SDSS DR10 \citep{Ahn_2014} to extend the unusual quasar search \citep{Meusinger_2016}. SOMs computed from the SDSS DR12 \citep{Alam_2015} are currently being analysed.  

Here we aim to create a selection of E+A galaxies from SDSS DR7 based 
on the SOM technique. We demonstrate the applicability of a large SOM for such a task and
describe the newly developed user interface that allows for easy exploration of huge maps for example
for comfortable visual examination, projecting input catalogues, tagging and collecting single or several objects, 
etc. The SOM and the user interface are made fully available for the astronomical community.
The second aim of this study is to analyse the properties of our E+A sample, where the focus is on 
the position in the colour-mass plane, the morphological distortions, and possible indications of AGNs.
The paper is structured as follows. The data sources are described in Sect.\, 2. The SOM, the user interface, and the selection method are presented in Sect.\,3. In the following Sect.\, 4, the properties of our E+A sample are discussed. Finally, summary and conclusions are given in Sect.\, 5. The cosmological parameters 
$\Omega_{\rm M} = 0.27, \Omega_\Lambda = 0.73$, and $H_0 = 70$ km\,s$^{-1}$\,Mpc$^{-1}$
were used throughout this paper.

%
\section{Data sources}

%
\subsection{Sloan Digital Sky Survey}
\label{sec:SDSS}

The original core science goal of the SDSS  was obtaining CCD imaging over roughly a quarter of the high-Galactic latitude sky and spectroscopy of a million galaxies and quasars. 
The imaging part covers more than 10\,000 deg$^2$ in the five broad bands u,g,r,i, and z, mostly taken under good seeing conditions in moonless photometric nights. It includes a deep survey by repeated imaging
in the Stripe 82 (S82) area along the Celestial Equator. 
The SDSS data have been made public in a series of cumulative data releases. The SDSS DR7 \citep{Abazajian_2009}
includes all data accumulated up to the end of the phase known as SDSS-II that marks the completion of the original goals of the SDSS. The spectra database contains spectra of 930\,000 galaxies and 120\,000 quasars.  
The spectra were taken with the 2.5 m SDSS telescope at Apache Point Observatory equipped with a pair of double fibre-fed   spectrographs. 
The wavelength range covered by the SDSS spectra is 3\,800 \AA\ to 9\,200 \AA\ with a resolution of $\sim    2\,000$ and sampling of $\sim 2.4$ pixels per resolution element. 
For a galaxy near the main sample limit, the typical signal-to-noise ratio (S/N) is $\sim 10$ per pixel. 
The spectra are calibrated in wavelength and flux and classified by a spectroscopic pipeline, including redshift determination. 

The primary data set for the present study consists of a subset of about one million spectra downloaded from the SDSS DR7. 
Most of the spectroscopic data used in the present study where extracted from the SDSS fits files of the spectra downloaded from DR7. 
This includes in particular the equivalent widths (EWs) of the spectral lines. 
However, contrary to the definition in that database, we followed the convention in the literature to indicate the EWs of absorption lines by positive and those of emission lines by negative values. 
Other data like redshift $z$ and object classification were taken from the SDSS DR12 \citep{Alam_2015}.

In addition, morphological data from Galaxy Zoo are used. In Sect.\,\ref{sec:morphology},
we make use of the data from the first Galaxy Zoo project \citep{Lintott_2011}, where $\sim 900\,000$ galaxies were included. 
Results from the complex classification system applied in Galaxy Zoo 2 \citep[GZ2;][]{Willett_2013} for $\sim 300\,000$ SDSS galaxies are discussed in Sect.\,\ref{sec:merger}. 

Further, we exploited the database of galaxy properties from the Portsmouth stellarMassStarFormingPort\footnote{http://www.sdss.org/dr12/spectro/galaxy\_portsmouth/} (sMSP), which is 
available from the SDSS DR12\footnote{http://www.sdss.org/dr12/spectro/galaxy/}.
The Portsmouth galaxy property computations deliver stellar masses and other properties by applying stellar 
population models \citep{Maraston_2013} to all objects that the SDSS spectroscopic pipeline classifies as a 
galaxy with a reliable and positive definite redshift. The stellar population models were used to perform a 
best fit to the observed ugriz magnitudes with the spectroscopic redshift determined by the pipeline.  The 
fit was carried out on extinction corrected model magnitudes that were scaled to the i band for two sets of 
models, a passively evolving galaxy or a galaxy with active star formation. The stellar mass, the SFR, 
and the age were computed from the best-fit spectral energy distribution \citep[SED;][]{Maraston_2006,Maraston_2009}.  

In Sect.\,\ref{sec:merger}, we make use of the deep imaging in the SDSS S82 \citep[][see next Sect. below]{Fliri_2016}.

\begin{figure*}[htbp]
  \centering
  \includegraphics[width=0.99\hsize]{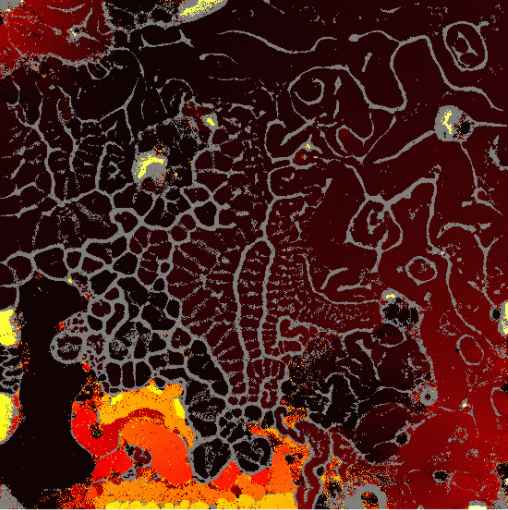}
  \vspace{0.5cm}
  \caption[Overview of the DR7 SOM]{Kohonen SOM of $\sim 10^6$ spectra from SDSS DR7 in its representation as a redshift map . Each pixel corresponds to one spectrum where the colour represents the redshift $z$ from 0 (dark) to 6 (bright yellow). The grey dots are empty neurons. (Resolution reduced in order to match the image size limitation of arXiv.)}
  \label{fig:overview of DR7 SOM}
\end{figure*}

%
\subsection{SDSS S82}
\label{sec:S82}

The SDSS S82 is the  275 deg$^2$ region of sky along the Celestial Equator 
in the southern Galactic cap at $\alpha = -50\degr \ldots 60\degr, \delta = -1\fdg25 \ldots +1\fdg25$.
Because of the combination of a high completeness level of SDSS spectroscopy and deep co-added images, this area is particularly attractive for the present study. 
In addition, the SDSS data in S82 are complemented by a broad multi-wavelength  coverage by existing and planned wide-field surveys. 
A second-epoch 1.4 GHz survey of S82, conducted with the Very Large Array, achieves an angular resolution of $1\farcs8$ and a median rms noise of 52 $\mu$Jy~beam$^{-1}$ over 92~deg$^2$ \citep{Hodge_2011}. 
Moreover, a wide-area X-ray survey endeavouring to achieve a survey area of $\sim 100$ deg$^2$ in S82
\citep{LaMassa_2013,LaMassa_2016} has detected 6\,181 unique X-ray sources so far.

Co-adding the multi-epoch observations in S82 leads to the construction of images that are considerably deeper than typical observations from the SDSS legacy survey.  
Co-added images were made available in the Data Archive Server (DAS) of a database called {\tt Stripe82} as part of the SDSS DR7 \citep{Abazajian_2009}.    
\citet{Annis_2014} combined about one third of all available SDSS runs in S82 to co-adds that are $\sim 1-2$ mag deeper than the regular SDSS images, depending on the band. 
In another approach, \citet{Jiang_2008} combined between 75 and 90\% of the data and reported that the resulting images are 0.3-0.5 mag deeper than the previous co-adds. 
A new reduction of the S82 data was provided recently by \citet{Fliri_2016}. Compared to the previous studies, these new images focus on the surface brightness depth rather than on faint point sources. 
The main intention was to prevent the destruction of low-surface brightness features in the process of co-addition by an optimal local sky brightness correction and to reduce the probability of confusing low-surface brightness features of galaxies and sky background.
Averaging the $g, r$ and $i$ co-adds yields another gain in depth by $\sim 0.2 \ldots 0.3$ mag in the so-called $r_{\rm deep}$ images.  
The co-adds reach $3\sigma$ surface brightness limits $\mu_{r} \sim 28.5$ mag arcsec$^{-2}$ with 50 \% completeness limits at (25, 26, 25.5, 25, 24) mag arcsec$^{-2}$ for $(u,g,r,i,z)$.

%
\subsection{Self-organising map of the SDSS DR7 spectra}
\label{sec:SOM}

In a previous paper \citep{inderAu_2012} we described the software tool ASPECT\footnote{http://www.tls-tautenburg.de/TLS/fileadmin/research/meus/ASPECT/ ASPECT.html}
that was developed to organise a large number of spectra by means of  their relative similarity in a topological map. 
Similarity maps are generated using self-organising maps (SOMs) as proposed by \citet{Kohonen_2001}. 
The SOM technique is an artificial neural network algorithm that uses unsupervised learning. 
The network consists of neurons represented by weight vectors, where the number $n$ of neurons must be at least equal to the number $k$ of source spectra. 
We found that good results are achieved for $n/k \sim 1.2$, that is about 20\% of the neurons are empty (i.e. not linked to spectra).

ASPECT maps spectra (dis-)similarity to position in the resulting SOM. 
For the bulk of the SDSS sources, the spectral properties vary more or less smoothly over the SOM. 
The spectra thus form coherent areas interspersed with small areas of `no men's land' that are often occupied by rare spectral types with pronounced spectral properties. 
Uncommon spectra often cluster to smaller structures between large coherent areas where the latter are occupied by the more common objects. 
For extragalactic objects, the shape of the observed spectra and the observed wavelengths of the characteristic spectral features change of course with redshift $z$ so that the appearance of several clusters for different $z$ intervals is natural. 
The very fact of such clustering properties makes the SOMs useful for efficiently selecting uncommon or rare objects from large data samples.

\begin{figure}[htbp]
\centering
\setlength\fboxsep{5pt}
\includegraphics[width=0.32\hsize]{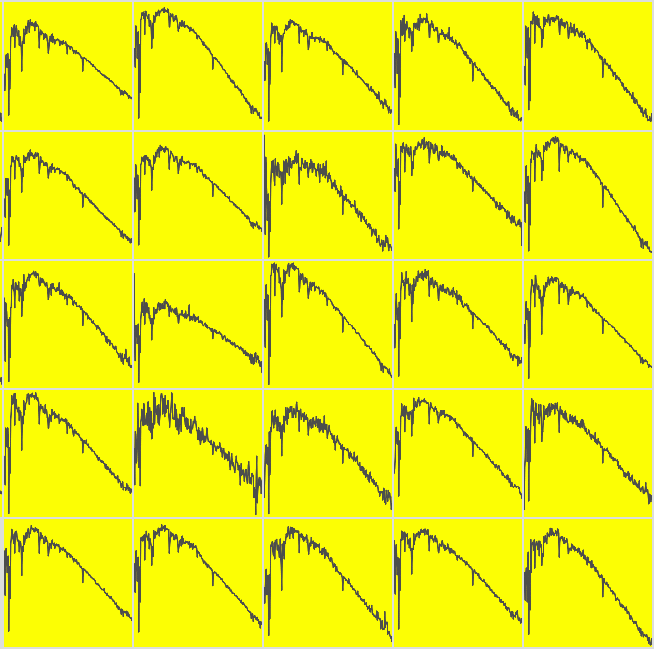}
\includegraphics[width=0.32\hsize]{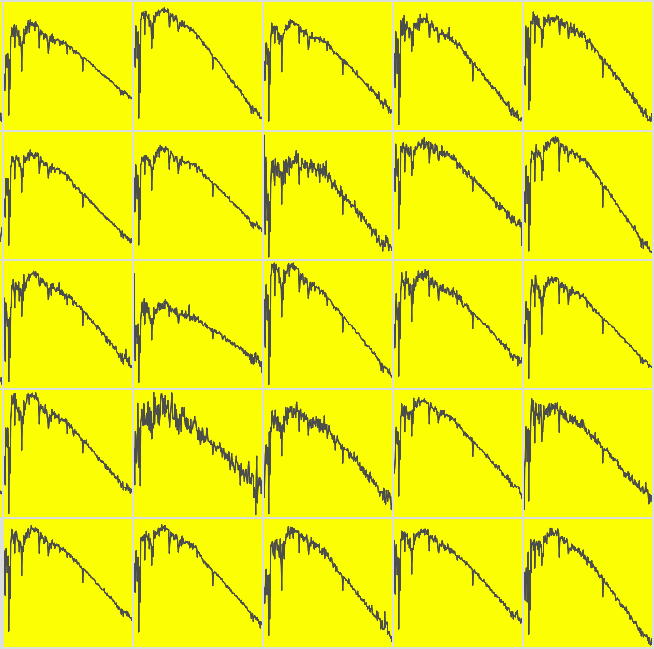}
\includegraphics[width=0.32\hsize]{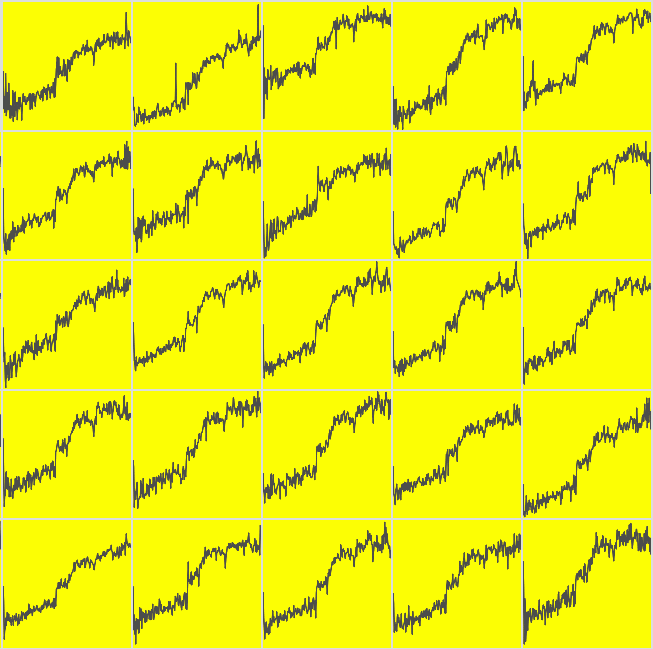}\\
\includegraphics[width=0.32\hsize]{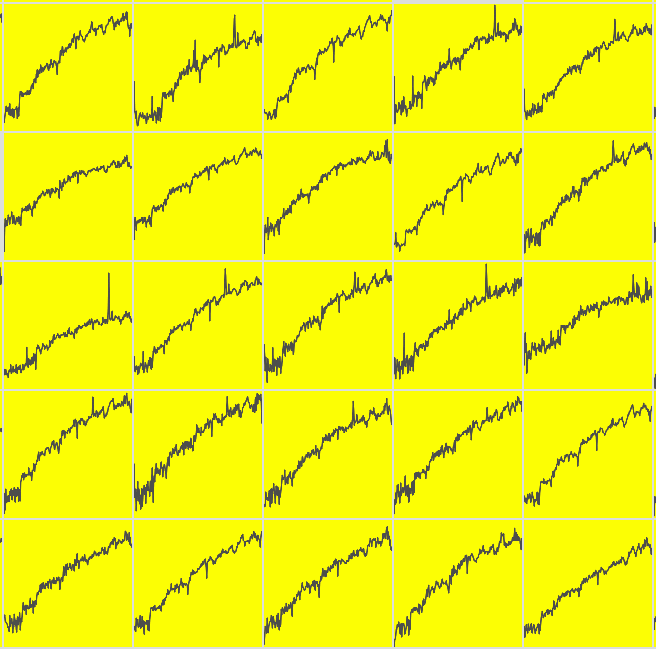}
\includegraphics[width=0.32\hsize]{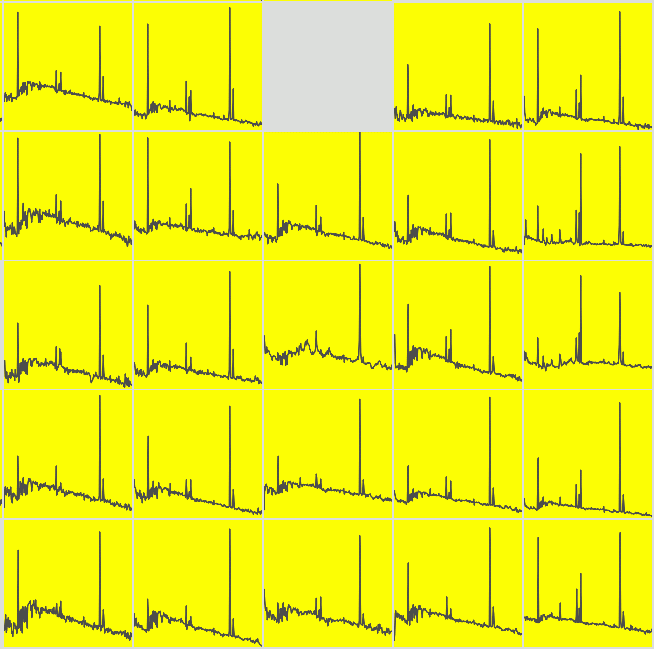}
\includegraphics[width=0.32\hsize]{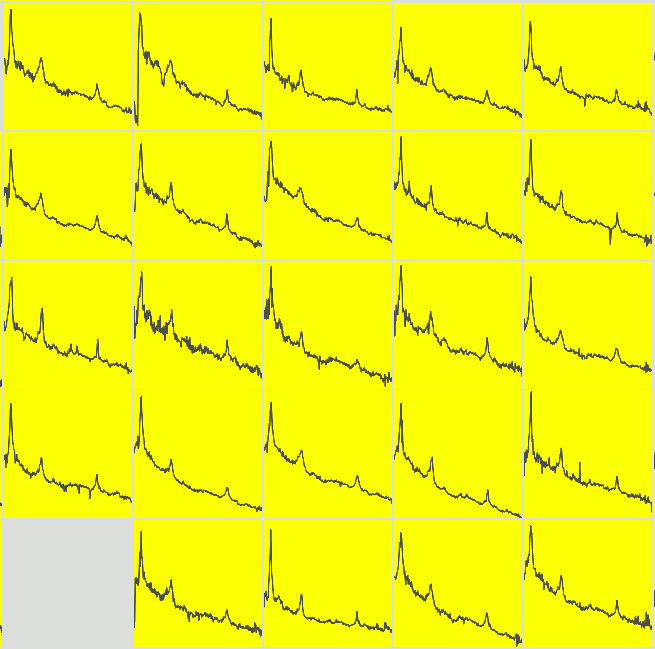}
\vspace{0.3cm}
\caption[Spectra in different parts of the SOM]{Six cutouts from different
parts of the icon map, each five by five pixels in size}
\label{fig:kohonen map spectra}
\end{figure}

In the previous paper \citep{inderAu_2012}, we discussed a SOM of $\sim 5\cdot10^5$ spectra from the SDSS DR4.
Afterwards, we computed and exploited a large number of smaller SOMs (several $10^4$ spectra each) for different data sets of stars and $z$-binned quasars and galaxies from DR7 to DR12 with the main aim to search for different types of unusual quasars \citep{Meusinger_2012,Meusinger_2016,Meusinger_2014}. 
However, ASPECT was developed in particular to compute SOMs of sizes that existing implementations of the algorithm where not able to cope with. Here we present, for the first time, a SOM of about one million SDSS spectra. 
The SOM contains all useful spectra from the SDSS DR7 that were available for download, regardless of spectral type, redshift, or signal-to-noise ratio (S/N).
The procedure and the parameters of the neural network are essentially the same as in \citet{inderAu_2012}, the computation time was about eight months on a state-of-the-art personal computer. 

Figure~\ref{fig:overview of DR7 SOM} shows a low-resolution image of the whole map. Each pixel corresponds to a neuron of the SOM. 
Colours indicate redshifts, the grey filamentary network represents empty neurons.  
In the resulting SOM, these empty neurons tend to settle at the borders of the coherently populated areas thus enhancing the clustering power of the method.
Figure~\ref{fig:kohonen map spectra} shows six cutouts from different regions of the icon map of the same SOM.
It illustrates how the algorithm implemented in ASPECT clusters spectra of the same type.
The icon map is a representation of the SOM where each pixel is represented by the SDSS spectrum at low spectral resolution (icon).

%
\subsection{SOM user interface}
\label{sec:Kohonen Map user interface}

The selection of E+A galaxies presented below (Sect.\,\ref{sec:Methods}) is based on the huge SOM of roughly one million spectra from SDSS DR7. 
The detailed analysis of such a large map is challenging. 
Several types of representations of the SOMs computed by ASPECT (e.g. U matrix, $z$ map, type map, icon map) were discussed by \citet{inderAu_2012}, in combination with different methods of their analysis. 
The present study is focused on the selection of spectra from the icon map by means of an input catalogue.

ASPECT saves the computed SOM to a HTML document that can be viewed in its icon map version using a web browser. 
The output is internally structured as HTML tables containing the spectra icons. 
Rendering these tables results in a representation as a map of sorted spectra that can be explored using standard scrolling and zooming functionality of the web browser. 
While this is a practical method for smaller maps with some thousand spectra it becomes a straining user experience when the browser has to render a whole huge SOM at once and to keep it in memory.
The latter becomes nearly impossible for SOMs consisting of several $10^5$ spectra because of technical restrictions.

We developed the ASPECT user interface (AUI) for the efficient work with even very large icon maps.
The AUI is publicly available together with the SDSS DR7 icon map\footnote{http://aspect-ui.de/sdssdr7/}.
So far, it is focused on the following features. Firstly, the map must be easily zoomable
for a convenient work flow, that is it should provide representations of the SOM in different detail levels. 
Secondly, for the selection of objects from the icon map, specific data from SDSS or other sources can be very helpful.
Therefore, it should be possible to overlay additional information. 
Thirdly, to compile lists of objects selected from the icon map, it should be 
possible to highlight already selected spectra and to tag spectra for later data export. 

\begin{figure}[htbp]
\centering
\includegraphics[width=\hsize]{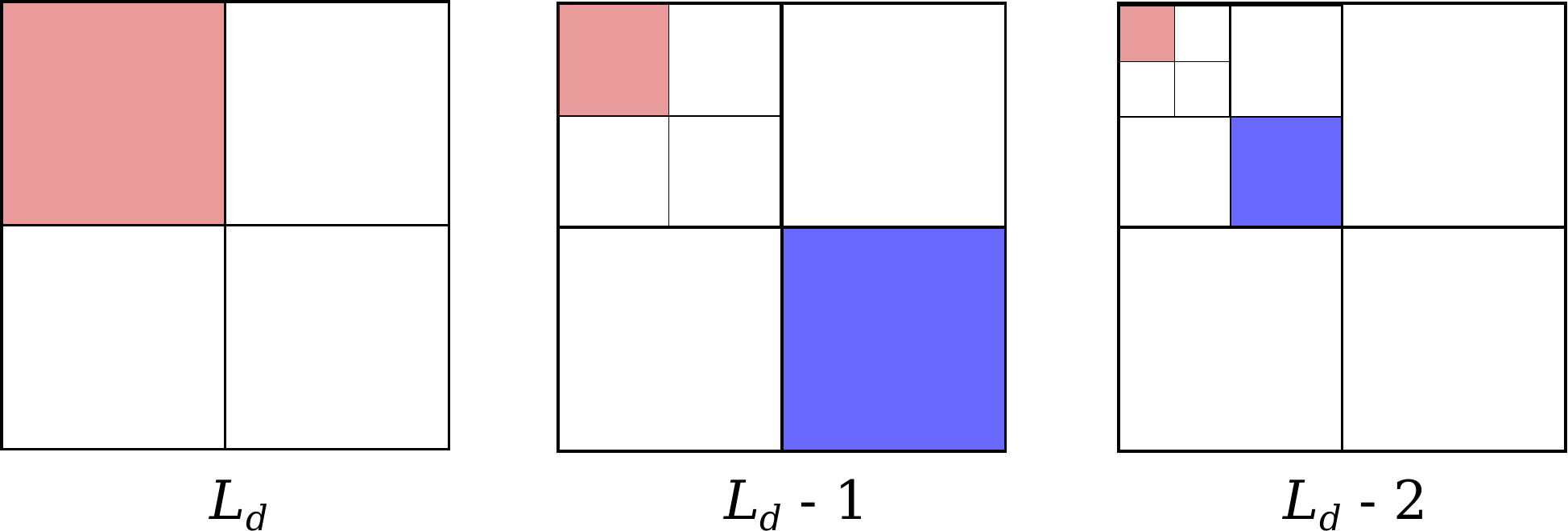}
\caption[Schematic of map tool zoom levels]{Schematic illustration of two lower-detailed levels derived from an
existing detail level $L_{\mathrm d}$}
\label{fig:generation of lower zoom levels}
\end{figure}

We applied modern web techniques and the framework {\tt leaflet.js}\footnote{http://leafletjs.com} to add data layers and interactivity to the rendered SOM. 
The original intention behind leaflet.js was to support development of online street maps. 
However, the principle of aligning smaller images to compile a large map is very well applicable to our use case. 
Speaking in terms of {\tt leaflet.js} we organise the spectra icons in adjacent tiles, each icon occupying one tile in the highest detail level. 
In order to have the map available for different zoom levels it is necessary to rescale and re-size the spectra plots computed by ASPECT. 
To this end we start with the static ASPECT output as the highest zoom respective detail level $L_d$, meaning that one spectra icon fills one tile. 
The composition of tiles of the next lower detail level $L_d -1$ is achieved by combining four tiles of $L_d$ into one new tile. 
In every such re-size step the number of tiles is divided by four compared to the predecessor zoom level. 
This process has to be repeated with lower detail levels as needed until all spectra are re-sized and rescaled into one remaining image tile at the lowest detail level. 
Figure~\ref{fig:generation of lower zoom levels} schematically illustrates the process. In the field of computational graphics, the approach to build such image pyramids is known as mipmapping. It was first described 
by \citet{Williams_1983}.

The application of the previously described steps to the icon map of about one million spectra provided us with the necessary tool for the search for E+A galaxies. 
We extended our basic map view with several features that can be toggled on and off by adding or removing layers to the view to prevent information overload (see also the AUI homepage for a live demonstration):

Firstly, map controls, such as a zoom level switcher, a data layer selectors, and marker tools are available. 

Secondly, additional data from the SDSS DR7 spectra FITS files, such as the object type or the equivalent widths for 
H$\alpha$ and H$\delta$, can be over plotted in each icon.

Thirdly, a click on an icon provides the SOM coordinates and hyperlinks to the SDSS explorer homepage of the object and, alternatively, to the redshift tool {\tt zshift}, which 
was inspired by the interactive spectra tool from the SDSS DR12 Science Archive Server and is used here mainly for manually checking the redshift in case of a doubtful result from the SDSS spectroscopic pipeline.

Furthermore, several spectra icons within an area can be marked and tagged for later export. 
This works like rectangular selection tools in common graphics software. 
The list of selected objects can be exported to a csv file containing the SOM coordinates and SDSS identifiers (plate, MJD, fibre ID). A SOM pixel (icon) marked previously with the selection tool is indicated by a coloured margin. 
Layers of different colours and tag names can be created in a separate menu.

Finally, the AUI provides the opportunity for the import of an input catalogue. Given their presence in the SOM, objects from the input catalogue (again identified by MJD, plate ID, fibre ID) can be mapped to and highlighted in the icon map.

Depending on requirements and resources the overlay and tagging information can be stored in more or less sophisticated storage back-ends. 
For the present application, we extracted most of the additional data from the FITS files of the SDSS spectra and stored them in a database management system. 
We chose an SQL database server as storage back-end. 
The spectra were stored as image files in plain file system.
Additional information was transferred from the SDSS DR7 1d spectra database\footnote{http://classic.sdss.org/dr7/dm/flatFiles/spSpec.html}, in particular the equivalent widths (EW) 
of spectral lines derived by the SDSS pipeline. 

In the previous paper \citep{inderAu_2012}, we demonstrated that the use of an input catalogue of known objects of a given
spectral type can be very useful for an efficient search of further objects of the same or similar types in a large SOM. 
Here, we illustrate this approach by another example.

%
\section{Selecting E+A galaxies}
\label{sec:Methods}

%
\subsection{Input catalogue}
\label{sec:Input}
The most comprehensive compilation of E+A galaxies was performed by
Tomotsugu Goto, published in several updates. \citet{Goto_2003} presented a catalogue of galaxies with strong H$\delta$ absorption from SDSS DR1. 
In the following, \citet{Goto_2005} provided a list of 266 E+A galaxies, picked from the SDSS DR2.
Subsequently, that number was roughly doubled with an update after SDSS DR5 \citep{Goto_2007a}
and was eventually increased again after the SDSS DR7. 
Goto's latest catalogue\footnote{http://www.phys.nthu.edu.tw/~tomo/research/ea\_dr7/}
compiles 837 E+A galaxies found in the SDSS DR7.  
The criteria for objects to qualify for the catalogue are 
EW(H$\delta$) $> 5$ \AA, EW(H$\alpha$) $>-3$ \AA, EW([\ion{O}{ii}])$> -2.5$ \AA,
and redshift not in the range $0.35<z<0.37$ in order to exclude intervening sky lines.
In the following, the objects from this catalogue will be referred to as `Goto galaxies'.
We identified Goto galaxies in our SOM by the plate--MJD--fiberID combination using the
skyserver\footnote{http://skyserver.sdss.org} links provided in the catalogue. 
All entries of the input catalogue could be mapped to spectra in our data base.
The arithmetic median composite rest-frame spectrum of the Goto sample is shown in Fig.\,\ref{fig:composites}\,b.

%
\subsection{Selection of new candidates}
\label{sec:goto galaxy clustering}

\begin{figure}[htbp]
\centering
\includegraphics[width=0.95\columnwidth]{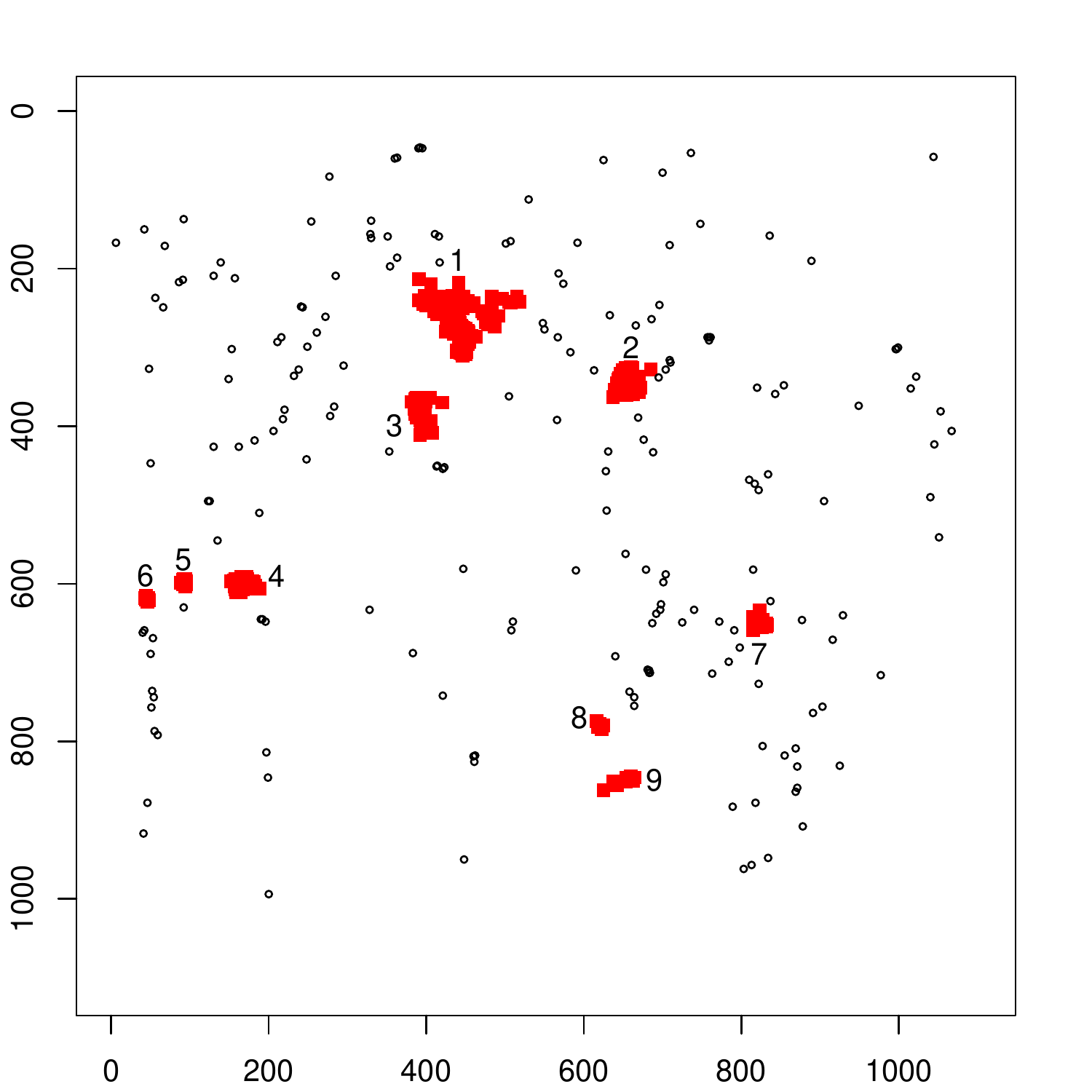}\\
\vspace{0.3cm}
\caption[Distribution of Goto galaxies in the SOM]
{Distribution of Goto galaxies (black open circles) over the SOM. 
The clusters identified by {\tt dbscan} clustering are shown as red squares and labelled
by the cluster ID. The axis are the pixel coordinates of the SOM.
}
\label{fig:goto_galaxies_in_SOM}
\end{figure}

The input catalogue is used as seed for the search for further E+A galaxies in the DR7 SOM.
Figure~\ref{fig:goto_galaxies_in_SOM} shows the distribution of the input galaxies over the map. 
The labelling of the axis indicates the coordinate system of the SOM with 1104 by 1104 neurons. Every black dot indicates the
position of a galaxy from the input catalogue, larger red symbols mark concentrations, referred to as `clusters' throughout this paper. 
On the one hand, the clustering is an eye-catching feature of the SOM. On the other hand, it is remarkable that E+A galaxies do not form a single cluster, a substantial part of the input sample is scattered across the map.

As a first step, we had to find out which of the aggregations of input galaxies could be considered to be clusters. 
In the next step, each cluster should be used to define an environment where the probability is high for finding further galaxies with similar spectral properties. 
We intended to inspect all spectra within the resulting areas for E+A features. To define a cluster, we applied the data clustering tool {\tt dbscan} 
\citep[density-based spatial clustering of applications with noise;][]{Ester_1996} in its implementation for the statistics software
R\footnote{https://www.r-project.org/}.

{\tt dbscan} is a commonly used clustering algorithm. Compared to simpler algorithms, the advantage is its ability to locate clusters of arbitrary shape. 
The basic principle is to find aggregations of a defined minimum number of objects with defined maximum distances and to recognise them as clusters:
Let $M \in \mathbb{N}$ and $\epsilon \in \mathbb{R}$ be the input parameters 
\citep[named $MinPts$ and $Eps$ respectively in][]{Ester_1996}. 
$M$ defines the minimum number of members constituting a cluster, whereas $\epsilon$ is the radius of an epsilon ball $N_{\epsilon}$ around points in the cluster space. 
Using these input parameters, {\tt dbscan} clusters objects in a given map by their relative distance and categorises them as directly density reachable, density reachable, density connected, and neither density reachable nor density connected.

To approach the problem of Goto galaxy clusters in the SOM, we assumed a two-dimensional grid.
Assuming that a subset of $n$ grid points is occupied by the objects $O_1, O_2, \ldots, O_n$ with the corresponding   two-dimensional position vectors $\vec{o_1}, \vec{o_2}, \ldots, \vec{o_n}$, we define for two objects $O_k$ and $O_l$ that
$O_k$ is directly density reachable from $O_l$ if $\vec{o_k}$ lies within
$N_{\epsilon}$ around $\vec{o_l}$ and the number of objects within this epsilon ball is greater or equal to $M$.

$O_k$ is density reachable from $O_l$ if there is a chain of grid points occupied by the elements of a subset $\{P_i\} \subset \{O_1,O_2 \dots O_n\}, i = 1, \dots, q$ connecting $O_k$ and $O_l$ such that $\forall i: P_{i+1}$ is directly density reachable from $P_i$ and $O_k = P_1$ and $O_l = P_q$. 

$O_k$ is density connected to $O_l$ if there is a grid point $\vec{g}$ such that $O_k$ and $O_l$ are both density reachable from an object at $\vec{g}$.

These definitions allow us to define a cluster as the set of density connected objects with maximised density reachability with respect to the parameters $M$ and $\epsilon$, which have to be fixed in advance \citep{Ester_1996}.
The Euclidean distance was used as distance function.
 
We ran {\tt dbscan} with a set of various combinations of $M$ and $\epsilon$.
The analysis of the results led to the conclusion that solutions with 8 to 12 clusters were meaningful. 
Solutions with less clusters missed some of the clearly visible accumulations. 
On the other hand, solutions with more than 12 clusters assigned the cluster status 
to very small groups merely scattered throughout the map, or even to single galaxies.
Considering the fraction of objects bound in clusters and the visual appearance of the solutions' plots, 
we finally chose the parameter values $\epsilon=18$ and $M=5$. The {\tt dbscan} run then resulted in
nine clusters of altogether 645 objects from the input catalogue (Fig.~\ref{fig:goto_galaxies_in_SOM}).
Another 192 galaxies were found to be scattered across the SOM and were not assigned to a cluster.
Following \citet{Ester_1996} we refer to the latter as `noise' in this context.

\begin{table}[h]\centering
\caption{Mean properties of the galaxies from the input catalogue clusters 1 to 9 and noise.}
\begin{tabular}{ccccc}
\hline\hline
C     &  $z$          &EW(H$\delta$)& $f_{\mathrm e/s}$ &S/N \\
\hline
1     &$0.144\pm0.044$&$6.60\pm1.16$&2.85               &$16.7\pm4.38$\\
2     &$0.209\pm0.036$&$6.66\pm1.14$&4.18               &$16.6\pm4.25$\\
3     &$0.088\pm0.048$&$6.32\pm1.22$&2.56               &$17.2\pm4.59$\\
4     &$0.070\pm0.037$&$8.81\pm13.3$&1.61               &$17.1\pm4.83$\\
5     &$0.104\pm0.031$&$6.82\pm1.13$&1.52               &$17.1\pm4.84$\\
6     &$0.069\pm0.020$&$6.10\pm0.79$&2.38               &$17.2\pm4.91$\\
7     &$0.283\pm0.058$&$7.61\pm1.97$&3.92               &$16.6\pm4.73$\\
8     &$0.225\pm0.058$&$7.38\pm1.29$&4.72               &$16.9\pm4.72$\\
9     &$0.052\pm0.029$&$5.01\pm1.56$&1.36               &$16.8\pm4.96$\\
noise &$0.131\pm0.086$&$5.72\pm1.96$&1.08               &$12.4\pm6.76$\\
\hline
\end{tabular}
\label{table:Goto_clusters_1}
\end{table}

Table~\ref{table:Goto_clusters_1} lists mean properties of the seed galaxies from the input catalogue in the nine 
clusters and the noise:
the sample-averaged redshift, the mean EW(H$\delta$), and the mean ratio $f_{\mathrm e/s} = P_{\mathrm e}/P_{\mathrm s}$, 
where $P_{\mathrm e}$ and $P_{\mathrm s}$ are the probabilities for being an elliptical galaxy or a spiral galaxy,    
respectively, from  the Galaxy Zoo project \citep[][see Sect.\,\ref{sec:morphology}]{Lintott_2011}. 
The three largest clusters correspond to three different $z$ intervals, though there is some overlap. 
The noise contains galaxies from all redshifts.  
As expected, the ratio $f_{\mathrm e/s}$ increases with $z$ due to the Malmquist bias.
The S/N was measured in the continuum around H$\delta$ at rest-frame wavelengths $\lambda = 4030 - 4080$ \AA\ and $\lambda = 4122 - 4170$ \AA.

\begin{figure}[htbp]
\begin{tabbing}
\fbox{\includegraphics[width=4.00cm]{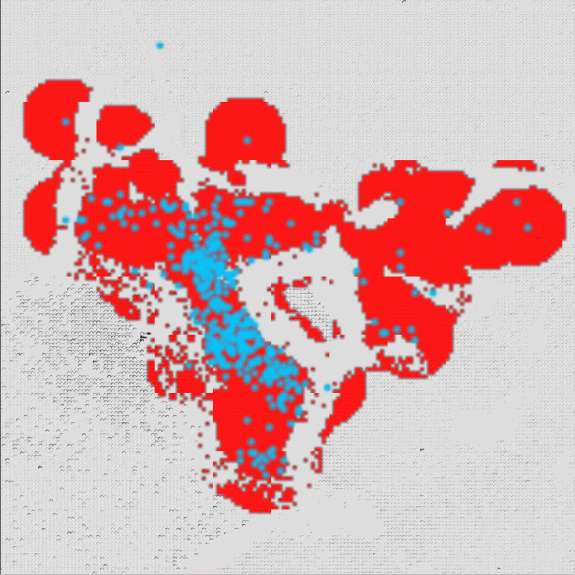}}\hfill \=
\fbox{\includegraphics[width=4.00cm]{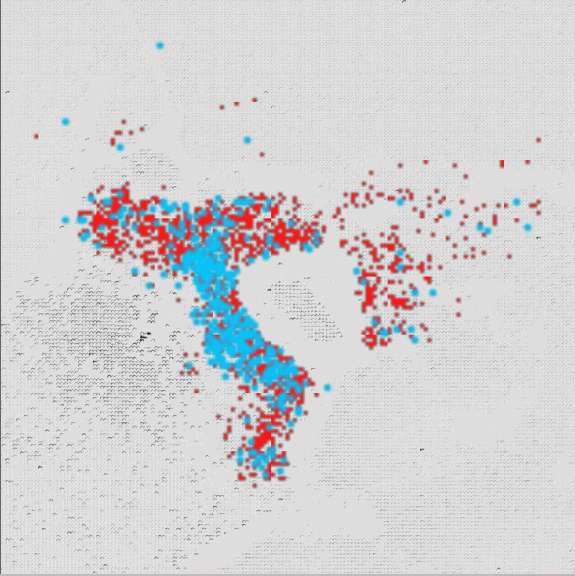}} \\ \\
\fbox{\includegraphics[width=4.00cm]{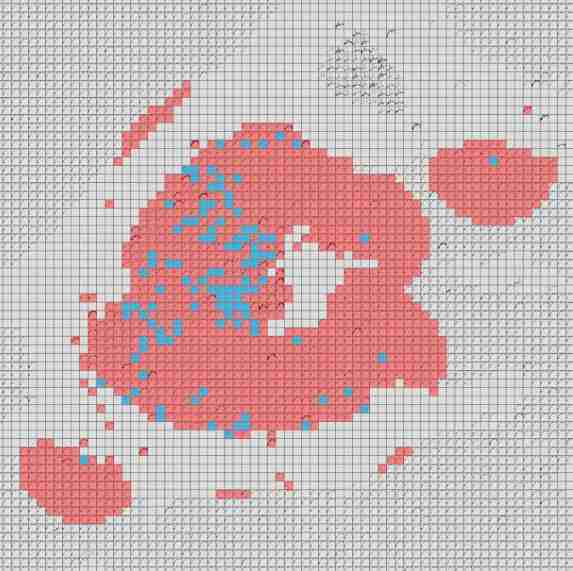}}\hfill \=
\fbox{\includegraphics[width=4.00cm]{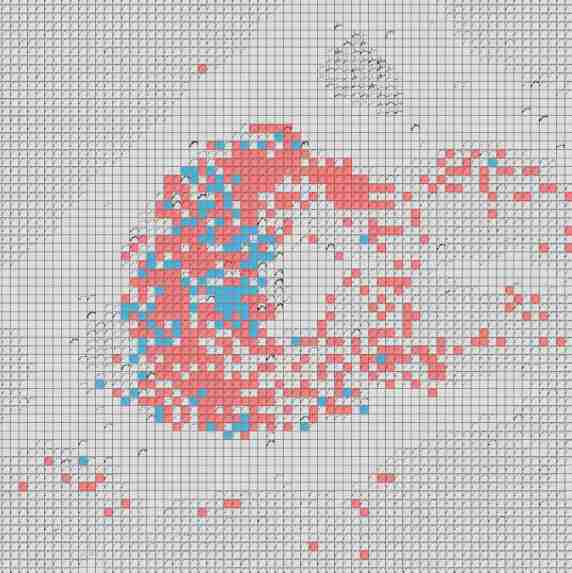}} \\ \\
\fbox{\includegraphics[width=4.00cm]{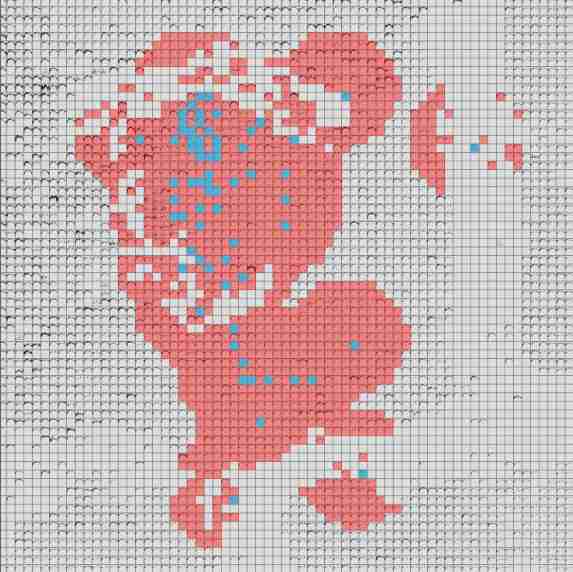}}\hfill \=
\fbox{\includegraphics[width=4.00cm]{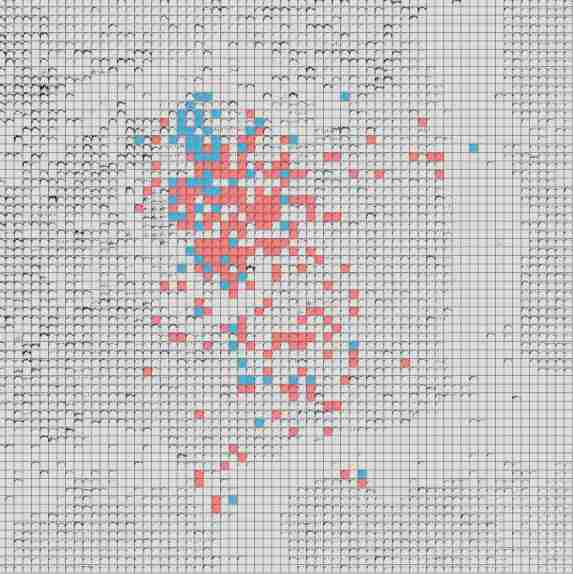}} \\ \\
\fbox{\includegraphics[width=4.00cm]{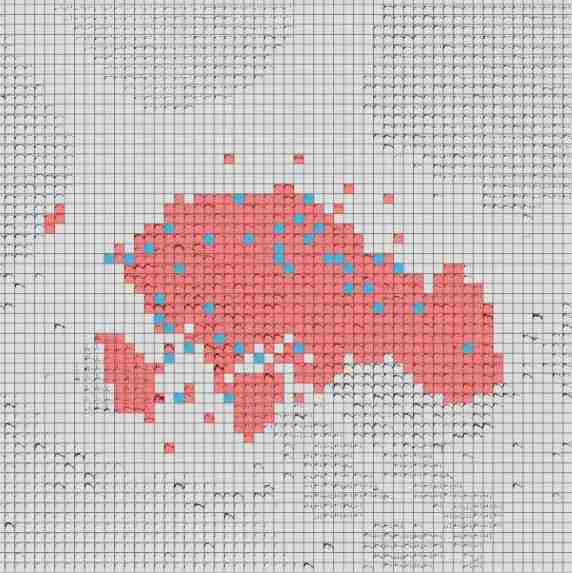}}\hfill \=
\fbox{\includegraphics[width=4.00cm]{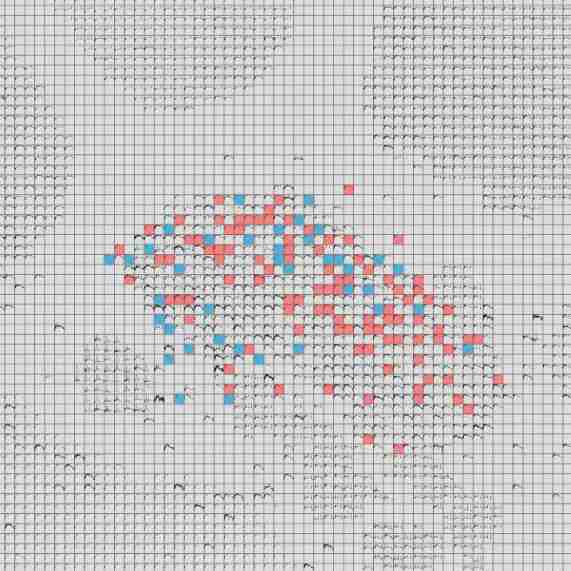}} \\ \\
\fbox{\includegraphics[width=4.00cm]{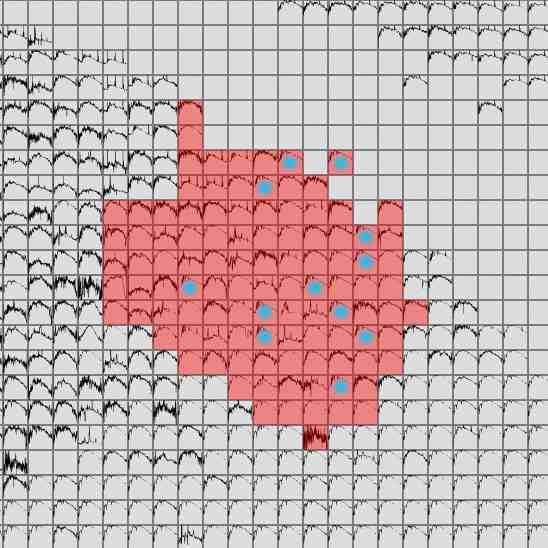}}\hfill \=
\fbox{\includegraphics[width=4.00cm]{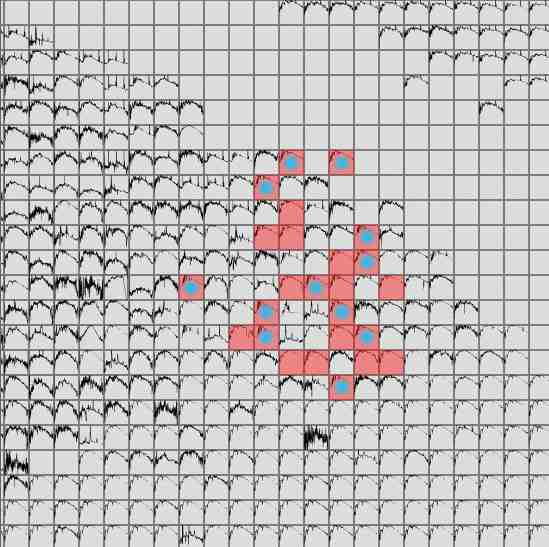}}
\end{tabbing}
\caption{Cutouts from the icon map containing the E+A clusters 1 to 5 (top to bottom; first selection on the left side, final selection on the right side). Galaxies from the input catalogue are marked blue. The red background colour indicates the newly selected E+A galaxies. }
\label{fig:clusters}
\end{figure}

The next step was the eyeball examination of the neighbourhood of the Goto clusters.
The properties of Kohonen maps imply that it is likely to find there more objects with similar spectra. 
To keep the effort at a manageable level we restricted the search area in the following way.
Assuming that
$\mathcal{O}$ is the set of objects $O_i$ that do not belong to any cluster and $\vec{o_i}$ 
are their corresponding position vectors in the map,
$\mathcal{C}_k$ is the set of the $n$ objects that belong to cluster $k$ and $\vec{c}_{k,l}$ are their corresponding vectors ($k=1 \ldots 9, l=1 \dots n$),
and \ $D_k = \max\left(\left\Vert \vec{c}_{k,p} - \vec{c}_{k,q} \right\Vert\right)_{p,q = 1 \dots n}$ is the largest distance between any two members of cluster $k$,
we defined the neighbourhood $\mathcal{N}_k \subset \mathcal{O}$ of cluster $k$ as the subset of all objects $O_i$ with $\left\Vert \vec{c}_{k,l} - \vec{o}_i \right\Vert \leq \sqrt{D_k}$ for at least one value of $l$. 
The resulting set $\mathcal{N}_k$ takes into account the varying sizes of clusters but prevents too large samples. 
Nevertheless, the joint set of all clusters and their neighbourhoods comprised a still fair number of 14\,828 
E+A galaxy candidates that had to be examined `manually'.
The panels on the left hand side of Fig.~\ref{fig:clusters} show the cutouts from the icon map for the clusters 1 to 5. 
The seed of input galaxies is marked blue, the cluster neighbourhood is red.

%
\subsection{Final selection}
\label{sec:final selection}

After the sampling of E+A galaxy candidates described above, we added a coloured overlay to
the Kohonen map indicating all objects from the input catalogue in blue and all objects added to the sample in the steps described in Sect.~\ref{sec:goto galaxy clustering} in red. 
Afterwards we made use of the tool set described in Sect.~\ref{sec:Kohonen Map user interface} to sort out contaminant spectra.
Roughly the following routine was applied to any single spectrum in the red area:
Firstly, are there spectral features indicative of E+A galaxies?
Secondly, exclude stars that can have spectra more or less similar to PSB galaxies.
(In fact, galaxies and stars are mixed in some areas of the SOM.)
Thirdly, a strong H$\alpha$ emission line is usually easy to spot. Is  EW(H$\alpha$) $\ga -3$?
If yes, it can be a candidate.
Next, click on the spectrum for more information. 
Is Balmer absorption dominant? 
Is the [\ion{O}{ii}] line weak in emission?
Check for H$\delta$ absorption. Is EW(H$\delta$) $\ga 3$?
Finally, if in doubt, open the link to the SDSS DR12 sky server and inspect the original spectrum.

The outcome of the `manual' selection and rejection process for each of the nine clusters from Fig.~\ref{fig:goto_galaxies_in_SOM} is the following:

Cluster~1: This is the biggest aggregation of input galaxies. 
Although a substantial number of further E+A galaxies were found, most had to be excluded. The shape of the final cluster resembles structures of the underlying SOM very well. 
What remains after selection can be seen as two distinct clusters. (Another $\epsilon$ value could probably have led to less work.)

Cluster~2: With an exception of a small area in the lower left region this cluster fits an island in the Kohonen landscape very well. Here we see that `ditches' of empty neurons are congruent with E+A cluster borders. 
The final relative outcome is quite large compared to the clusters 3 through 9.

Cluster~3: Compared to the most other clusters, the area occupied by the outcome for this cluster is less clearly constrained by the Kohonen map landscape.

Cluster~4: This cluster provides a rather small outcome. However it is well observable that the E+A galaxies strongly cluster in the SOM.

Cluster~5: The small distances between the seed objects led to a small cluster neighbourhood, and
only a few of the pre-selected candidates remained after closer examination.

Cluster~6: The outcome is similar to that of cluster 5, which also lies in the same region of the SOM.

Cluster~7: The large $M$ and small $\epsilon$ parameters resulted in a
relatively large cluster and included a large neighbourhood compared to the actual seed size.

Cluster~8: This cluster can be seen as a false positive.
It has the minimum number of Goto galaxies as seed to meet the {\tt dbscan} $\epsilon$ parameter.
After examination the Goto galaxies and only one additional E+A galaxy were left.

Cluster~9: After detailed inspection of the preselected galaxies, almost nothing remained.

\begin{table}[h]\centering
\caption{Result of {\tt dbscan} clustering of the galaxies from the input catalogue, number of rejected galaxies, and
final numbers per cluster and noise.}
\begin{tabular}{crrrrc}
\hline\hline
C     &         \multicolumn{5}{c}{number}\\
\cline{2-6}
      &$n_i$      &$n_{\rm D}$&$n_{\rm r}$&$n_{\rm f}$&$(n_{\rm f}-n_i)/n_i$\\
\hline
1     &       376 &      8569 &      7474 &      1471 & 2.9\\
2     &       110 &      1977 &      1503 &       584 & 4.3\\
3     &        67 &      1682 &      1487 &       262 & 2.9\\
4     &        38 &       961 &       871 &       233 & 2.5\\
5     &        12 &       124 &       111 &        25 & 1.1\\
6     &        12 &        83 &        72 &        23 & 0.9\\
7     &        16 &       584 &       431 &       169 & 9.6\\
8     &         5 &       163 &       162 &         6 & 0.2\\
9     &         9 &       685 &       655 &        39 & 3.3\\
noise &       192 &       161 &        -  &       353 & 0.8\\
\hline
\end{tabular}
\tablefoot{
$n_{\rm i}$: number of galaxies from the input catalogue; 
$n_{\rm D}$: added by the $\sqrt{D_k}$ criterion (except for noise);
$n_{\rm r}$: removed manually;
$n_{\rm f}$: final number.
}
\label{table:Goto_clusters_2}
\end{table}

The results of the visual inspection of clusters 1 to 5 are illustrated in the panels on the right hand side of
Fig.~\ref{fig:clusters} for the clusters 1 to 5. As in the panels on the left hand side,
galaxies from the input catalogue are marked blue, selected E+A galaxies from the present study are marked red. 
The clusters are shown at different zoom levels because of the huge differences in cluster sizes. 
It is clearly visible that the neighbourhood of at least some Goto clusters includes a substantial number of further E+A galaxies that were not selected as such before. 
Table\,\ref{table:Goto_clusters_2} lists the corresponding numbers for the nine clusters and the noise: galaxies from the input catalogue, selected by the $\sqrt{D_k}$ criterion, manually rejected, and final number.

%
\subsection{Selection effects}
\label{sec:Selection_effects}

Selection effects are induced by various processes, mainly the target selection, plate definition, and fibre spectroscopy in the SDSS \citep{Stoughton_2002}, the definition of the input sample of Goto galaxies, the selection from the SOM, and the use of the EWs from the SDSS spectroscopic pipeline for the final selection. The biases in the final E+A sample are complex and a quantitative description is barely achievable. Here, we focus on the effects from the SDSS and from the SOMs

\begin{figure}[h]
\includegraphics[viewport= 50 20 460 780,width=4.7cm,angle=270]{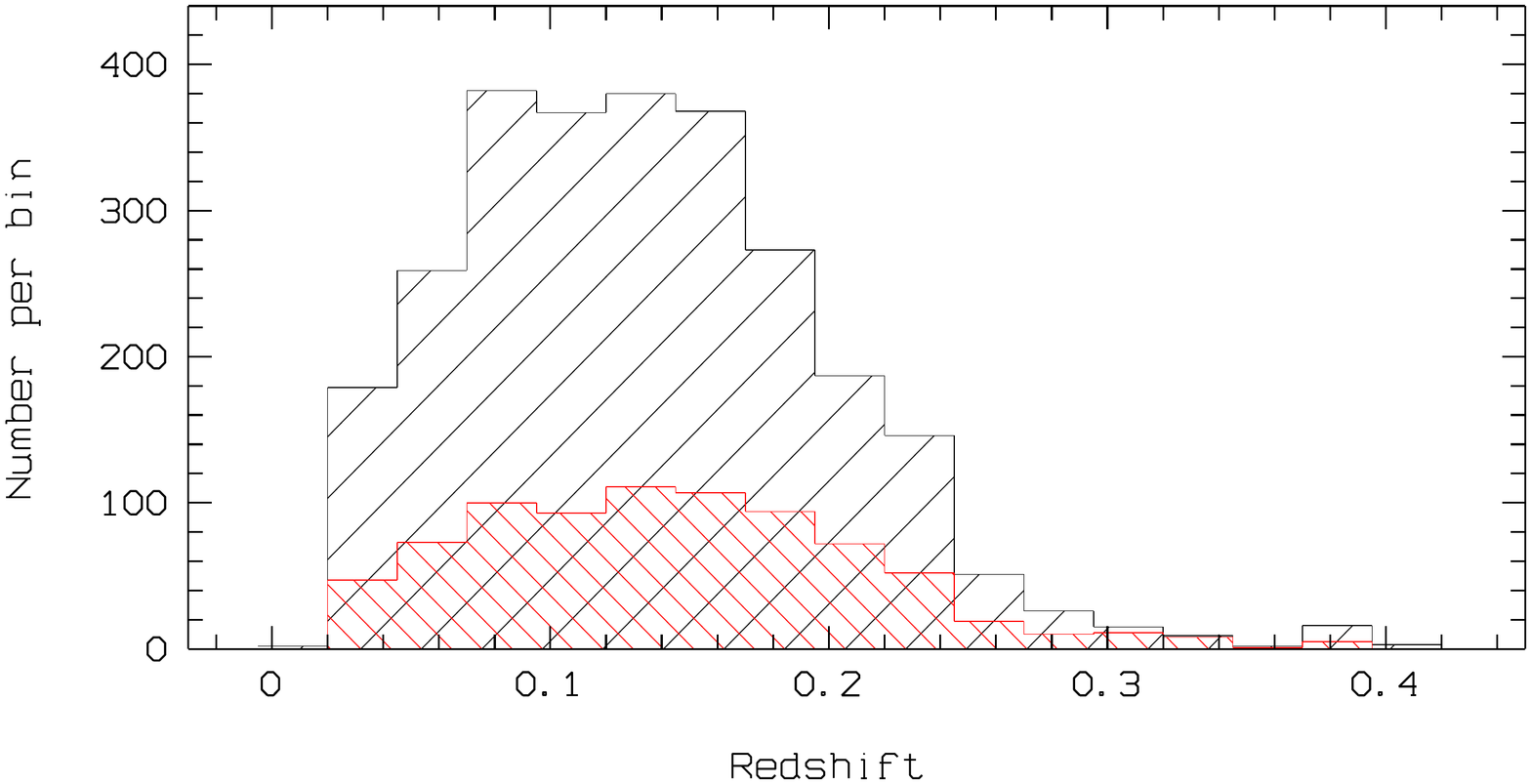}\\
\includegraphics[viewport= 50 20 460 780,width=4.7cm,angle=270]{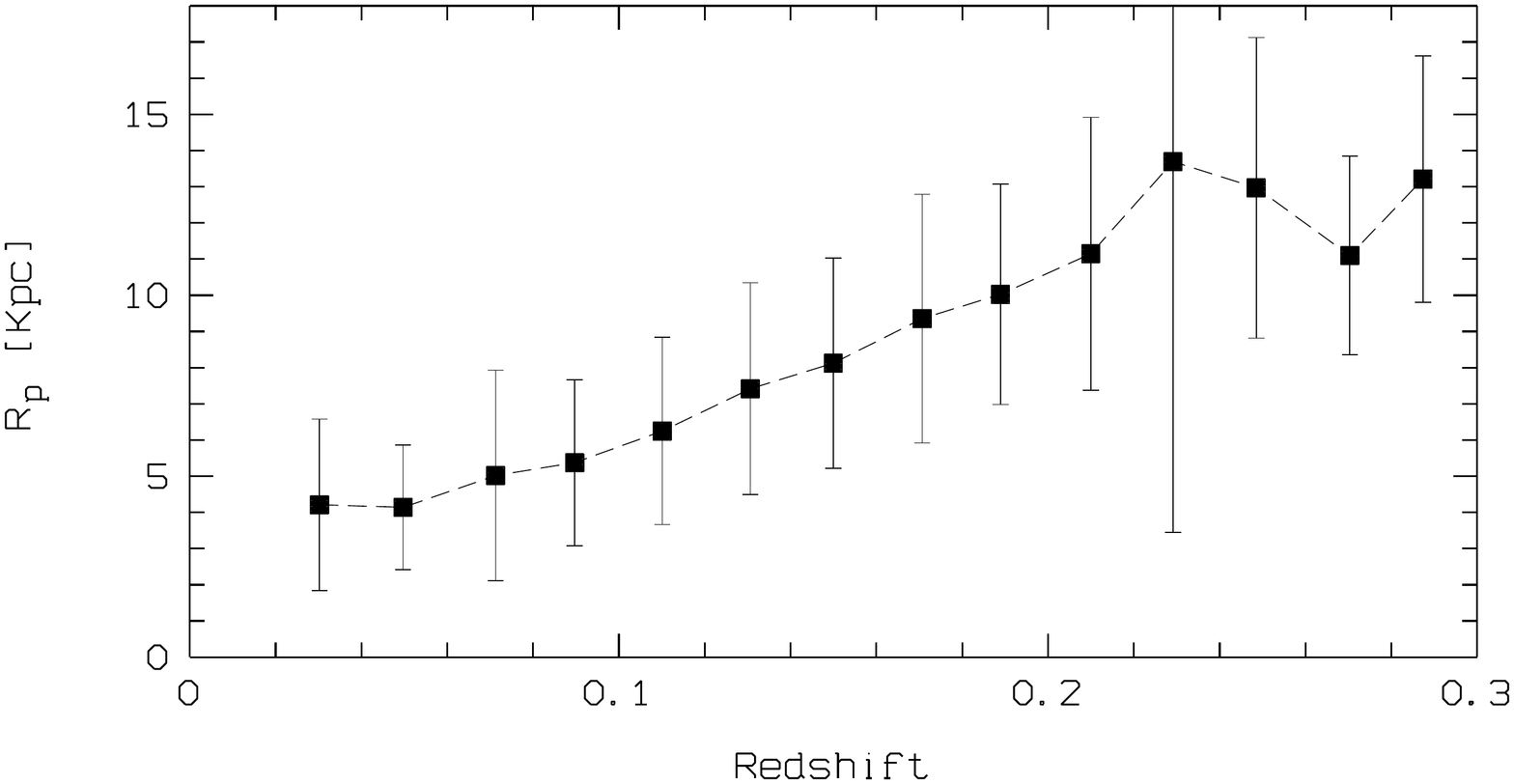}\\
\includegraphics[viewport= 50 20 460 780,width=4.7cm,angle=270]{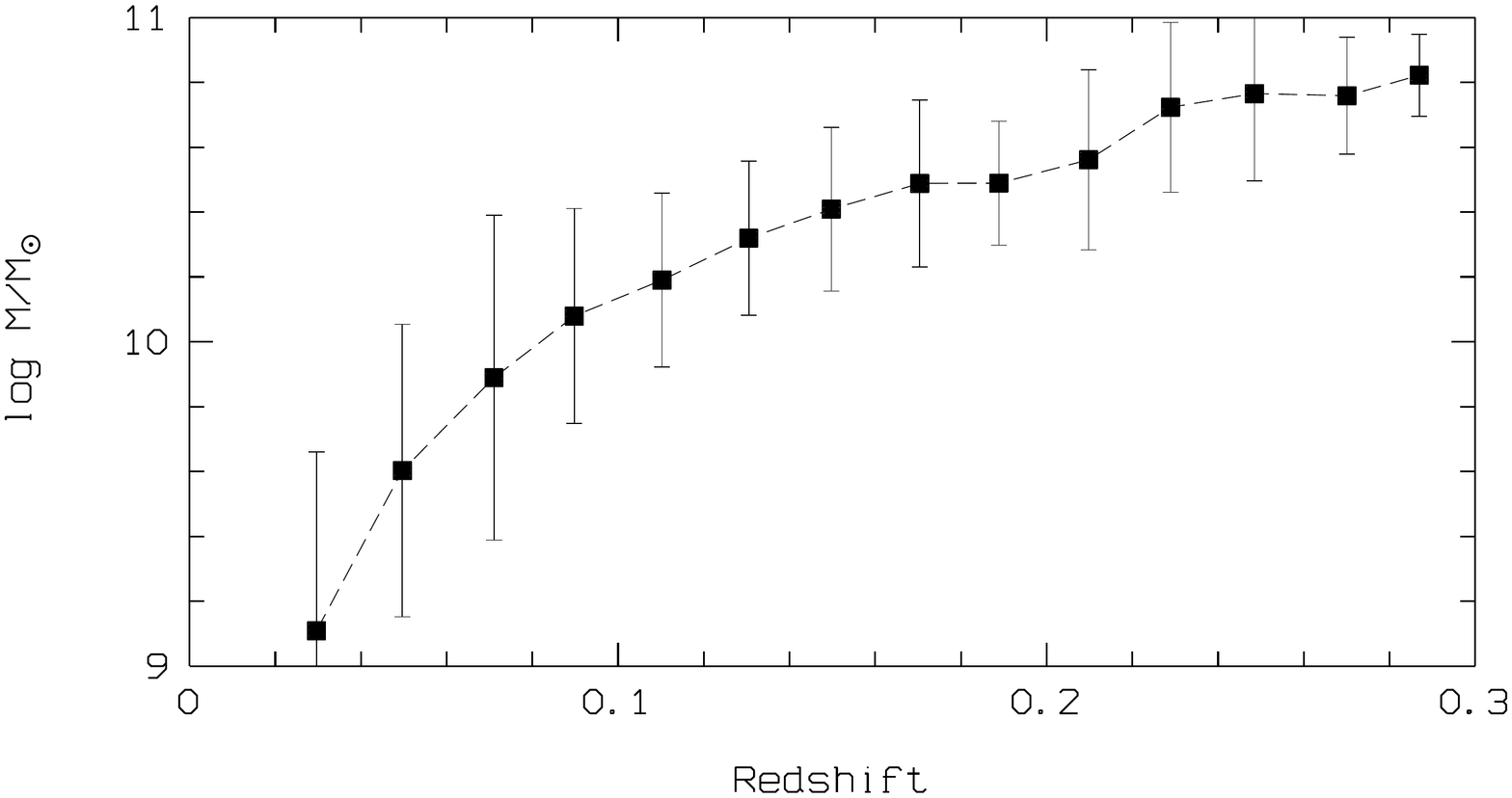}\\
\vspace{0.3cm}
\caption
{
Top: Histogram distribution of redshifts for the final E+A sample (black) and the input sample (red).  
Middle: Mean Petrosion radius in redshift bins.
Bottom: Mean stellar mass in redshift bins. 
}
\label{fig:hist_z}
\end{figure}

\begin{figure}[h]
\includegraphics[viewport= 50 20 460 780,width=4.7cm,angle=270]{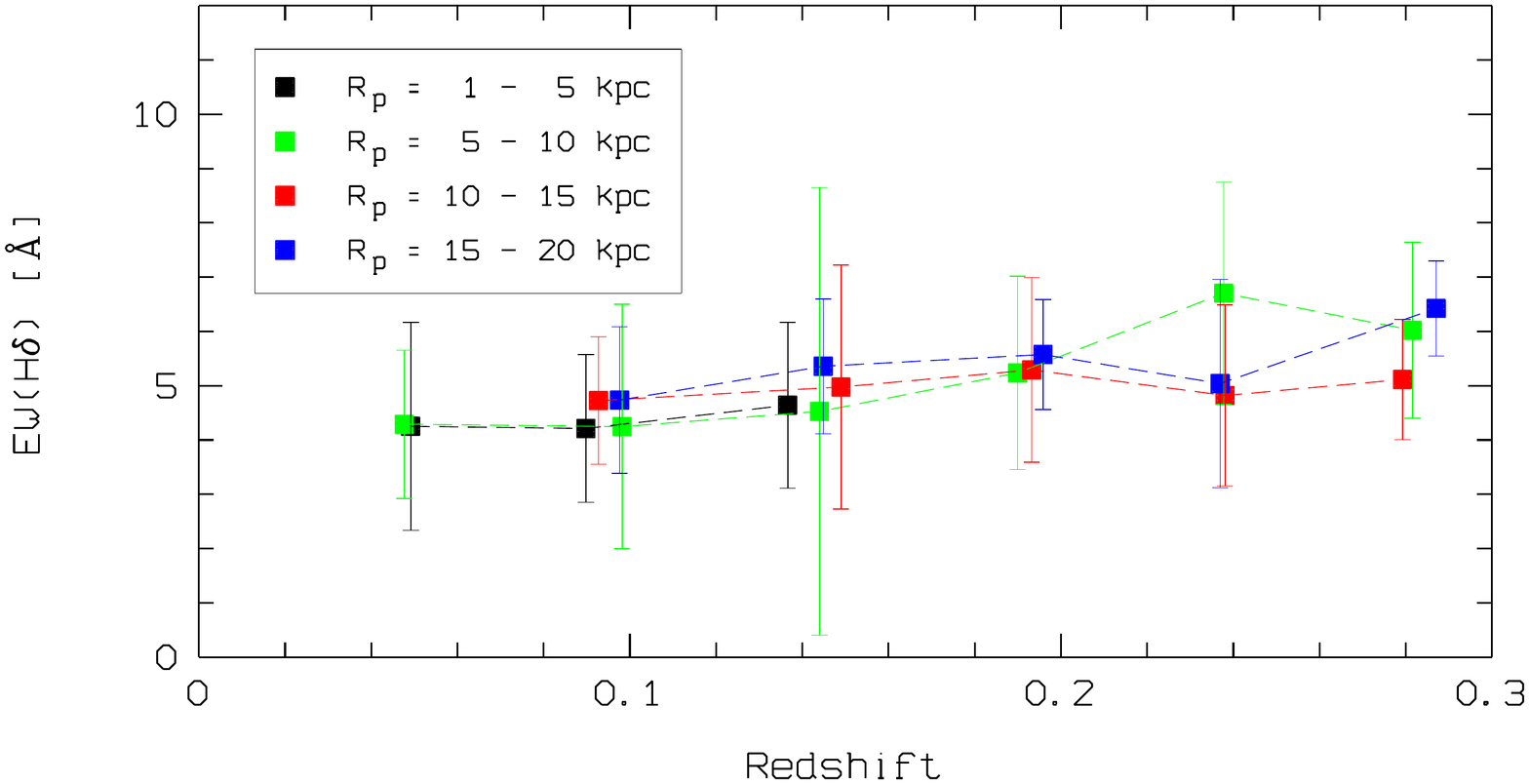}\\
\includegraphics[viewport= 50 20 460 780,width=4.7cm,angle=270]{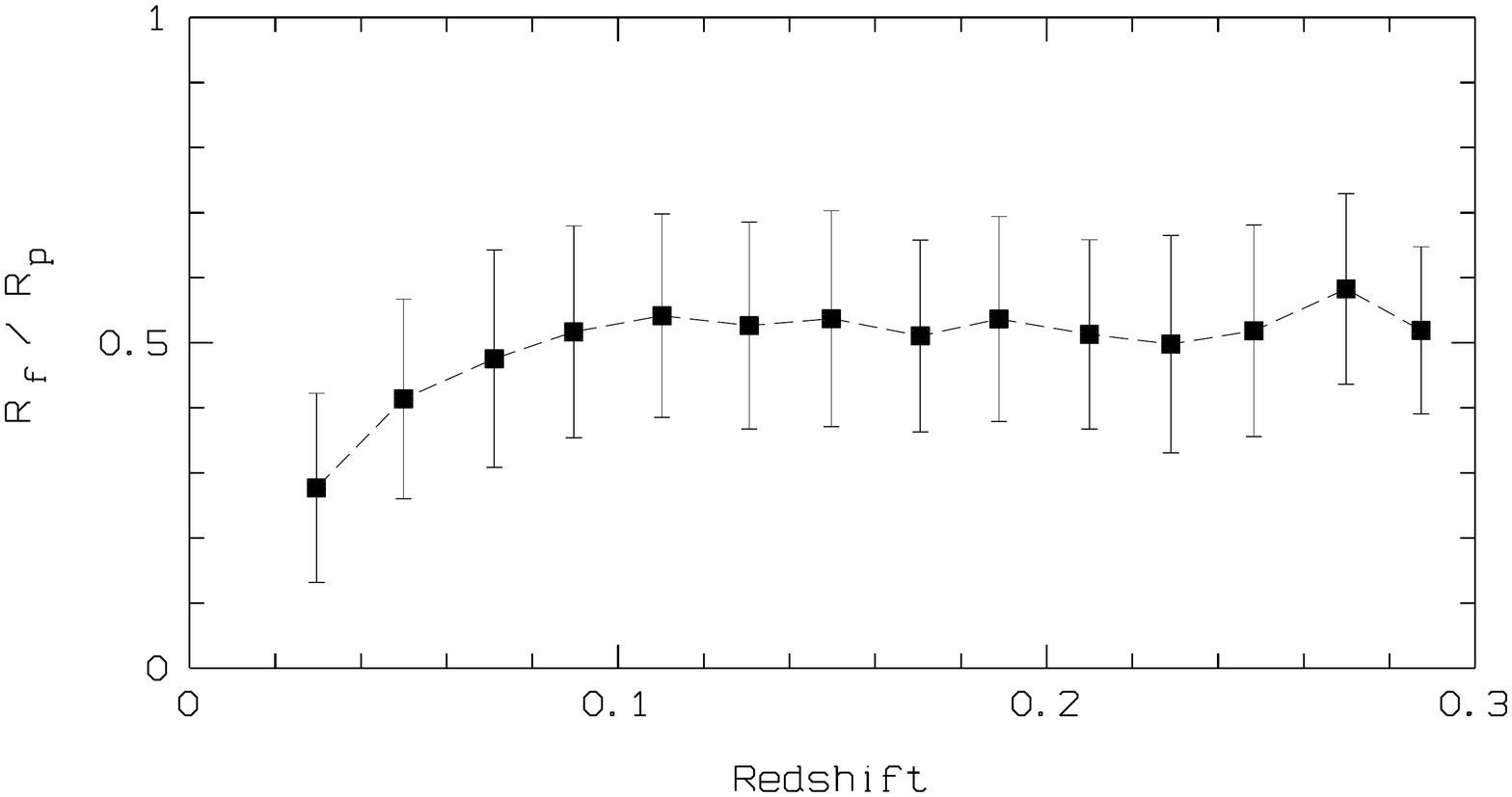}\\
\includegraphics[viewport= 50 20 460 780,width=4.7cm,angle=270]{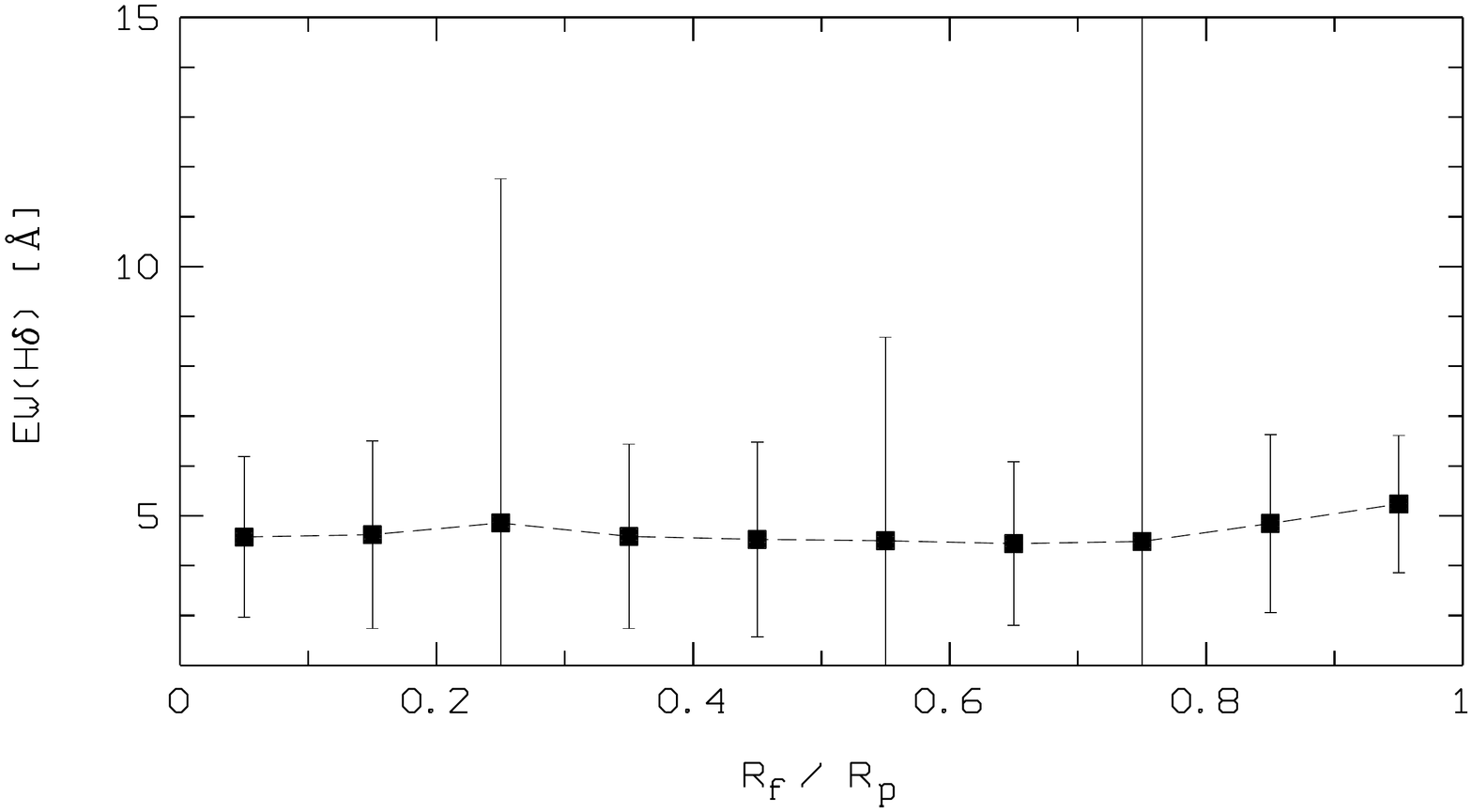}\\
\vspace{0.3cm}
\caption
{
Top: EW(H$\delta$) as a function of redshift in four different size categories.
Middle: Mean coverage factor $R_{\rm f}/R_{\rm P}$ for the E+A galaxies in redshift intervals. 
Bottom: Median of EW(H$\delta$) in different intervals of the coverage factor.
}
\label{fig:aperture_bias}
\end{figure}

\subsubsection{SDSS}

The redshift distribution of the E+A sample is shown in the top panel of Fig.\,\ref{fig:hist_z}. As a direct consequence of the flux limitation of the SDSS in combination with the $z$ distribution, the galaxy sample suffers from the Malmquist bias: At each redshift, the galaxy luminosities show a broad distribution with a lower limit increasing with increasing $z$. Properties correlated with the luminosity must also show a trend with $z$, such as the mean size (Fig.\,\ref{fig:hist_z}, middle) and the mean stellar mass (Fig.\,\ref{fig:hist_z}, bottom). The size is expressed here by the Petrosian radius, $R_{\rm P}$, in the r band from the SDSS Photometric Catalog, DR7 \citep{Abazajian_2009}. The Petrosian radius is the radial distance $R$ from the centre of a galaxy where the mean local surface brightness in an annulus at $R$ is equal to 20 per cent of the mean surface brightness within $R$. Theoretically, $R_{\rm P}$ recovers essentially all of the flux of a galaxy with an exponential profile and about 80\% for a de Vaucouleurs profile.
The stellar mass is taken from the Portsmouth sMSP data base (see Sect.\,\ref{sec:colour_mass}). To take account of the Malmquist bias we will compare our sample with a control sample of the same $z$ distribution in the Sects.\,\ref{sec:merger} and \ref{sec:AGN} below, or we consider the sample in different $z$ bins separately (Sect.\,\ref{sec:colour_mass}).

There is still another selection effect caused by the range of redshifts observed with a fibre spectrograph, as in the SDSS. As a consequence of the fixed size of the entrance aperture of the fibre, the observed spectrum probes different parts of the galaxies at different $z$ \citep[e.g.][]{Brinchmann_2004,Bergvall_2016}.   
For the SDSS fibre ($3\arcsec$), the linear radius $R_{\rm f}$ of the covered field changes from $\approx$ 2\,kpc at lowest redshifts to 10\,kpc at $z = 0.2$. If the starburst is strongly concentrated in a small central region of a more or less constant size the post-starburst spectrum is expected to be more and more diluted by the light from the 
underlying stellar population in the galaxy with increasing $z$. \citet{Bergvall_2016} analysed how EW(H$\alpha$) changes with $z$ in a sample of local starburst and PSB galaxies. They found a significant trend at $z < 0.02$ and argued that a sample with a lower redshift limit $z_{\rm \, low} = 0.02$ is less affected by aperture losses. It should be noted that the lower limit is 0.02 in our sample. The top panel of Fig.\,\ref{fig:hist_z} shows EW(H$\delta$) as a function of $z$ for four different size catagories of our E+A galaxies. We do not see a significant trend.  

The Malmquist effect leads to a selection bias against smaller galaxies with increasing $z$. The middle panel of Fig.\,\ref{fig:hist_z} shows the mean ratio $R_{\rm f}/R_{\rm P}$ averaged in $z$ bins of the width 0.02 as a function of $z$. 
The ratio $R_{\rm f}/R_{\rm P}$ is a measure of the coverage of the galaxy by the aperture. There is a moderate increase at lowest redshifts, but the ratio changes only weakly over the interval $0.06 \la z \la 0.3$ that contains 90\%  of the E+A sample. If the starburst is not strictly confined to the nuclear region \citep{Swinbank_2012} and if the size of the starburst region scales with the galaxy size, the aperture effect may thus be essentially compensated by the Malmquist bias. The bottom panel of Fig.\,\ref{fig:hist_z} indicates that there is no significant trend of EW(H$\delta$) with the coverage ratio $R_{\rm f}/R_{\rm P}$.

\begin{figure*}[htbp]
\begin{tabbing}
\includegraphics[width=9.0cm]{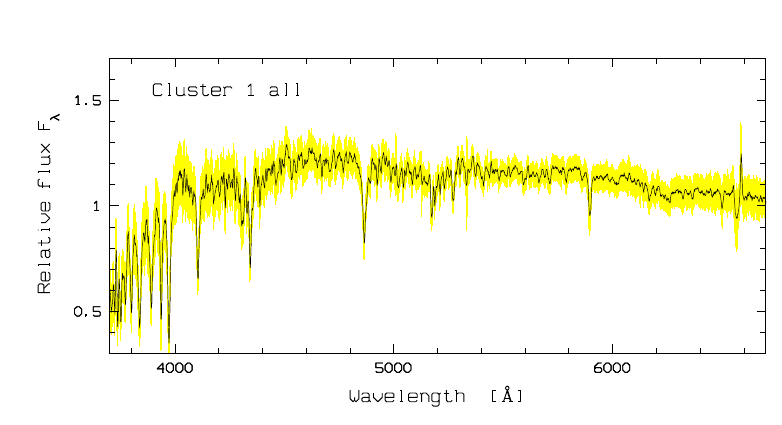} \hfill \
\includegraphics[width=9.0cm]{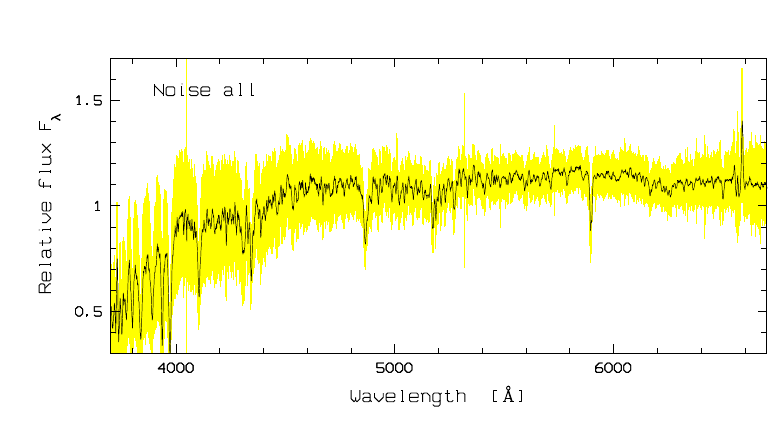} \\
\includegraphics[width=9.0cm]{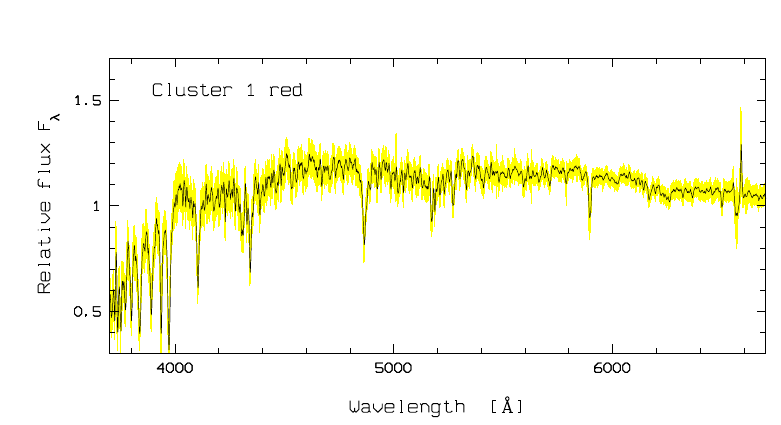} \hfill \
\includegraphics[width=9.0cm]{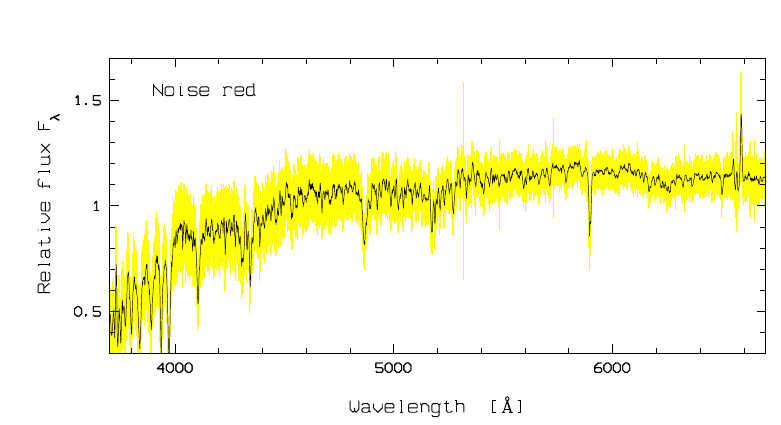} \\
\includegraphics[width=9.0cm]{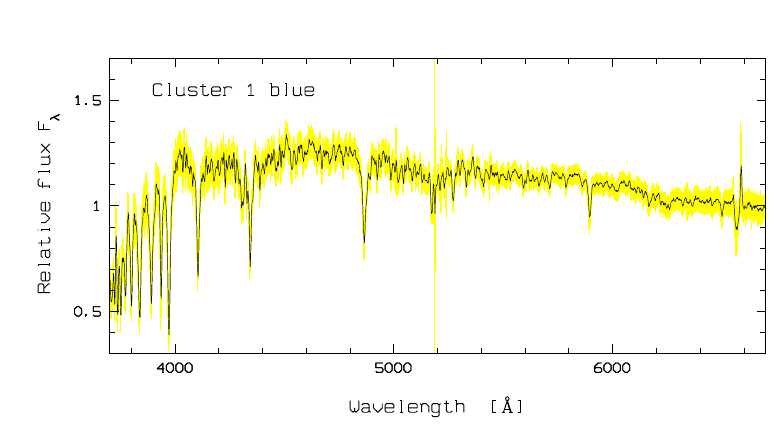} \hfill \
\includegraphics[width=9.0cm]{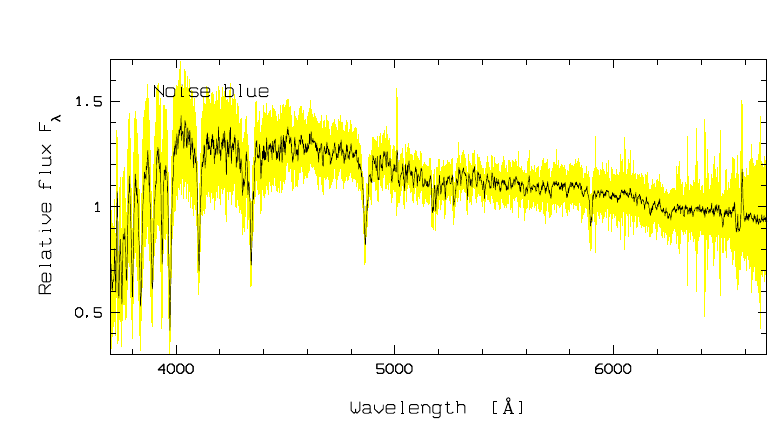} \\
\end{tabbing}
\caption{
Median composite (black) and standard deviations (yellow) from the rest-frame spectra of
the Goto galaxies in cluster 1 (left) and in the noise (right).
}
\label{fig:composites_cl_noise}
\end{figure*}

\subsubsection{SOM}

The Kohonen method is expected to produce selection biases mainly because the clustering power depends on several properties of the spectra, such as the strength of the characterising spectral features, the diversity of the underlying spectral components, the redshift distribution, and the S/N.  

A first way to check the selection effects from the SOM is to compare the redshift distribution of the final sample with that of the input sample.  The top panel of Fig.\,\ref{fig:hist_z} clearly indicates that the two distributions are very similar. As for the input sample the redshift range is from $z \approx$ 0.02 to 0.4, with 96\% below $z =$ 0.25. 

As described in detail in the ASPECT paper \citep{inderAu_2012}, the computation of a SOM of this size requires the reduction of the overall size of the data pool to a necessary minimum. The spectra had to be smoothed and rebinned in order to reduce the number of pixels. Initial tests have shown that the reduction of the spectral resolution caused by the rebinning does not significantly affect the quality of the clustering results as long as spectral features are considered that are clearly broader than the spectral resolution of the original SDSS spectra, for example quasar broad absorption lines. Compared to such spectra, the characterising spectral features of E+A galaxies are relatively narrow. For weak and narrow features, the clustering is of course stronger dominated by the underlying spectral components and is thus less efficient. One solution would be trading spectral coverage against spectral resolution. However, though the main intention for the construction of the SOM described in Sect.\,\ref{sec:SOM} was to search for extreme BAL quasars \citep{Meusinger_2012,Meusinger_2016}, it was originally not designed for any special application. In addition, the selection of E+A galaxies requires the wide wavelength coverage from \ion{O}{ii} to (redshifted) H$\alpha$.  

The clustering strength of the E+A galaxies depends on the EW of the H$\delta$ line, the S/N of the continuum near H$\delta$, and the spectrum of the underlying stellar population. Table\,\ref{table:Goto_clusters_1} compares mean properties of the nine clusters and the noise. While the mean redshift of the noise is very similar to that of the richest cluster, the other properties are different. The lowest mean EW(H$\delta$) is found for the (poor) cluster 9 and the noise. The mean S/N is smallest and its scatter is largest for the noise. 
In addition, the composite spectra from the noise and the rich clusters show some differences.  The top panels of Fig.\,\ref{fig:composites_cl_noise} compare the median rest-frame composite spectra of the Goto galaxies from cluster 1 on the left-hand side with that of the noise galaxies on the right-hand side. The noise composite is redder and shows a 
much larger standard deviation. We subdivided each of the two samples into a blue and a red subsample defined by 
$F_\lambda(4030-4080\, \mbox{\AA})/F_\lambda(5200-5800\, \mbox{\AA}) \ge 1$ or \ $<1$, respectively, and comptuted the composites for these subsamples. The comparison shows (middle and bottom raw of  Fig.\,\ref{fig:composites_cl_noise}) that the composite of the red noise spectra is redder, that of the blue ones is bluer, and in both cases the scatter is larger than for the cluster. This means that the noise galaxies show a considerably larger variety of spectral slopes. If dust obscuration is one of the reasons for these differences, this could mean that our sample is biased against dusty E+A galaxies.

The last column of Table\,\ref{table:Goto_clusters_2} lists the ratio of the number of newly selected galaxies to the number of seed galaxies, which is a proxy for the efficiency of the search. For the seven clusters with $n_i > 10$, the mean ratio is 3.4, compared to 1.0 for the combination of the noise and the clusters 8 and 9. If we assume complete selection around the rich clusters, that means each seed galaxy corresponds, on average, to 3.4 galaxies in our sample, we find that 482 galaxies are missed from the poor clusters and the noise. This corresponds to $\sim 15$\% of the entire sample, with the largest part lost in the noise.

%
\section{Properties of the E+A sample}

%
\subsection{The final catalogue}
\label{sec:catalogue}

As a result of the thorough examination of the SOM neighbourhood of the input galaxies, both of the nine clusters and the 
noise (Fig.\,\ref{fig:goto_galaxies_in_SOM}),  we selected altogether 3\,060 E+A galaxies. 
The subsample fulfilling Goto's  selection criteria (Sect.\,\ref{sec:Input}) consists of 774 galaxies, among them 539 Goto galaxies. 
The lack of 35\% of the Goto galaxies in this subsample is caused by small differences in the values for the EWs from 
the data used by Goto, most likely due to differences in the output of the spectroscopic pipeline of the SDSS from rerun to rerun.

The final selection was used to compile a catalogue of 2\,665 galaxies defined by the following properties: EW(H$\delta$) $> 3$ \AA, EW(H$\alpha$) $>-5$ \AA, EW([\ion{O}{ii}])$> -5$ \AA, $z < 0.35$ or $>0.37$. 
This sample is referred to as our E+A sample throughout this paper.
The catalogue will be published in electronic form in the Vizier service of the CDS Strasbourg.
The galaxy redshifts cover the range $z \approx 0.02 - 0.4$, with a mean value at 0.13. The vast majority (84\%) of the galaxies have $z < 0.2$. 
The catalogue includes 803 galaxies from the input catalogue (96\%).  
Applying Goto's stronger selection criteria to the line equivalent width data used in the present study leads
to a reduced sample of 916 E+A galaxies, among them 615 Goto galaxies.

\begin{figure}[htbp]
\includegraphics[width=9.0cm,angle=0]{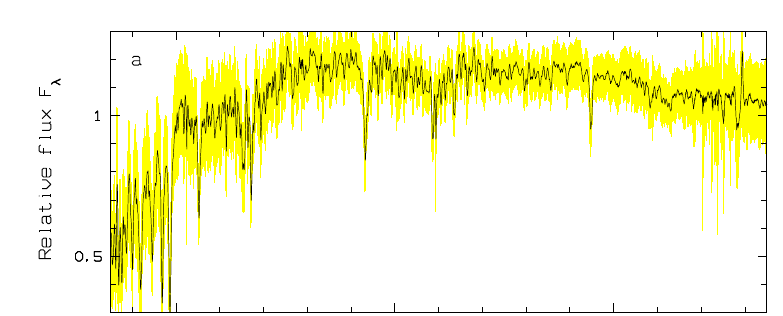} \\
\includegraphics[width=9.0cm,angle=0]{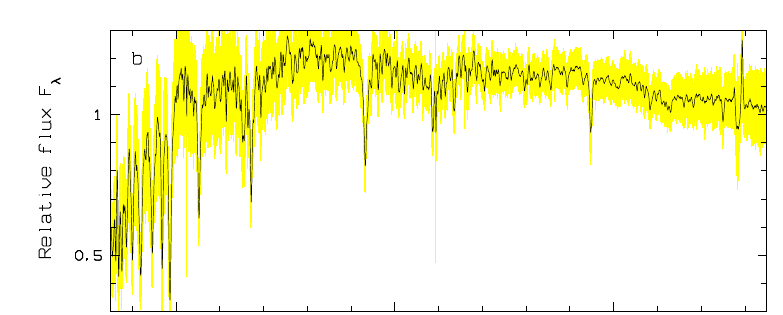} \\
\includegraphics[width=9.0cm,angle=0]{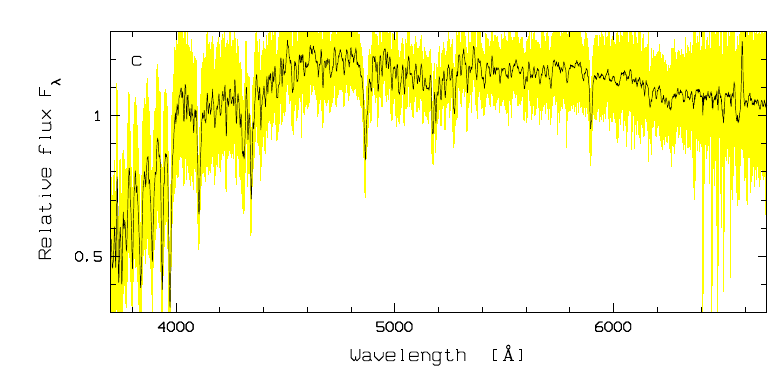} \\
\vspace{0.0cm}
\caption{Median composite (black) and standard deviations (yellow) from the rest-frame spectra of
(a) the whole set of the E+A galaxies from the present study with  EW(H$\delta$) $>3$ \AA,
(b) the subset of the galaxies from the input catalogue,
(c) the subset of newly selected galaxies fulfilling the selection criteria of the input catalogue
EW(H$\delta$) $>5$ \AA, EW(H$\alpha$) $>-3$ \AA, and EW(\ion{O}{ii}) $>-2.5$ \AA.
}
\label{fig:composites}
\end{figure}

\begin{figure}[htbp]
\includegraphics[width=0.8\columnwidth]{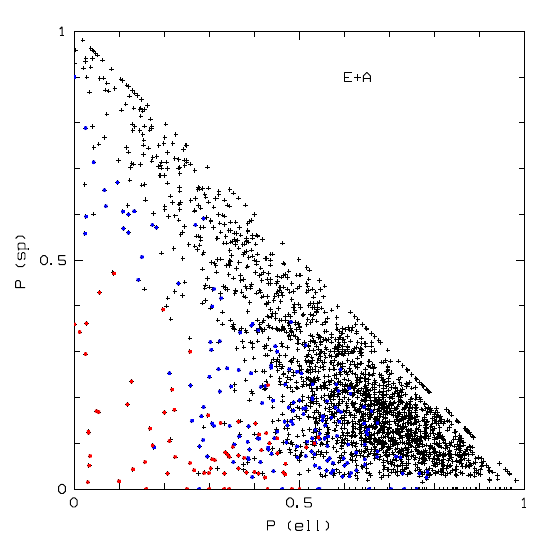}
\includegraphics[width=0.8\columnwidth]{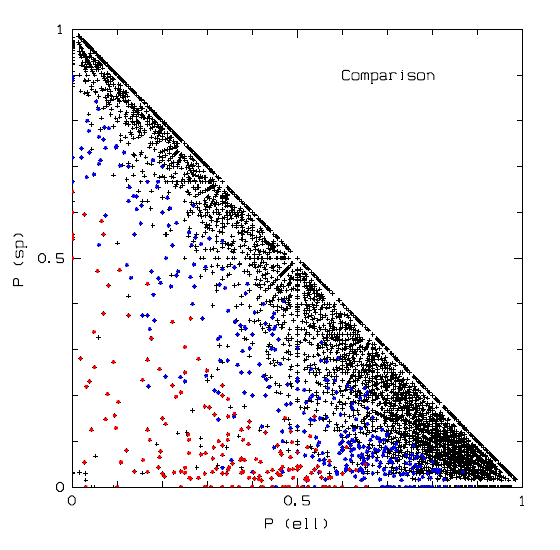}
\vspace{0.5cm}
\caption{
Morphological type probabilities from Galaxy Zoo for the E+A sample (top) and the comparison sample (bottom).
The colours indicate different merger probablilities: $P_{\rm m} < 0.1$ (black), $P_{\rm m} = 0.1 \ldots 0.3$ (blue),             and $P_{\rm m} > 0.3$ (red). }
\label{fig:GalaxyZoo}
\end{figure}

In Fig.\,\ref{fig:composites}, we compare the rest-frame arithmetic median composite spectrum of the E+A sample (upper panel) 
with that of the galaxies from the input catalogue (middle panel). 
In addition, the bottom panel shows the composite of the new E+A galaxies fulfilling the selection criteria of the input catalogue. 
There is no substantial difference between the three spectra. 
The weak emission line close to the right margin is \ion{N}{ii}$\lambda$6584, H$\alpha$ is seen as a weak absorption line.

%
\subsection{Morphological types}
\label{sec:morphology}

Morphological type classification of the SDSS galaxies is available from the Galaxy Zoo project. 
The first version of Galaxy Zoo \citep{Lintott_2011} provided the most fundamental morphological types 
(spiral or elliptical systems) for nearly $9\cdot 10^5$ galaxies based on the contributions from more than 
$2\cdot 10^5$ online volunteer citizen scientists. 
More detailed morphological information is available from Galaxy Zoo 2 \citep[GZ2;][]{Willett_2013}
for the brightest 25 per cent of the resolved SDSS DR7 galaxies in the north Galactic cap region. At this point, 
we restrict the discussion to the fundamental morphological types from the original Galaxy Zoo
because only one third of our E+A galaxies were identified in the GZ2 sample. The results of the classifications are expressed by weighted probabilities $P_{\rm e},  P_{\rm s},  P_{\rm m}$ for being an elliptical galaxy, a spiral, or a galaxy merger, respectively. The fraction of gravitationally distorted galaxies and mergers will be the subject of Sect.\,\ref{sec:merger} where we will discuss, among others, classifications from GZ2.

The morphological type probabilities are available for 94\% of the galaxies from our sample. 
The top panel of Fig.\,\ref{fig:GalaxyZoo}
shows $P_{\rm s}$ versus $P_{\rm e}$  as black symbols for galaxies with 
merger probabilities $P_{\rm m} < 0.1$, galaxies with $P_{\rm m} = 0.1 \ldots 0.3$ are highlighted blue, those with $P_{\rm m} > 0.3$ are red. 
Although there is a concentration towards large $P_{\rm e}$ and small $P_{\rm s}$, only a small fraction are 
clearly attributed to one of the two types. Usually a minimum 80\% majority agreement is required for a clean classification. 
This is found for only 12\% of the galaxies in the sample, where $\sim 11$\% are classified as ellipticals and $\sim 1$\% as spirals. The overwhelming majority of the E+A galaxies have intermediate morphologies.  The median values for the three probabilities are $(P_{\rm e},P_{\rm s},P_{\rm m}) = (0.62,0.17,0.00)$.

For comparison we created a randomly selected control sample of `normal' galaxies from the SDSS DR7. The control sample 
has the same redshift distribution as the E+A sample but is twice as large. The distribution of the comparison galaxies in the 
$P_{\rm s}$ versus $P_{\rm e}$ diagram (bottom panel of Fig.\,\ref{fig:GalaxyZoo}) appears similar to that of the E+A galaxies and the median values are also similar (0.65,0.20,0.00). However, the percentage of comparison galaxies with a minimum 80\% majority agreement is clearly larger: 28\% have $P_{\rm e} \ge 0.8$ and 10\% have $P_{\rm s} \ge 0.8$.

%
\subsection{Colour-mass diagram}
\label{sec:colour_mass}

\citet{Wong_2012} analysed the distribution of a local sample of 80 PSB galaxies with $z= 0.02 - 0.05$ in the  
$u-r$ colour-mass diagram. They found that the majority reside within the lower mass part of the `green valley' 
between the red sequence and the blue cloud. 
Here, we exploited the data from the Portsmouth sMSP (Sect.\,\ref{sec:SDSS}) to study the distribution of the galaxies from our E+A sample on the colour-mass plane. 
We used the data set portsmouth\_stellarmass\_starforming\_krou-26-sub for the galaxies from SDSS DR8 adopting
the Kroupa initial mass function. 
We identified 96\% of the galaxies from our E+A sample in the Portsmouth data base. 

\begin{figure*}[htbp]
\begin{tabbing}
\includegraphics[width=6.0cm]{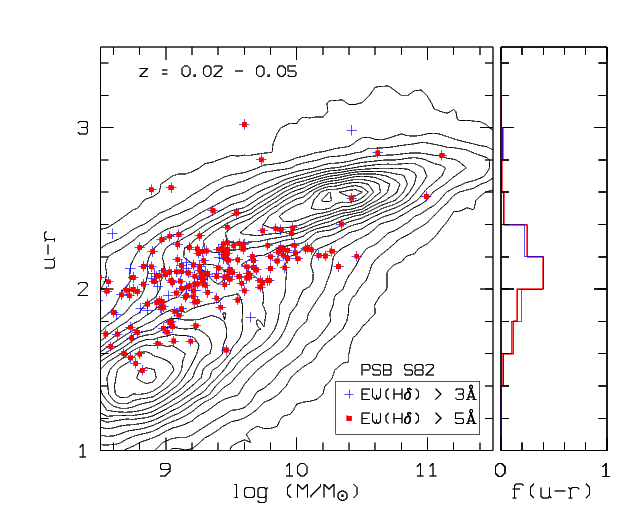}\hfill \=
\includegraphics[width=6.0cm]{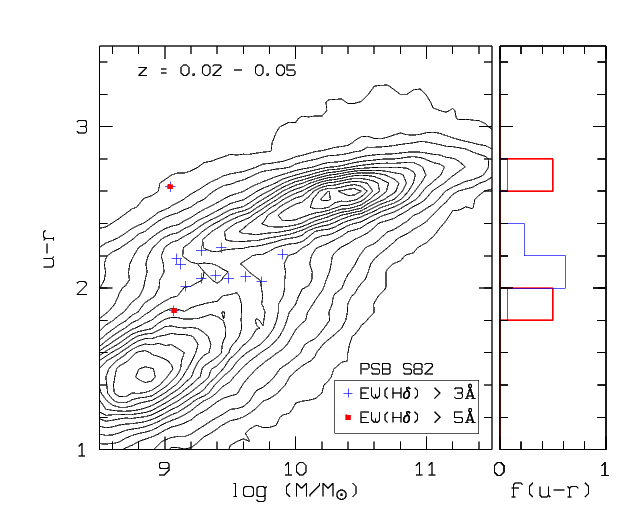}\hfill \=
\includegraphics[width=6.0cm]{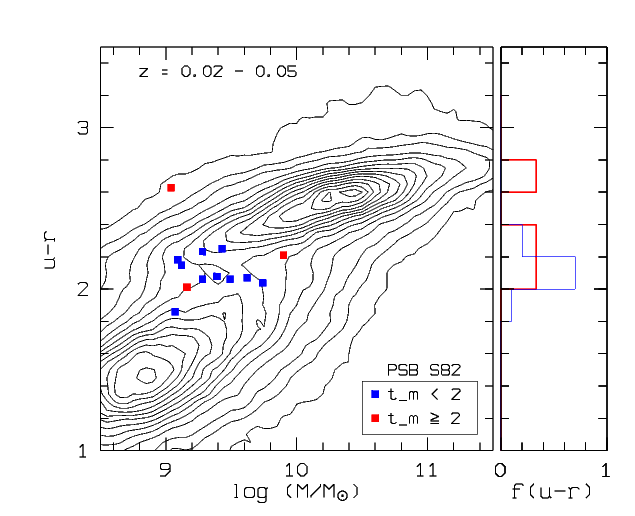}\\
\includegraphics[width=6.0cm]{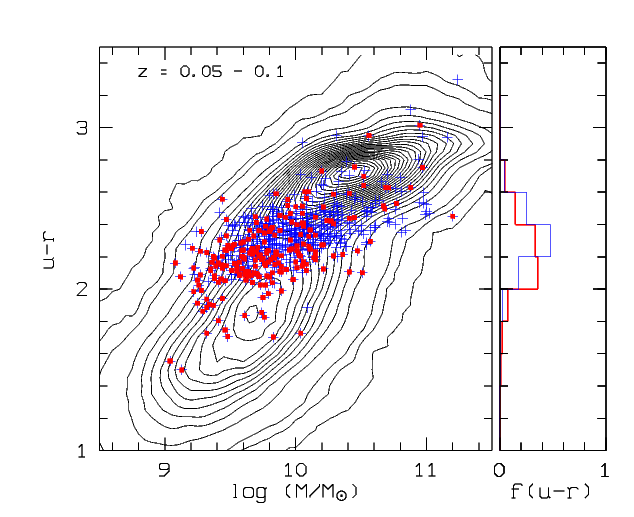}\hfill \=
\includegraphics[width=6.0cm]{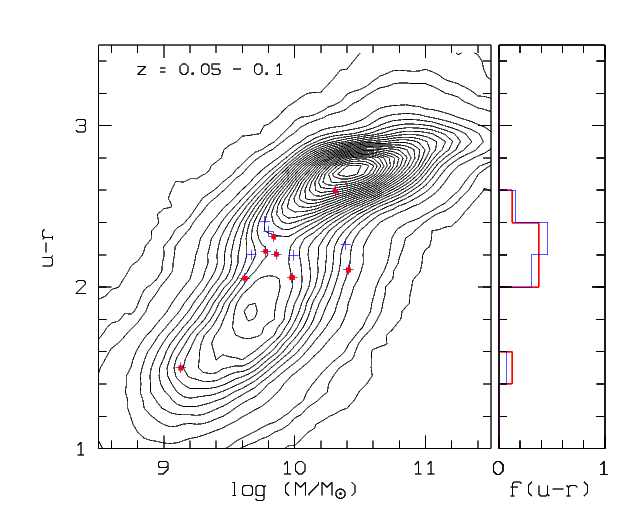}\hfill \=
\includegraphics[width=6.0cm]{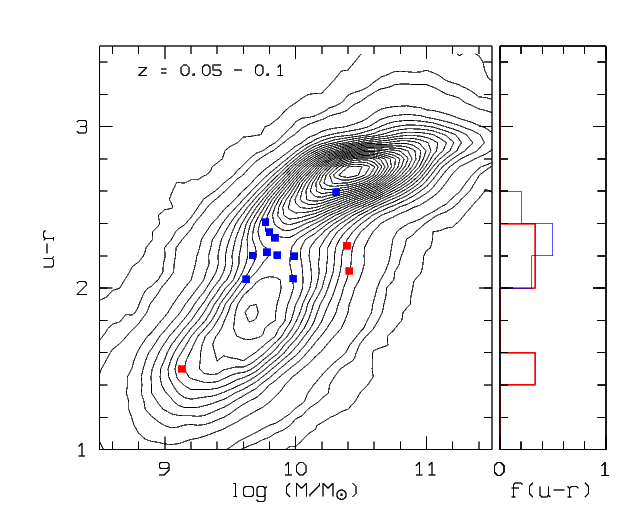}\\
\includegraphics[width=6.0cm]{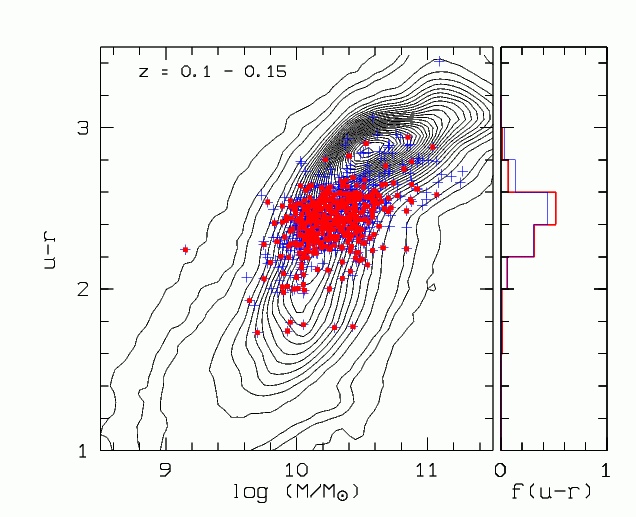}\hfill \=
\includegraphics[width=6.0cm]{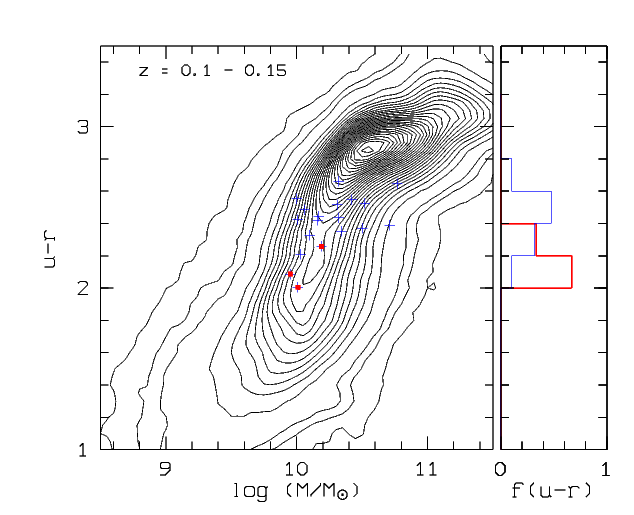}\hfill \=
\includegraphics[width=6.0cm]{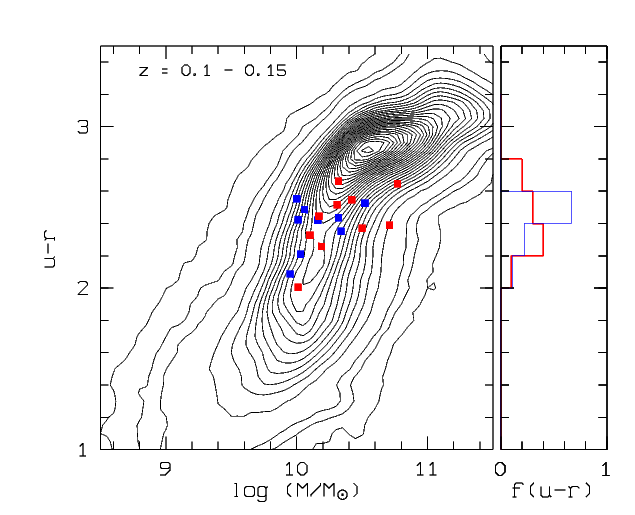}\\
\includegraphics[width=6.0cm]{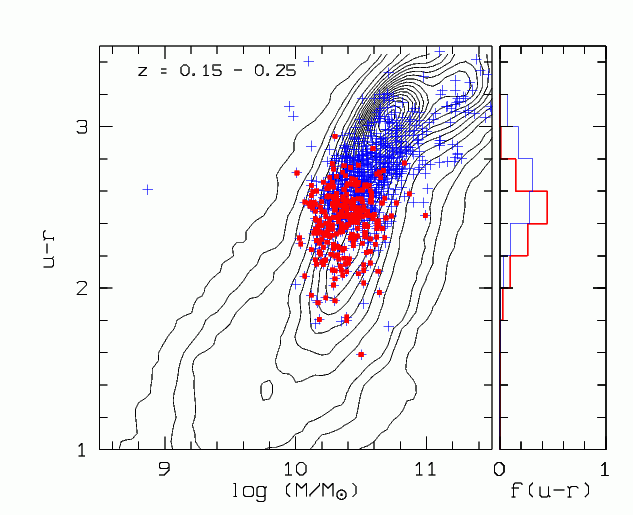}\hfill \=
\includegraphics[width=6.0cm]{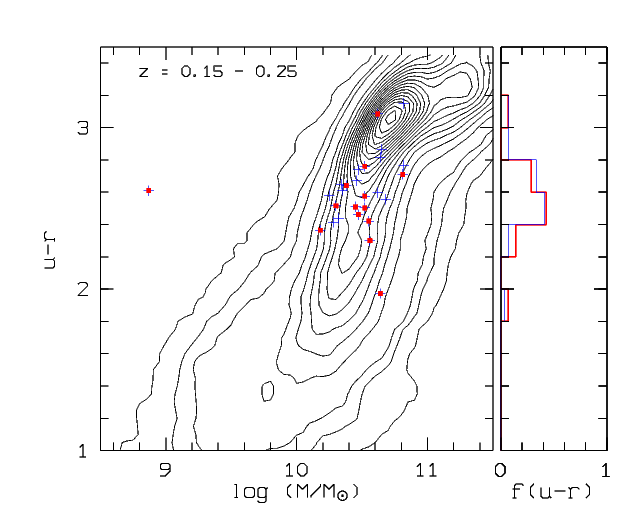}\hfill \=
\includegraphics[width=6.0cm]{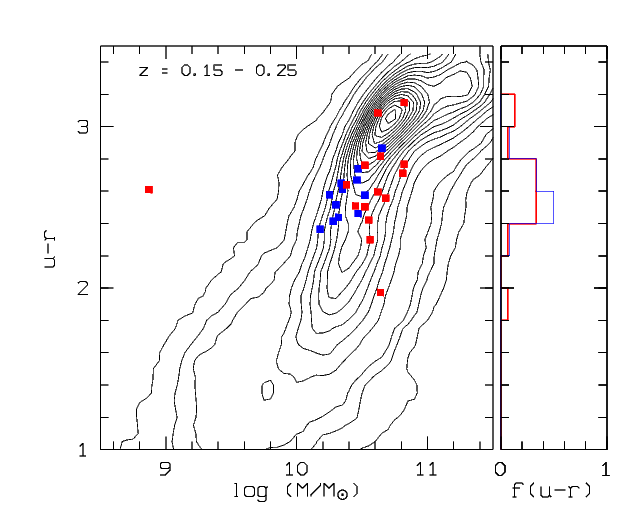}\\
\end{tabbing}
\caption{
Colour-mass diagram for the galaxies from the SDSS DR8 (contours) and our E+A galaxies (symbols)
for four different redshift ranges (top to bottom). The redshifts are given in the top left corner of each panel.
Left column: entire E+A sample, middle and right columns: S82 E+A sample. The histogram on the right-hand side of each panel shows the distribution of $u-r$. The colours are
explained in the insets in the top raw.
}
\label{fig:colour_mass}
\end{figure*}

Figure\,\ref{fig:colour_mass} shows the distribution of all SDSS galaxies in the $u-r$ versus log $M$ diagram
represented by equally-spaced local point density contours. Because of the heavy Malmquist bias we plotted the diagrams for four different $z$ bins (top to bottom): $z = 0.02-0.05,\ 0.05-0.10,\ 0.10-0.15,\ 0.15-0.25$. In the panels on the left-hand side, the galaxies from the E+A sample are over-plotted as coloured symbols with H$\delta$-strong galaxies (EW(H$\delta) > 5$\AA) in red, the others (EW(H$\delta) = 3 - 5$\AA) in blue. The mean uncertainties are $\sim 0.1$ mag in $u-r$ and $<0.3$ in $\log M$. E+A galaxies are scattered over a relatively wide area of the colour-mass plane. The main conclusion from Fig.\,\ref{fig:colour_mass} is, however, that there is clearly a strong concentration of the E+A sample in the region between the blue cloud and the red sequence, independent of $z$. 
This is in agreement with \citet{Wong_2012}, who argued that their finding is consistent with the idea that E+A galaxies are in the transformation stage between actively evolving galaxies from the blue cloud into passively evolving members of the red sequence. In fact, there is a tendency for stronger H$\delta$ absorption galaxies to be on average somewhat closer to the blue cloud and those with weaker H$\delta$ lines to  be closer to the red sequence. However, this trend is significant only in the highest $z$ bin and not seen at lowest $z$.

%
\subsection{Merger fraction}
\label{sec:merger}

\subsubsection{Classification of morphological distortions}
\label{sec:merger_class}

Galaxy-wide shock fronts in merging gas-rich disks are thought to be a major trigger of starburst activity. 
According to the Galaxy Zoo data, the fraction of E+A galaxies classified as mergers is remarkably low, only 3\% have $P_{\rm m} > 0.3$. However, detecting the fine structures that most unambiguously indicate a 
gravitational perturbation induced by a close encounter or merger event in the past history of a galaxy requires deep imaging. The characteristics of such fine structures depend on the properties of the parent galaxies and the details of the encounter \citep[for details see e.g.][]{Duc_2013}. Extended tidal structures,
remarkable anisotropies, shells, rings, or simply close galaxy pairs are usually taken as indicators of ongoing or past collisions. Tidal tails are considered the most direct signpost of a previous merger, particularly of major mergers of disk galaxies. Such tails are short lived with fall-back times between a few hundred Myr and a few Gyr and can be of very low surface brightness. Generally, a deep surface brightness limit is a basic requirement for the evaluation of such faint morphological structures. The SDSS S82 is perfectly suited for such a task
thanks to the combination of the remarkable depth of the co-adds from the multi-epoch imaging, the large field, and the dense spectroscopic coverage (see Sect.\,\ref{sec:S82}).

\begin{figure}[h]
\includegraphics[width=9.0cm]{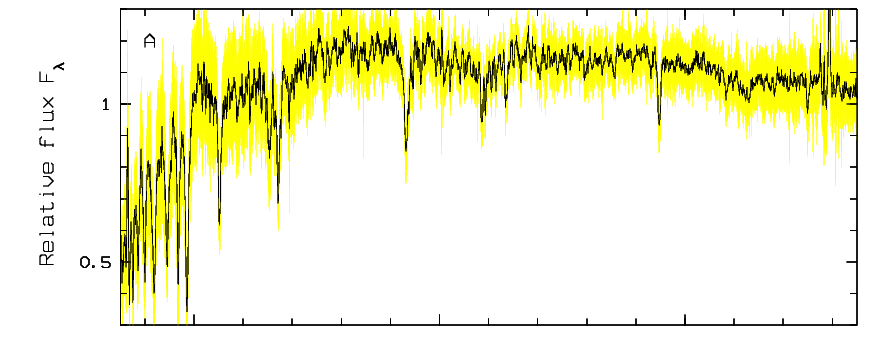} \\
\includegraphics[width=9.0cm]{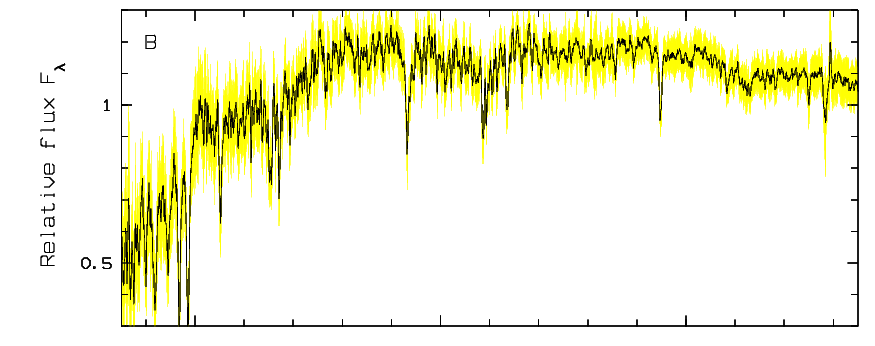} \\
\includegraphics[width=9.0cm]{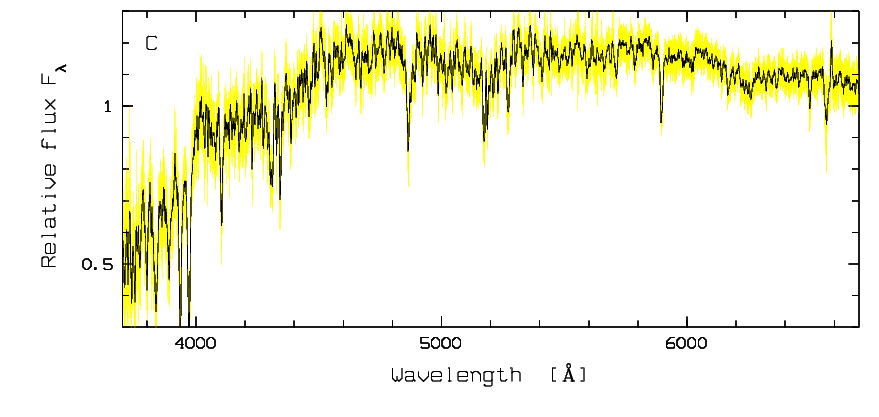}
\caption{Median composite (black) and standard deviations (yellow) from the rest-frame spectra of
the subsamples A, B, and C (top to bottom) from the PB-S82 sample.
}
\label{fig:composites_S82}
\end{figure}

Our catalogue of E+A galaxies contains 74 systems in S82. The redshifts are between $z = 0.02$ and 0.22. Throughout this section, this sample is referred to as the S82 E+A sample. Among them are 18 galaxies from the input catalogue, that is 77\% of the sample comes from the present search. The arithmetic median composite spectrum is very similar to the composite spectrum from the input catalogue (Fig.\,\ref{fig:composites_S82}).  We subdivided this sample into the three subsamples with EW(H$\delta$) $> 5$ \AA\ (sample A), $4 \ldots 5$ \AA\ (sample B), and $3 \ldots 4$ \AA\ (sample C). The threshold for EW(H$\delta$) in sample A corresponds to the stronger selection criterion used by \citet{Goto_2007a}, but it is worth mentioning that only 17 of the 28 galaxies in that subsample are also in the Goto catalogue, 11 galaxies (39\%) are new. Table~\ref{table:S82} lists the galaxies in the three subsamples sorted by increasing $z$.

\begin{figure}[h]
\begin{tabbing}
\fbox{\includegraphics[width=4.2cm,height=4.2cm]{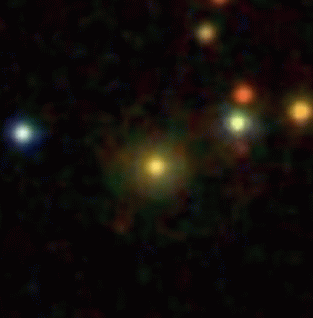}}\hfill \=
\fbox{\includegraphics[width=4.2cm,height=4.2cm]{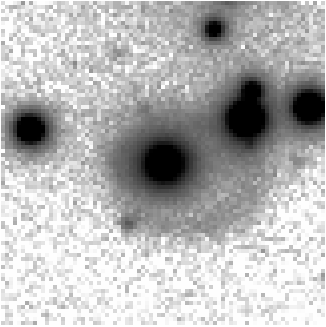}}\\
\fbox{\includegraphics[width=4.2cm,height=4.2cm]{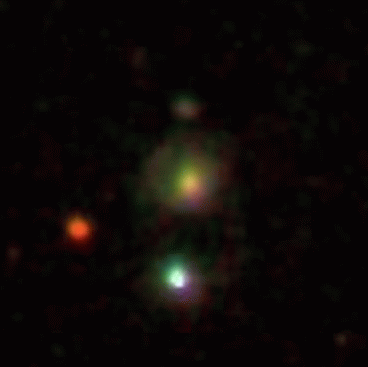}}\hfill \=
\fbox{\includegraphics[width=4.2cm,height=4.2cm]{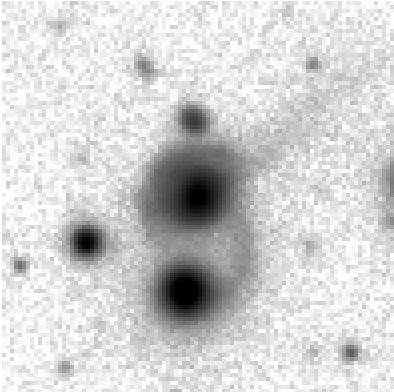}}\\
\fbox{\includegraphics[width=4.2cm,height=4.2cm]{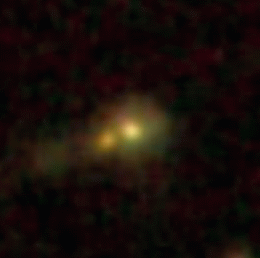}}\hfill \=
\fbox{\includegraphics[width=4.2cm,height=4.2cm]{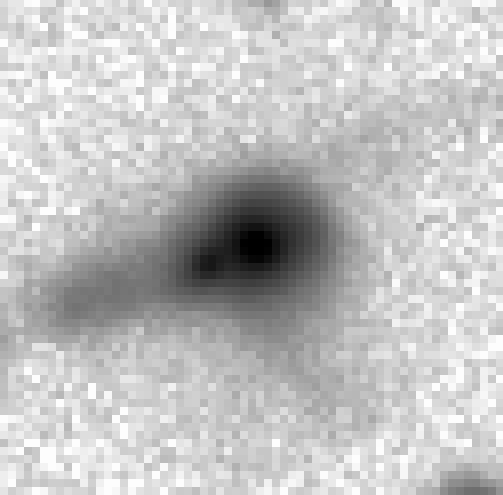}}\\
\fbox{\includegraphics[width=4.2cm,height=4.2cm]{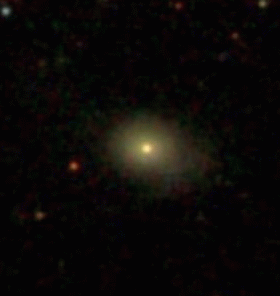}}\hfill \=
\fbox{\includegraphics[width=4.2cm,height=4.2cm]{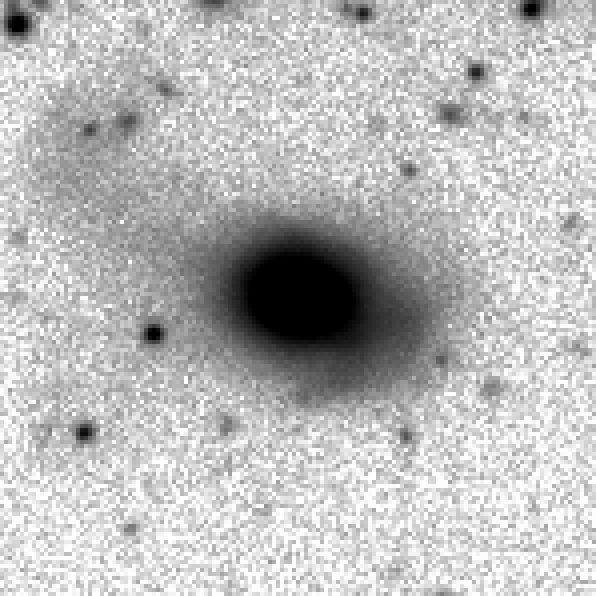}}\\
\end{tabbing}
\caption{
Cutouts from the normal-depth images provided by the SDSS DR12 navigator (left)
and from the co-adds provided by \citet{Fliri_2016} (right, inverted grey scale) for four E+A galaxies .}
\label{fig:comp_FT_SDSS}
\end{figure}

We analysed the morphology on the cutouts from the sky-rectified co-added $r_{\rm deep}$ images provided by \citet{Fliri_2016}
(see Sect.\,\ref{sec:S82}). The size of the cutouts was set to $n\cdot R_{p}$, where $R_{p}$ is the Petrosian radius of the galaxy and $n = 4$ turned out to be a good choice in most cases. 
Figure\,\ref{fig:comp_FT_SDSS} shows four examples of E+A galaxies with faint tidal structures to
illustrate the gain in surface brightness depth by the co-addition.

In some studies \citep[e.g.][]{Reichard_2009}, methods for automatic measurements of  morphological asymmetries were applied to huge numbers of galaxies. However, if the data set is not too large, the human visual system still remains the most efficient and complete system for pattern recognition.
To determine the fraction of mergers or otherwise morphologically distorted galaxies in the E+A sample,
we classified the galaxies according to morphological peculiarities recognised by a simple visual inspection.

\begin{table}[htbp]
 \centering
 \caption{Peculiarity types.}
 \begin{tabular}{lc}
 \hline\hline
 Peculiarities                             & Type $t_{\rm m}$ \\
 \hline
uncertain                                  &  $-1$\\
no peculiarities                           &  0 \\
M51-like, unusually strong bar, outer ring & 1\\
weak streamers, lopsidedness               & 2 \\
strong streamers, lopsidedness             & 3 \\ 
 \hline
 \end{tabular}
\label{tab:pec_types}
\end{table}

As a result of the inspection each galaxy was assigned to one of the types listed in Table\,\ref{tab:pec_types} and coded by the flag $t_{\rm m}$. A nearby neighbour galaxy is considered indicative of a perturbation only if the two galaxies are either embedded in a common halo or connected with each other by a light bridge. Types 2 and 3 are the best candidates for mergers. In a few cases  (e.g. J211230.60$-$005022.4 and J223006.83$-$004031.3 in sample A, and J015012.99+000504.8 and J215738.85+000416.9 in sample B), the intermediate stage of a major merger is clearly indicated. 
Other galaxies classified as type $t_{\rm m} \ge 2$ may represent either minor mergers or older major mergers. In general it is difficult to distinguish faint stellar streams in minor mergers from faded tidal debris in major mergers. 
Figures\,\ref{fig:images_merger_A} to \ref{fig:images_merger_C}
show the contrast-enhanced image cutouts for the galaxies assigned to types $t_{\rm m} \ge 2$ from the 
S82 E+A samples A, B, and C. 
The structure in the bright inner parts of the galaxies is indicated by
isophotes equally spaced on a logarithmic scale.

\begin{table}[htbp]
\centering
\caption{
Mean weighted fractions from GZ2 for different peculiarity types.}
\begin{tabular}{lrcccccc}
\hline\hline
&&&\\
Type              & $N$& $a14$& $a21$& $a22$& $a23$& $a24$& $a31$-1 \\
\hline
\multicolumn{6}{l}{S82 co-add1} \\
$t_{\rm m} = -1$  &  9 & 0.17 & 0.40 & 0.18 & 0.31 & 0.11 & 0.50\\
$t_{\rm m} = 0$   & 12 & 0.35 & 0.00 & 0.00 & 0.33 & 0.26 & 0.00\\
$t_{\rm m} = 1$   &  3 & 0.18 & 0.50 & 0.50 & 0.00 & 0.00 & 0.00\\
$t_{\rm m} = 2$   &  6 & 0.19 & 0.25 & 0.17 & 0.17 & 0.25 & 0.00\\
$t_{\rm m} = 3$   & 33 & 0.42 & 0.07 & 0.23 & 0.28 & 0.15 & 0.19\\
\hline
\multicolumn{4}{l}{S82 co-add2} \\
$t_{\rm m} = -1$  &  9 & 0.34 & 0.25 & 0.42 & 0.12 & 0.14 & 0.50\\
$t_{\rm m} = 0$   & 12 & 0.37 & 0.04 & 0.07 & 0.46 & 0.25 & 0.00\\
$t_{\rm m} = 1$   &  3 & 0.09 & 0.00 & 0.00 & 0.00 & 0.00 & 0.00\\
$t_{\rm m} = 2$   &  6 & 0.26 & 0.00 & 0.33 & 0.61 & 0.06 & 0.00\\
$t_{\rm m} = 3$   & 33 & 0.57 & 0.14 & 0.17 & 0.28 & 0.21 & 0.29\\
\hline
\end{tabular}
\label{tab:S82_statistics_0}
\end{table}

The median values of the morphological type probabilities from the first Galaxy Zoo project \citep{Lintott_2011} are $(P_{\rm e},P_{\rm s},P_{\rm m}) = (0.64,0.16,0.00)$ for the S82 E+A sample. These values are very similar to those for the entire E+A sample (Sect.\,\ref{sec:morphology}). Because of its higher completeness in SDSS S82, compared to the SDSS Legacy survey, we can make use of the more detailed classifications from Galaxy Zoo 2 (GZ2) \citep{Willett_2013} for the S82 sample. GZ2 includes classifications from co-added images in S82 in addition to the normal-depth images, where two different sets of co-adds were used (co-add1, co-add2). 
The GZ2 decision tree consists of 11 classification tasks with a total of 37 possible responses. Most relevant for the classification of distorted morphology are 
task 6 (Is there anything odd?),
task 8 (Is the odd feature a ring, or is the galaxy disturbed or irregular?), 
and task 11 (How many spiral arms are there?). 

We used the VizieR catalogue search page at CDS\footnote{vizier.u-strasbg.fr/viz-bin/VizieR} to match the S82 E+A sample
with the catalogue of morphological types from GZ2. We identified 59 galaxies (80\%) from the S82 E+A sample. We restricted the analysis to the following weighted fractions of votes:

$a$14 `something odd'

$a$21 `odd feature is a disturbed galaxy'

$a$22 `odd feature is an irregular galaxy'

$a$23 `odd feature is something else'

$a$24 `odd feature is a merger' 

$a$31-1 `one spiral arm'.

Table\,\ref{tab:S82_statistics_0} lists the mean weighted fractions for the five different types $t_{\rm m}$ from the present study, defined in Table\,\ref{tab:pec_types}. There is a trend of increasing $a14$ from $t_{\rm m} = 1$ to $t_{\rm m} =  3$, as expected. The trend is stronger for co-add2 than for co-add1. \citet{Willett_2013} found that either set of co-adds can likely be used for science, but they recommend using co-add2 if choosing between them. In the following we will use co-add2 data only. Individual values for $a14$ to $a21$ are listed in Table\,\ref{table:S82} along with $t_{\rm m}$ from the present analysis of the Fliri \& Trujillo co-adds. 

No clear-cut correlation with $t_{\rm m}$ is seen for the other GZ2 parameters. In addition, the values of $a14,\ a21,\ a22$, and $a24$ for the $t_{\rm m} = 3$ subsample are surprisingly small. For the four galaxies shown in Fig.\,\ref{fig:comp_FT_SDSS}, we find the mean co-add2 values $(a14,a21,a22,a24) = (0.48,0.16,0.18,0.20)$. Similarly, the mean values for the four intermediate-stage mergers mentioned above in Sect.\,\ref{sec:merger_class} are $(a14,a21,a22,a24) = (0.66,0.12,0.12,0.38)$. We conclude that the faint morphological distortions found in the E+A galaxies are not adequately reflected by the GZ2 data for a substantial fraction of our sample. The database from the Galaxy Zoo project  is definitely of outstanding importance for statistical studies in a wide area of applications, particularly for large samples. For the analysis of low-surface brightness features in small samples, on the other hand, it can eventually not completely supersede a target-oriented detailed visual inspection. The following analysis will be focused on the $t_{\rm m}$ classification.

\subsubsection{E+A sample versus control sample}
\label{sec:Merger_EA_vs_CS}

A limitation of the `eyeball' classification of morphology is of course its subjective nature. Drawing conclusions on the relative frequency of distorted morphology of E+A galaxies requires therefore the comparison with a control sample.  
We defined a control sample of 149 galaxies randomly selected from the SOM in such a way that its $z$ distribution is the same as for the E+A sample. To reduce the risk of a subjective bias towards overestimating peculiarities in the E+A sample
it is important to guarantee an unprejudiced evaluation of the images from both samples.
Therefore, the image cutouts of the E+A galaxies and the comparison galaxies were put into the same archive and then randomly selected in such a way that the examiner did not know to which sample a galaxy belongs when its image was inspected.

\begin{figure}[htbp]
\includegraphics[viewport= 10 10 550 790,width=0.70\hsize,angle=270]{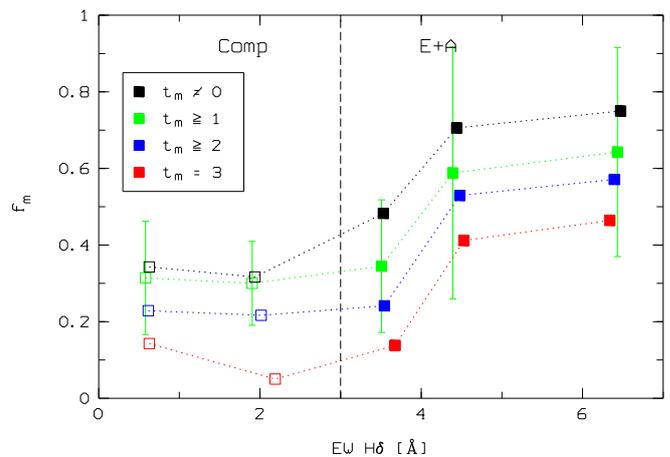}  
\vspace{0.5cm}
\caption{
Merger fractions $f_{{\rm m},i}$ in EW(H$\delta$) bins for the E+A sample (filled squares) and the control sample (open squares). Vertical bars indicate the errors from the Poisson statistics; due to clarity reasons, error bars are given only for the case $t_{\rm m} \ge 1$. The colour coding is described in the inset. 
}
\label{fig:Merger_age_Hd_1}
\end{figure}

After the classification of the morphological peculiarities, we computed relative `merger fractions' 
$f_{\rm m} = N_{\rm m}/N_{\rm tot}$, where $N_{\rm m}$ is the number of mergers and $N_{\rm tot}$ 
is the total number of galaxies in the sample. 
To discuss trends with EW(H$\delta$) we subdivide also the comparison sample in Comp A with EW(H$\delta) > 1$\AA\ and 
Comp B with EW(H$\delta) < 1$\AA.
Table\,\ref{tab:S82_statistics_2} lists statistical properties of the three S82 E+A samples and the comparison samples: the number $N$ of galaxies and the mean values (with standard deviation) of redshift $z$, 
stellar mass $\log M_\ast/M_\odot$, and stellar age $\tau_\ast$. 
The latter two quantities were taken from the Portsmouth sMSP data base (see Sect.\,\ref{sec:colour_mass}).  
In the bottom part of Table\,\ref{tab:S82_statistics_2}, the different measurements of the relative frequency of mergers $f_{\rm m}$ are listed. 

In Fig.\,\ref{fig:Merger_age_Hd_1}, we plotted $f_{{\rm m},i}$ ($i=1,2,3$ and 1,ul) versus the mean EW(H$\delta$) for the five samples from Table\,\ref{tab:S82_statistics_2}. The data points were interconnected just to guide the eye. The main result is that E+A galaxies with  EW(H$\delta) > 4$ \AA\ have a higher merger fraction
$f_{\rm m}$, though the formal error bars are large.
To check whether the difference is significant or not we applied the statistical test for the comparison of two relative frequencies described by  \citet{Sachs_1982}.
The null hypothesis, $H^{\, 0}: f_{\rm m}^{\rm \, E+A} = f_{\rm m}^{\rm \, C}$ is tested against the alternative hypothesis 
$H^{\rm \, A}: f_{\rm m}^{\, E+A} > f_{\rm m}^{\rm \, C}$ at an error probability $\alpha$. The null hypothesis
$H^{\, 0}$ is rejected in favour of $H^{\rm \, A}$\ if the test statistics 
\begin{equation}
\hat{z} =  \frac{f_{\rm m}^{\rm \, E+A}-f_{\rm m}^{\rm \, C}}{\sqrt{f_{\rm m}^{\rm \, T}\cdot 
      \Big[1-f_{\rm m}^{\rm \, T}\Big]}}\cdot \sqrt{\frac{N_{\rm m}^{\rm \, E+A}\cdot N_{\rm m}^{\rm C}}{N_{\rm m}^{\rm \, T}}},
\end{equation}\label{eq:statistics}
is larger than a critical value $\hat{z}_\alpha$.
The upper indices indicate the sample with C = comparison and T = total (i.e. E+A plus comparison sample).
For the statistical test we excluded the subsample C and considered only E+A galaxies with EW(H$\delta)>4$\AA. We
combined the S82 E+A subsamples A and B and compared that test sample with the control sample. 

Table\,\ref{tab:S82_statistics_1} lists the relative fractions $f_{\rm m}$ and the test statistics $\hat{z}$ 
based on $t_{\rm m}$ in the upper part and on GZ2 S82 co-add2 weighted fractions in the bottom part.
In the upper part, the number of distorted galaxies (`mergers') is defined as $N_{\rm m} = N(t_{\rm m}\ge i)$. 
As an upper limit  for the observed relative merger frequency we consider the case 
$i = 1$;ul with $N_{\rm m} = N(t_{\rm m} \ge 1) + N(t_{\rm m} = -1)$. 
For the GZ2 data we used $N_{\rm m}(a) = N(a > a_{\rm crit})$  with the liberal criterion $a_{\rm crit} = 0.5$.
The relative frequency of mergers is $f_{\rm m} = f_{{\rm m},i} = N(t_{\rm m}\ge i)/N_{\rm tot}$ where $i = 1$;ul means $f_{\rm m,1; ul} = [N(t_{\rm m}\ge 1)+N(t_{\rm m}=-1)]/N_{\rm tot}$, which can be considered as a kind of an upper limit for the observed relative merger frequency. The relative merger frequencies from the GZ2 data are defined as $f_{\rm m}(a) = N_{\rm m}(a>a_{\rm crit})/N_{\rm tot}$ using the moderate criterion $a_{\rm crit} = 0.5$.
We found $\hat{z} > 3$ for the four selections based on $t_{\rm m}$ and also for the GZ2 parameters $a14$ and $a22$.
These $\hat{z}$ values are clearly larger than $\hat{z}_\alpha = 2.58$ for the (low) error probability $\alpha = 0.01$. 
Consequently, $H^{\, 0}$ has to be rejected in favour of $H^{\rm \, A}$ at an error probability less than 1\% for these tests.
We conclude that both our $t_{\rm m}$ classifications and the GZ2 data support the view that the 
E+A sample has a significantly higher merger fraction than the control sample.

\begin{table}[htbp]
\centering
\caption{ 
Comparison of relative fractions of morphological distortions in the E+A samples A and B with
the control sample C.}
\begin{tabular}{lrrr}
\hline\hline
&&&\\
Measure of distortion  & $f_{\rm m}^{\rm \ E+A}$ & $f_{\rm m}^{\rm \ C}$ & $\hat{z}\ \ \ $ \\
&&&\\
\hline
\multicolumn{4}{l}{Present study} \\
$i=1;$ul \ $(t_{\rm m} \ne 0)$ & 0.76 & 0.36 & 4.36 \\
$i=1     \ (t_{\rm m} \ge 1)$  & 0.63 & 0.35 & 3.12 \\
$i=2     \ (t_{\rm m} \ge 2)$  & 0.58 & 0.26 & 3.71 \\
$i=3     \ (t_{\rm m} = 3)$    & 0.47 & 0.11 & 5.11 \\
\hline
\multicolumn{4}{l}{GZ2 S82 co-add2} \\
$a$14                   & 0.32 & 0.10 & 3.33 \\
$a$21                   & 0.00 & 0.05 &-1.40 \\ 
$a$22                   & 0.13 & 0.02 & 3.18 \\
$a$23                   & 0.13 & 0.20 & -0.9 \\
$a$24                   & 0.08 & 0.07 & 0.23 \\
$a$31-1                 & 0.08 & 0.02 & 1.65 \\
\hline
\end{tabular}
\label{tab:S82_statistics_1}
\end{table}

\begin{table*}[htbp]
 \centering
 \caption{Statistical properties in dependence on EW(H$\delta$) for the S82 E+A samples (PSB) and the comparison sample (Comp).}
 \begin{tabular}{lcccccc}
 \hline\hline
Sample             &PSB A
                   &PSB B
                   &PSB C
                   &Comp A
                   &Comp B\\
EW(H$\delta$)\ [\AA] &$> 5$
                   &$4.0\ldots5.0$
                   &$3.0\ldots4.0$
                   &$>1$
                   &$<1$\\
(mean value $\pm$ SD)&($6.30\pm1.01$)
                   &($4.36\pm0.31$)
                   &($3.37\pm0.29$)
                   &($2.43\pm1.29$)
                   &($0.09\pm0.70$)\\
 \hline
 $N$               &28
                   &17
                   &29
                   &81
                   &68\\
$z$           &0.140 $\pm0.066$
                   &0.117 $\pm0.054$
                   &0.116 $\pm0.060$
                   &0.125 $\pm0.060$
                   &0.128 $\pm0.059$\\
age [Gyr]     &0.93  $\pm0.72$
                   &1.14  $\pm0.59$
                   &1.46  $\pm0.59$
                   &2.64  $\pm2.56$
                   &3.58  $\pm3.47$\\
log $M_\ast/M_\odot$&10.10 $\pm0.54$
                   &10.11 $\pm0.50$
                   &10.09 $\pm0.49$
                   &9.95 $\pm2.06$
                   &9.84 $\pm2.56$\\                   
$u-r$ [mag]        &2.39  $\pm 0.38$
                   &2.41  $\pm 0.27$
                   &2.47  $\pm 0.27$
                   &
                   & \\
\hline                   
$f_{\rm m,3}$      &0.46 
                   &0.41 
                   &0.14 
                   &0.09 
                   &0.15 \\                   
$f_{\rm m,2}$      &0.57 
                   &0.53 
                   &0.24 
                   &0.25 
                   &0.26 \\                   
$f_{\rm m,1}$      &0.64
                   &0.59
                   &0.34
                   &0.33
                   &0.35\\                                   
$f_{\rm m,1; ul}$  &0.75
                   &0.71
                   &0.48
                   &0.35
                   &0.37\\                                       
 \hline
 \end{tabular}
 \label{tab:S82_statistics_2}
\end{table*}

\subsubsection{Trends with age}

In a simple one-generation starburst population, the H$\delta$ absorption line reaches a maximum at an age of 0.1 - 1 Gyr when the continuum is dominated by the A stars. The strength of the H$\delta$ line is thus an age indicator \citep[e.g.][]{Reichard_2009}.
The Portsmouth sMSP database provides information on the ages of the stellar population in the SDSS galaxies
derived by the fit of stellar population models to the observed SDSS photometry.
On the basis of a set of single stellar populations, composite population models
were created adopting different star formation histories. Exponentially declining SFRs are parametrised as 
$SFR(t) = SFR(t_0) \times \exp{-(t-t_0)/\tau}$, where the SFR  starts at the time $t_0$.
At a cosmic time $t$, the age derived for a galaxy from the best fit is then defined as 
the time $t - t_0$ elapsed since the beginning of star formation in the best-fitting population model. This value is not necessarily identical with the true age of the galaxy, but can be taken as a measure of the age of the stellar population dominating the optical luminosity. As in Sect.\,\ref{sec:colour_mass}, we used the data set portsmouth\_stellarmass\_starforming\_krou-26-sub  for the galaxies from SDSS DR8.

\begin{figure}[htbp]
\includegraphics[viewport= 10  5 550 790,width=0.70\hsize,angle=270]{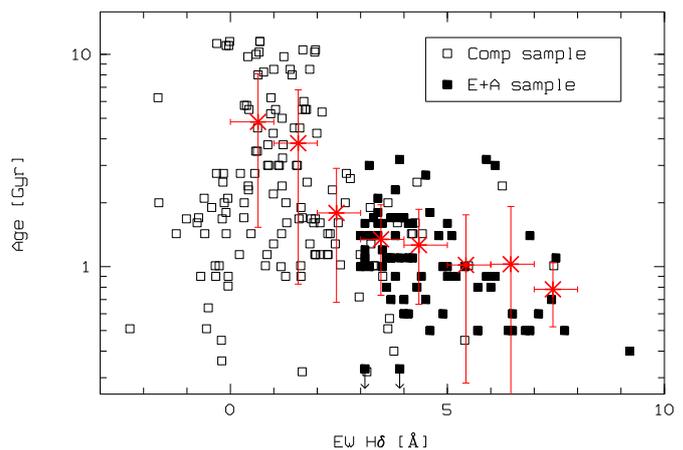} 
\vspace{0.5cm}
\caption{
{
Stellar age versus EW(H$\delta$) for the galaxies from the E+A sample (filled squares) and from the control sample (open squares). The two filled squares with downward arrows indicate ages $< 0.25$ Gyr. The red asterisks are mean ages in the EW(H$\delta$) bins indicated by the horizontal bars, the vertical bars are 1$\sigma$ standard deviations.
}
}
\label{fig:Merger_age_Hd_2}
\end{figure}

\begin{figure}[htbp]
\includegraphics[viewport= 10 10 550 790,width=0.70\hsize,angle=270]{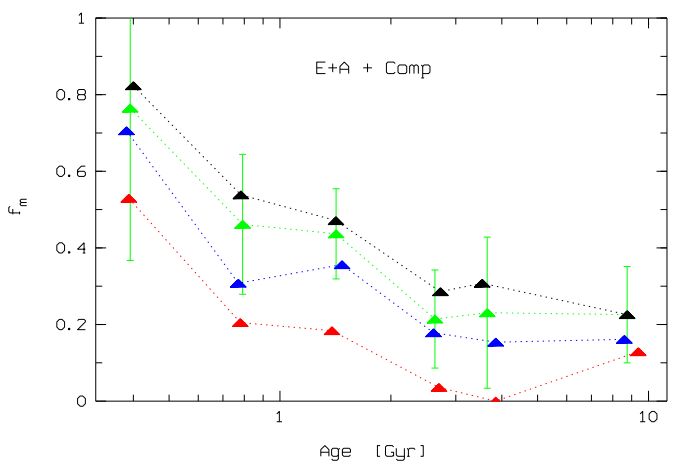}
\vspace{0.5cm}
\caption{
{
Merger fraction $f_{{\rm m},i}$ in age bins for the combined S82 sample (E+A sample plus comparison sample).
Colours and error bars as in Fig.\,\ref{fig:Merger_age_Hd_1}.
}
}
\label{fig:Merger_age_Hd_3}
\end{figure}

Figure\,\ref{fig:Merger_age_Hd_2} shows the sMSP age as a function of EW(H$\delta)$ for the combined sample of E+A galaxies plus the galaxies from the comparison sample. Most of the E+A galaxies have ages $< 2$ Gyr. For the vast majority in the S82 E+A sample A, the age is slightly smaller than 1 Gyr, which is consistent with previous results \citep{Poggianti_1999,Goto_2007b}. For the galaxies with EW(H$\delta) = 0 - 3$ \AA\  from the comparison sample, on the other hand, the scatter is much larger. This sample includes both systems dominated by old stars and such systems dominated by a young stellar population. The mean age increases from $0.93\pm0.72$ Gyr for sample A, to $1.14\pm0.59$ Gyr for sample B, $1.46\pm0.59$ Gyr for sample C, and more than 3 Gyr for the comparison sample (Table \ref{tab:S82_statistics_2}). In 
Fig.\,\ref{fig:Merger_age_Hd_2}, we over plotted the ages averaged in EW(H$\delta)$ bins of the width 1 \AA, starting at
EW(H$\delta) = 0.5 \pm 0.5$. The mean ages continuously decrease with EW(H$\delta)$ from $\sim 5$ Gyr to $\sim 800$ Myr, though the scatter is large.

We used the sMSP ages to study the dependence of the merger fraction on the stellar population age.
Figure\,\ref{fig:Merger_age_Hd_3} shows $f_{\rm m}$  in six age bins, again for the 
combined sample of S82 E+A plus comparison galaxies. The merger fraction increases with decreasing age for ages $\la 2$ Gyr.
The coincidence with the characteristic relaxation time of a galaxy after a merger \citep[e.g.][]{Bournaud_2005,Duc_2013} hints at a causal connection between the starbursts and the morphological perturbations. 
In the youngest age bin ($< 0.5$ Gyr), the percentage of distorted galaxies reaches 71\% (82\%) for $t_{\rm m} \ge 2$ 
\ ($t_{\rm m} \ne 0$), compared to $\la$ 20\% (30\%) for the old ( $> 2$ Gyr) subsample that is dominated by the galaxies from the comparison sample.  
There are 17 galaxies with age $< 0.5$ Gyr in the combined sample, among them 11 E+A galaxies (65\%). For the E+A galaxies with age $< 0.5$ Gyr we find 73\% (91\%), which is comparable with 75\% found by \citet{Sell_2014} for a sample of 12 massive, young PSB galaxies at $z \sim 0.6$ studied on {\it HST} images. The high merger fraction of 67\% (67\%) for the comparison galaxies in that age bin is not surprising, given that a young sMSP age may indicate an ongoing starburst. 

The panels in the middle row and on the right-hand side of Fig.\,\ref{fig:colour_mass} show the locations of the S82 E+A galaxies in the colour-mass plane. With very few exceptions, the galaxies populate the area between the blue cloud and the red cloud, as was found already for the entire E+A sample. In the middle row, sample A galaxies are shown in red, sample B and C in blue. There is a weak trend for galaxies with stronger H$\delta$ lines to be on average slightly bluer. 
Because of the correlation of EW(H$\delta$) with the age of the stellar population (Fig.\,\ref{fig:Merger_age_Hd_2}),
this is consistent with the idea that E+A galaxies are in a rapid transition phase from the blue cloud towards the red cloud. No such trend is seen in the morphological distortions on the right-hand side of Fig.\,\ref{fig:colour_mass}.

The Portsmouth sMSP also provides estimates of the SFR. 
However, only three galaxies from the S82 E+A sample have SFR $> 0$ (J$020505.99-004345.1$, J$025850.52+003458.7$, J$225506.79+005840.0$), all three belong to sample A. 
Their specific star formation rates sSFR = SFR/($M_\ast/M_\odot) = 0.04, 0.003,$ and 0.025 Gyr$^{-1}$ place them into the transition region between the blue cloud of star forming galaxies and the red sequence in the sSFR-$M_\ast$ diagram \citep[see][Fig.2]{Heckman_2014}. 
For $\log\,M_\ast/M_\odot \approx 10.1$ (Table\,\ref{tab:S82_statistics_2}), this transition occurs at  
sSFR $\approx 0.02$\,Gyr$^{-1}$, corresponding to a SFR $\approx 0.2\,M_\odot$ yr$^{-1}$. 
The zero SFR from the Portsmouth data for the vast majority of the galaxies in our sample 
is thus in agreement with the idea that we selected galaxies where the star formation is quenched.

\vspace{0.5cm}

%
\subsection{Hidden AGNs}
\label{sec:AGN}

Interactions and mergers are believed to provide an important channel for fuelling the central supermassive black hole and triggering AGNs. The energetic feedback from the AGN is thought to play a role in the co-evolution of galaxies and supermassive black holes \citep[see for reviews][]{Fabian_2012,Heckman_2014}. 
There are two categories of AGN feedback: the radiative mode and the jet mode.

A classical radiative-mode AGN of type 1 is characterised by a blue continuum and strong and broad emission lines. As a consequence of the selection criteria (Sect.\,\ref{sec:catalogue}), the E+A sample is lacking luminous type 1 AGNs. However, the central region of a galaxy can be hidden by dust, especially when the sight line is near to the plane of the putative obscuring torus (type 2 AGN) or crosses dense molecular clouds in the host galaxy. The latter case is particularly relevant in the context of the merger driven AGN scenario
with `wet' mergers where lots of gas and dust are concentrated towards the central region \citep{Sanders_1988,Hopkins_2006,Bennert_2008}. 
The integrated energy density of the cosmic X-ray background suggests that supermassive black holes grow mostly during phases when the AGN is obscured \citep{Fabian_1999}.
Highly obscured AGNs seem to prefer infrared luminous galaxies \citep{Symeonidis_2013}
and were found in morphologically disturbed hosts \citep{Urrutia_2008,Koss_2011}. 
   
In this Section, the final catalogue of all E+A galaxies is used to search for candidates of optically hidden AGNs.

\subsubsection{Mid infrared selection}

The presence of an optically hidden AGN can be indicated by the thermal emission in the infrared. 
Luminous AGNs are robustly differentiated from galaxies and stars by their red mid-infrared (MIR) colours. 
We exploited the database from the Wide-field Infrared Survey Explorer \citep[WISE;][]{Wright_2010}.
WISE performed an all-sky survey with images in four broad bands W1 to W4 around the central wavelengths 3.4, 4.6, 12, and 22 $\mu$m.  The difference of the magnitudes $W1$ and $W2$  is typically $\sim 1$ mag for low-redshift quasars. An efficient selection threshold for quasars is provided by the colour criterion $W1-W2 \ge 0.8$ \citep{Assef_2010,Stern_2012}. 
For a low-luminosity AGN the spectrum is diluted by stellar radiation from the host galaxy and the  
MIR colours depend on the AGN-to-host ratio. For early-type hosts $W1-W2$ becomes smaller with a decreasing ratio. A modestly extincted AGN in an early-type galaxy of the same luminosity is still expected to produce $W1-W2 \approx 0.7$ at $z \la 0.3$ \citep[see Figs. 1 and 2 in][]{Stern_2012}. However, if the flux from the AGN is less than half of the host flux in the WISE bands, the integrated colour index will hardly exceed $W1-W2 \approx 0.5$.

\begin{figure}[h]
\includegraphics[width=1.0\hsize,angle=0]{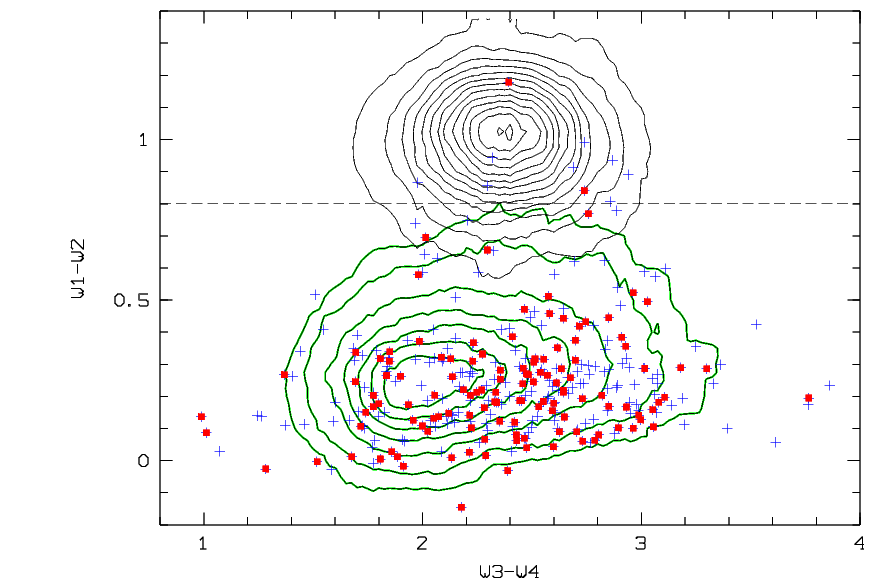}\\
\caption{WISE mid infrared two-colour diagram for the E+A galaxies
(blue crosses for EW(H$\delta)=3-5$\AA, red squares for EW(H$\delta)>5$\AA).
The contour curves indicate the distributions of 9\,468 SDSS galaxies with $0.02 \le z \le 0.25$ (green) and of 2\,373 SDSS quasars with $z < 0.5$ (black), respectively. Dashed vertical lines: AGN selection criterion thresholds.
}
\label{fig:WISE_4colours}
\end{figure}

We identified counterparts in the AllWISE Source Catalogue for 95\% of the E+A galaxies from our catalogue
within a search radius of 3 arcsec. The size of the WISE detected subsample is strongly reduced, however, when a good 
S/N is required.
Figure\,\ref{fig:WISE_4colours} shows the $W1-W2$ versus $W3-W4$ two-colour diagram
for the 310 galaxies with S/N $>5$ for $W1, W2, W3$, and $>2$ for $W4$.
The E+A galaxies from our catalogue are plotted as blue plus signs, filled red squares represent the 126 E+A galaxies fulfilling the stronger selection criteria from Goto's catalogue (in particular EW(H$\delta) > 5$\AA). The distribution of quasars and normal galaxies is over-plotted for comparison. Among the 6\,231 quasars with $z<0.5$ from the SDSS DR7 quasar catalogue \citep{Schneider_2010,Shen_2011}, 2\,373 have reliably measurements in W3 and W4 (sigma $<$ 0.3 mag). Their locus on the two-colour plane is shown by the black equally spaced local point density contours. We identified WISE counterparts of 83\,844 SDSS galaxies with $z=0.02 - 0.25$ from the Portsmouth sMSP database. The green contour curves show the 9\,468 galaxies with S/N $>5$ for $W1, W2, W3$, and $>3$ for $W4$. The dashed vertical line indicates the demarcation $W1-W2 = 0.8$. 

The vast majority of our E+A galaxies with reliable WISE measurements populate the region below the AGN demarcation line, the mean colour is $W1-W2 = 0.27 \pm 0.20$.  Only 10 sources are classified as AGN candidates based on the $W1-W2$ criterion corresponding to 3\% of the plotted E+A galaxies, 
their mean colour is $W1-W2 = 0.92\pm0.11$.
The WISE E+A sample becomes much larger (2\,569 galaxies) when there are no restrictions made on S/N in the W3 and W4 bands, which are not necessary for the AGN selection. However, there is no additional E+A galaxy with $W1-W2 > 0.8$ in this  larger sample, and the corresponding fraction of AGN candidates becomes as small as 0.4\%. There is no obvious trend of $W1-W2$ with EW(H$\delta$). None of the AGN candidates is lying in S82. 

\begin{figure}[h]
\includegraphics[width=1.0\hsize]{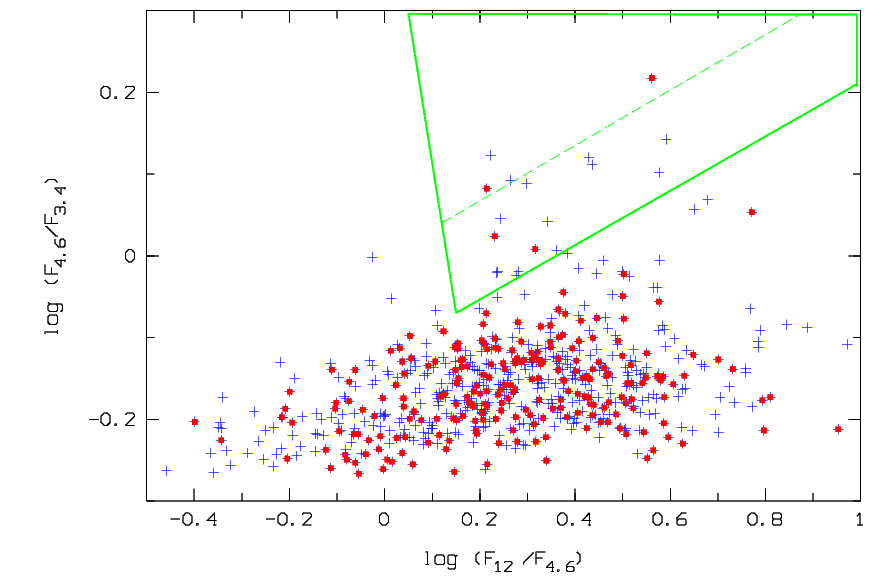}
\caption{WISE mid infrared two-colour diagram for the E+A galaxies based on the three WISE bands at
3.4, 4.6, and 12 $\mu$m.  Symbols as in Fig.\,\ref{fig:WISE_4colours}. The green lines indicate the AGN selection suggested by \citet{Mateos_2012}.}
\label{fig:WISE_3colours}
\end{figure}

The depth of WISE at 22 $\mu$m is significantly shallower than in the other bands. Non-detection and low S/N in the W4 band are the main reasons for the strong reduction of the galaxy sample in Fig.\,\ref{fig:WISE_4colours}. 
\citet{Mateos_2012} presented an AGN selection criterion based on the first three WISE bands. It was  
defined using a large complete flux-limited sample of bright hard-X-ray-selected AGNs of types 1 and 2.
Figure\,\ref{fig:WISE_3colours} shows the E+A sample in the log($F_{\rm 4.6 \mu m}/F_{\rm 3.4 \mu m})$ versus
log($F_{\rm 12 \mu m}/F_{\rm 4.6 \mu m})$ two-colour diagram in the same style as presented by \citet{Mateos_2012},
where $F=F_\nu$ is the monochromatic flux density per frequency interval.
The selection criteria for the E+A galaxies are the same as in Fig.\,\ref{fig:WISE_4colours}, but without restriction on S/N in the W4 band. The number of galaxies is now 596 (blue plus signs), among them 232 fulfilling the stronger Goto selection criteria (red squares). The `AGN wedge' is indicated by the green solid lines. 18 galaxies (3\%) are located  in the AGN wedge.
The dashed line illustrates the MIR power-law locus for different spectral indices 
\citep[taken from][Fig.\,6]{Mateos_2012}. \citet{Mateos_2012} found a strong dependence of log($F_{\rm 4.6 \mu m}/F_{\rm 3.4 \mu m})$ on the X-ray luminosity. Luminous quasars, 
expected to dominate the MIR emission, are preferentially located above the power-law locus. Less powerful AGNs, where the contribution from the host galaxy emission becomes significant, have bluer colours at the shortest WISE wavelengths, consistent with normal galaxies. As found by \citet{Mateos_2012} for their sample of X-ray AGNs, the log($F_{\rm 12 \mu m}/F_{\rm 4.6 \mu m})$ colour of the E+A sample shows a large scatter at blue log($F_{\rm 4.6 \mu m}/F_{\rm 3.4 \mu m})$ colours. They argued that such a broad range is expected for AGNs of lower luminosities with an important host galaxy contribution.

\begin{figure}[ht]
\includegraphics[viewport= 40 0 600 780,width=0.95\hsize,angle=270]{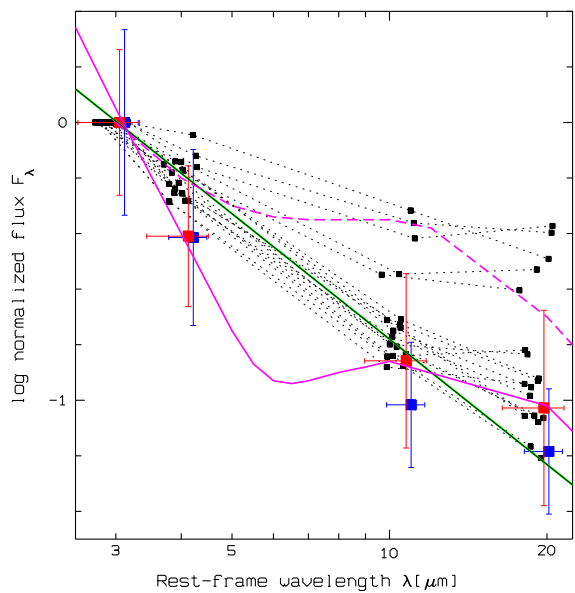}
\caption{
SED of E+A samples with EW(H$\delta) > 5$\AA\ (red) and 3-5 \AA\ (blue),
and of AGNs from WISE (black). Magenta curves: two-component model from \citet{Melnick_2013} for
PSB galaxies (solid) and three-component model from \citet{Melnick_2015} for post-starburst quasars (dashed).
Green line: power law $F_\lambda \propto \lambda^{-1.5}$.
}
\label{fig:WISE_SED}
\end{figure}

The distribution of the E+A galaxies in the $W1-W2$ versus $W3-W4$ diagram (Fig.\,\ref{fig:WISE_4colours}) is shifted towards redder $W3-W4$ compared to the distribution of the typical SDSS galaxies. 
\citet{Melnick_2013} found that all galaxies in their PSB sample show significant excess at MIR wavelengths, in particular in W3 and W4, compared to the galaxy model SED.
Figure \,\ref{fig:WISE_SED} shows
the composite rest-frame MIR SEDs of our E+A subsamples with EW(H$\delta) > 5$\AA\ (red) and             EW(H$\delta) = 3-5$\AA\ (blue), normalised at the de-redshifted wavelength of the W1 band. The coloured squares with vertical error bars are the mean values and 1$\sigma$ standard deviations in the four WISE bands, the horizontal bars indicate the wavelength range covered by the rest-frame central wavelengths in each band. The selection criteria are the same as in Fig.\,\ref{fig:WISE_4colours}. We also plotted the individual SEDs of the 18 AGN candidates from the WISE three-band selection, where the data points in the four bands where interconnected solely to guide the eye. 
For comparison we over-plotted the MIR part of the best-fit two-component model SEDs used by \citet{Melnick_2013} to analyse a sample of PSB galaxies with EW(H$\delta) > 5$\AA\ (solid magenta) and the three-component model presented by \citet{Melnick_2015} to fit the SED of PSB quasars (dashed magenta). The former invokes, in addition to the modelled star light, a 300 K blackbody from hot dust to fit the MIR excess. The dust may be heated by an embedded AGN, though other sources are possible as well (young star clusters or asymptotic giant branch stars).
For the PSB quasars \citet{Melnick_2015} added
a power law component with $F_\lambda \propto \lambda^{-1.5}$ (green diagonal line) to match the flatter MIR SED. The best-fit two-component model perfectly fits the WISE data of our E+A subsample with stronger H$\delta$ (red). The E+A galaxies with EW(H$\delta) = 3-5$ \AA\ (blue) have on average a weaker MIR bump. The situation is less clear
for the AGN candidates. At short WISE wavelengths, the decline is weaker because of the selection criterion, and the SED is fitted by the three-component model, which is dominated there by the power-law component. 
Contrary to the whole E+A sample, the majority ($\sim 72$\%) of the AGN candidates do obviously not require a substantial blackbody component.

\begin{figure}[ht]
\begin{tabbing}
\includegraphics[viewport= 50 0 600 790,width=0.53\hsize,angle=0]{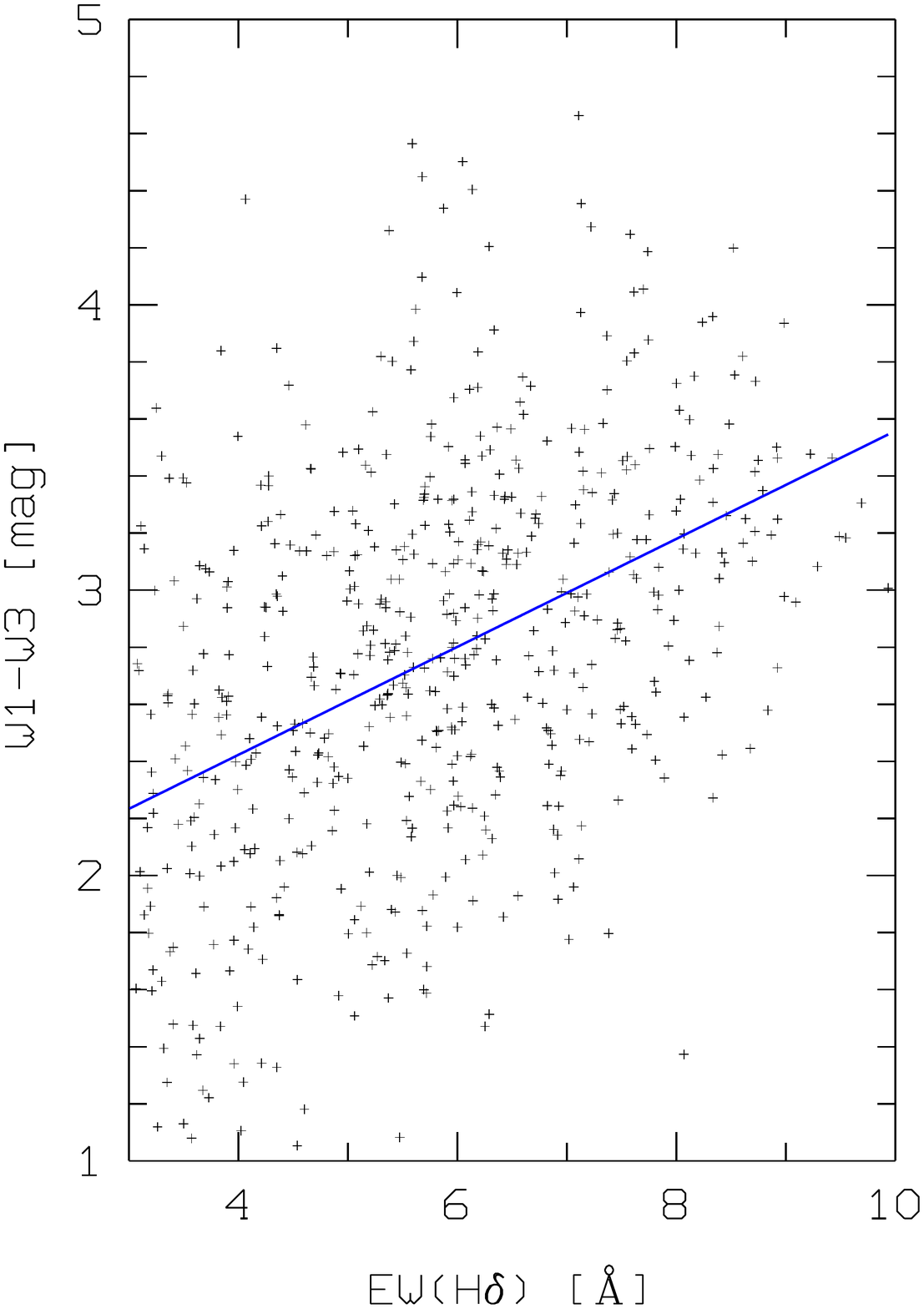}\hfill \=
\includegraphics[viewport= 50 0 600 790,width=0.53\hsize,angle=0]{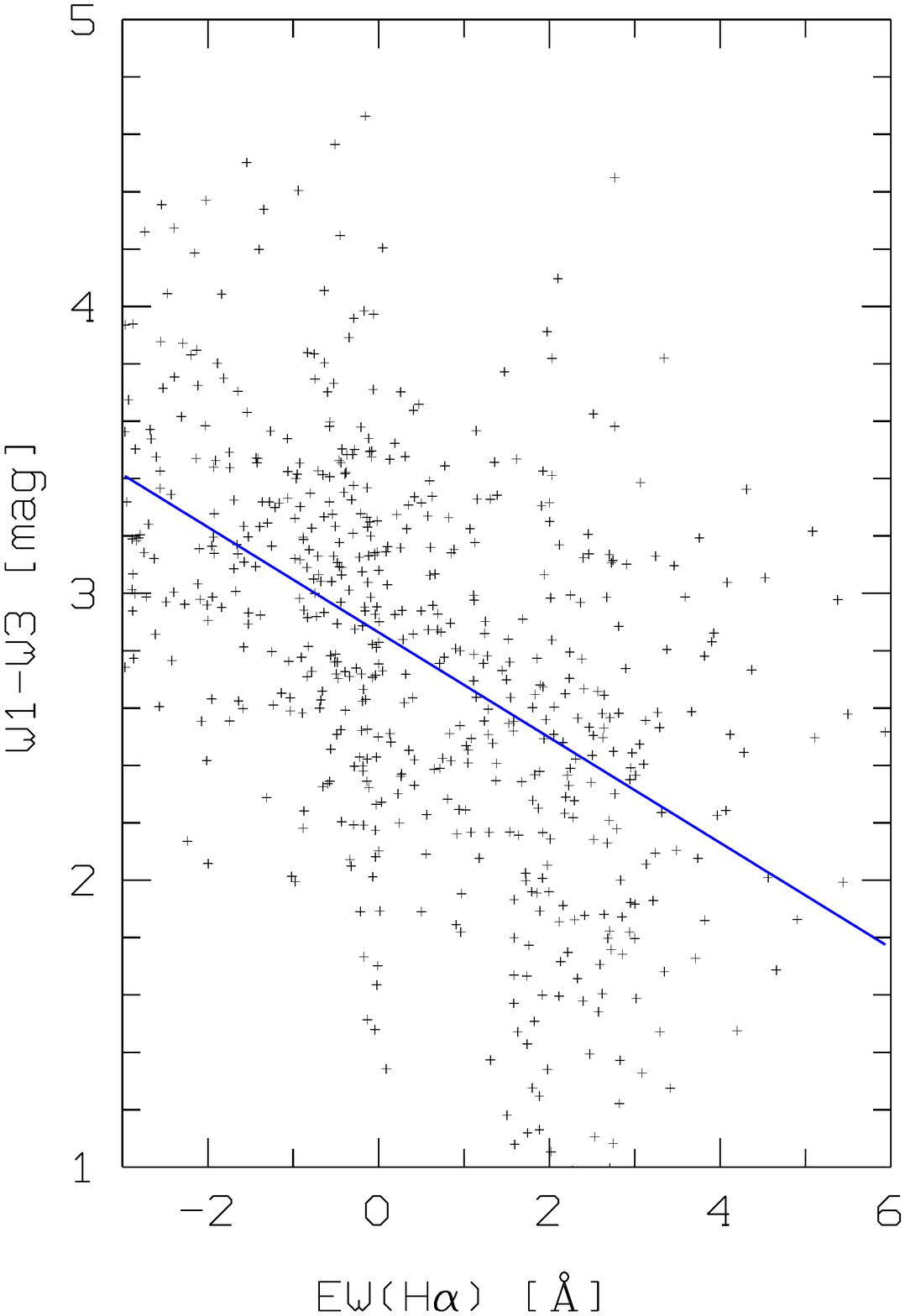}
\end{tabbing}
\caption{
WISE colour index $W1-W3$ as a function of the equivalent width of H$\delta$ (left) and H$\alpha$ (right.) The blue diagonal lines are the linear regression curves.
}
\label{fig:WISE_MIR_excess}
\end{figure}

\citet{Melnick_2013} found that the MIR excess is correlated with the intrinsic reddening. In Fig.\,\ref{fig:WISE_MIR_excess}, we show that the excess, expressed by the colour index $W1-W3$, is also correlated
with the strength of the optical lines. Stronger excesses are observed in galaxies with larger EW(H$\delta$) and
smaller EW(H$\alpha$). These correlations can be explained by a reddened emission line component 
(AGN or starburst) in addition to the PSB stellar component.

\subsubsection{Radio selection}

For a jet-mode AGN optical emission lines can be weak or absent, but the AGN can still be detected at moderate or weak radio luminosity. We exploited the source catalogue from the {\it Faint Images of the Radio Sky at Twenty Centimeters}
\citep[FIRST;][]{Becker_1995,Helfand_2015}. There was no E+A galaxy in Stripe 82 identified with a FIRST source. Also the inspection of the radio images from the {\it High-Resolution VLA Imaging of the SDSS Stripe 82} \citep{Hodge_2011} did not reveal any radio counterpart. 

Next we matched the entire E+A catalogue with the FIRST source catalogue. Using a search radius of 5\arcsec\ we identified FIRST counterparts of 47 E+A galaxies. All identified FIRST galaxies appear unresolved in the FIRST image cutouts. Because of the (by definition) weak or missing emission lines in the spectra of E+A galaxies, it is difficult to distinguish AGNs from starburst origin of the radio emission in terms of diagnostic line ratios. In a statistical sense, the different radio luminosity functions of AGNs and star forming galaxies can be used for the discrimination, because AGNs become more frequent with increasing radio luminosity. 
We used the integrated FIRST fluxes $F_{\rm 1.4, int}$ to compute the monochromatic luminosity at 
1.4 GHz (rest frame)
\begin{equation}
L_{1.4} = 4\pi\, D_{\rm L}^2\, F_{\rm 1.4, int}\, (1+z)^{-1+\alpha}
\end{equation}
\citep[see e.g.][]{Morrison_2003,Nielsen_2012},
where a power law $F_\nu \propto \nu^{\,-\alpha}$ was adopted with $\alpha$ = 0.8 in the radio domain;  
$D_{\rm L}$ is the luminosity distance.

\begin{figure}[h]
\includegraphics[viewport= 20 20 580 790,width=0.73\hsize,angle=270]{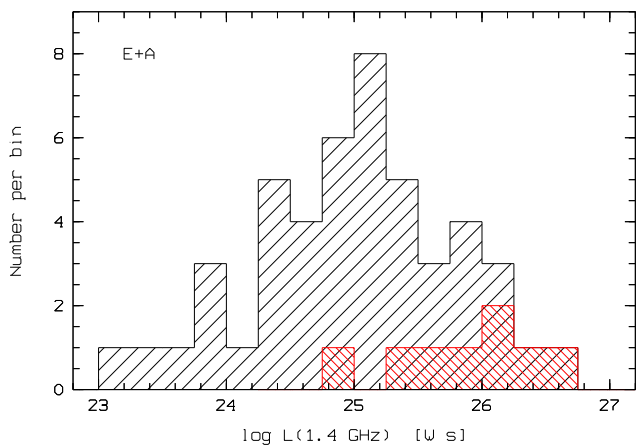}
\caption{Histogram distribution of the monochromatic 1.4 GHz luminosities (black: all, red: radio-loud) for the E+A sample.}
\label{fig:hist_1p4}
\end{figure}

The distribution of $L_{1.4}$ is shown in Fig.\,\ref{fig:hist_1p4}. 
All FIRST E+A galaxies have $\log L_{1.4} > 23$, the mean value is $24.9\pm0.7$, where $L_{1.4}$ is in Ws.
Eight galaxies (17\%) are classified as radio-loud when the criterion $R_i >1$ is applied to the radio-loudness parameter $R_i$ from \citet{Ivezic_2002}\footnote{$R_i$ is the logarithm of the ratio of the 1.4 GHz flux to the flux in the SDSS i band.}. Their radio luminosity distribution is clearly shifted to higher values (mean value $\log L_{1.4} = 25.6\pm0.5$).
Following \citet{Sadler_2002}, AGNs dominate the radio luminosity distribution for $L_{1.4} > 10^{23}$ Ws.
All 47 FIRST E+A galaxies have luminosities above that threshold. This result remains qualitatively unchanged when we correct for the different cosmological model used by \citet{Sadler_2002}. We conclude that the majority of the FIRST 
E+A galaxies likely host AGNs.

\subsubsection{X-ray selection}

For completeness we briefly mention that X-rays  provide  an  alternative  way  to identify AGNs,  complementing the optical, MIR, and radio identification techniques. X-ray surveys have identified thousands of AGNs, contributing significantly to our knowledge of this object class and may lead to the discovery of otherwise hidden AGNs \citep[e.g.][]{Mateos_2012,Civano_2014,Paggi_2016}. 
With the exception of the most heavily obscured AGNs, surveys with the X-ray satellite missions {\it XMM-Newton} and {\it Chandra} are sensitive to all types of AGNs.
Based on the analysis of XMM-Newton and Chandra data overlapping $\sim 16.5$  deg$^2$ of S82, \citet{LaMassa_2013} presented a compilation of 
3\,362 unique X-ray sources detected at high significance. We matched our S82 E+A list to this X-ray catalogue with a search radius of 10 arcsec. In no case an X-ray counterpart was found.

\subsubsection{E+A sample versus control sample}

The relative fractions of AGN candidates in the E+A sample have to be compared with those in a control sample. We constructed a comparison sample in the following way. For each galaxy $i$ from the E+A sample with redshift $z_i$ and mass $\log M_i$ we selected a galaxy in the Portsmouth sMSP catalogue with $z \in (z_i-0.015,z_i+0.015)$ and 
$\log M  \in (\log M_i - 0.1,\log M_i + 0.1)$. Thus, the E+A sample and the control sample have the same distribution on the 
$z - \log M$ plane. The control sample was matched with the AllWISE Source Catalogue and with the FIRST catalogue in the same way as the E+A sample.  

We applied the statistical test from Sect.\,\ref{sec:Merger_EA_vs_CS} to compare the relative AGN frequencies from the E+A sample with the control sample . The results are listed in Table\,\ref{tab:AGN_EA_comp}. The null hypothesis, 
$H^{\, 0}: f_{\rm AGN}^{\rm \, E+A} = f_{\rm AGN}^{\, C}$ is tested against the alternative hypothesis 
$H^{\rm \, A}: f_{\rm AGN}^{\rm \, E+A} > f_{\rm AGN}^{\rm \, C}$ at an error probability $\alpha = 0.01$. The null hypothesis
$H^{\, 0}$ is retained for the WISE two-band selection. On the other hand, $H^{\, 0}$ is rejected in favour 
of $H^{\rm \, A}$ for the WISE three-band selection. The fraction of AGN candidates selected by the two-colour AGN wedge in Fig.\,\ref{fig:WISE_3colours} is significantly larger for the E+A galaxies.

\begin{table}[htbp]
\centering
\caption{ 
Comparison of relative AGN fractions in the E+A sample and the control sample.}
\begin{tabular}{lrrr}
\hline\hline
&&&\\
Selection    & $f_{\rm AGN}^{\rm \ E+A}$ & $f_{\rm AGN}^{\rm \ C}$ & $\hat{z}\ \ \ $ \\
&&&\\
\hline
WISE 2 bands & 0.0042          &  0.0040       & 0.01  \\
WISE 3 bands & 0.0284          &  0.0073       & 0.48  \\ 
FIRST        & 0.0183          &  0.0387       &-0.69  \\
FIRST rl\tablefootmark{\ a}     & 0.1702          &  0.1939       &-0.14  \\
\hline
\end{tabular}
\label{tab:AGN_EA_comp}
\tablefoot{
\tablefoottext{a}{Here, $f_{\rm AGN}$ means the fraction of radio-loud AGNs among the FIRST detected sources.}
}
\end{table}

\begin{figure}[h]
\includegraphics[viewport= 20 20 580 790,width=0.73\hsize,angle=270]{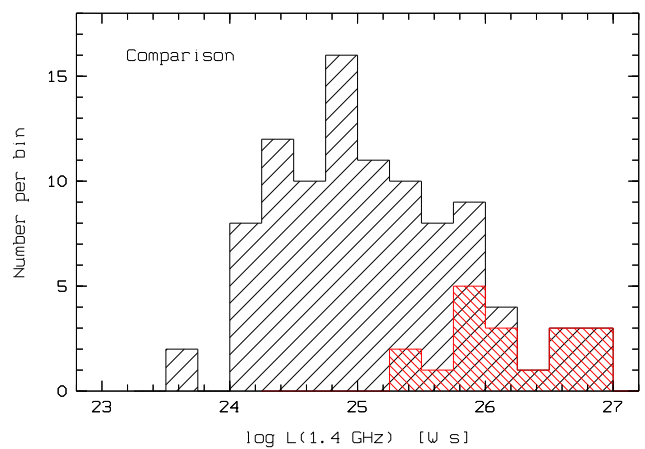}
\caption{
As Fig.\,\ref{fig:hist_1p4}, but for the comparison sample.
}
\label{fig:hist_1p4_CS}
\end{figure}

The opposite is found for the radio selection: The control sample shows a significantly larger relative fraction of FIRST sources than the E+A sample. For the brightness range of our sample (z-band magnitude $\la 18$),  \citet{Ivezic_2002} found that FIRST galaxies represent about 5\% of all SDSS galaxies, in agreement with $\sim 4\%$ found here for the control sample.
The fraction of 1.8\% FIRST galaxies among the E+A galaxies is thus rather small. However, the direct comparison of these values is difficult because of the different nature of the samples. FIRST SDSS galaxies are dominated by galaxies from the red sequence \citep[$u-g \approx 3$,][]{Ivezic_2002}, whereas the E+A galaxies typically lie between the blue cloud and the red sequence (Fig.\,\ref{fig:colour_mass}). The mean age of the FIRST E+A galaxies is 1.7 Gyr compared to 2.3 Gyr for FIRST galaxies in the control sample.  On the other side, the distribution of the radio luminosities is very similar for the two samples              (compare Figs.\,\ref{fig:hist_1p4} and \ref{fig:hist_1p4_CS}). The difference between the radio-loudness fractions is statistically insignificant (Table\,\ref{tab:AGN_EA_comp}).

\begin{table}[h]\centering
\caption{Statistical properties of the AGN candidates from FIRST and WISE compared with 
the entire E+A sample.}
\begin{tabular}{lccc}
\hline\hline
&&&\\
                      &   WISE 3 bands   & FIRST            &   E+A (all)     \\
\hline
number                & 18               & 47               & 2\,665          \\
$z$                   & $ 0.16 \pm 0.05$ & $ 0.24 \pm 0.10$ & $ 0.14 \pm 0.06$\\
age [Gyr]             & $ 1.28 \pm 0.73$ & $ 1.72 \pm 1.71$ & $ 1.55 \pm 1.45$\\
$\log M_\ast/M_\odot$ & $10.40 \pm 0.26$ & $10.39 \pm 0.49$ & $10.22 \pm 0.51$\\
$u-r$ [mag]           & $ 2.50 \pm 0.29$ & $ 2.51 \pm 0.30$ & $ 2.48 \pm 0.33$\\
EW(H$\delta$) [\AA]   & $ 5.74 \pm 1.77$ & $ 5.78 \pm 2.59$ & $ 5.25 \pm 4.49$\\
EW(H$\alpha$) [\AA]   & $ 0.26 \pm 3.12$ & $ 1.55 \pm 2.43$ & $ 1.93 \pm11.81$\\
EW(\ion{O}{ii}) [\AA] & $-2.07 \pm 1.11$ & $-1.20 \pm 1.56$ & $-0.82 \pm 3.21$\\
$P_{\rm e}$           & $ 0.52 \pm 0.22$ & $ 0.57 \pm 0.22$ & $ 0.59 \pm 0.19$\\
$P_{\rm s}$           & $ 0.27 \pm 0.17$ & $ 0.23 \pm 0.20$ & $ 0.23 \pm 0.19$\\
\hline
\end{tabular}
\label{tab:AGN}
\end{table}

\begin{figure}[h]
\includegraphics[width=1.0\hsize]{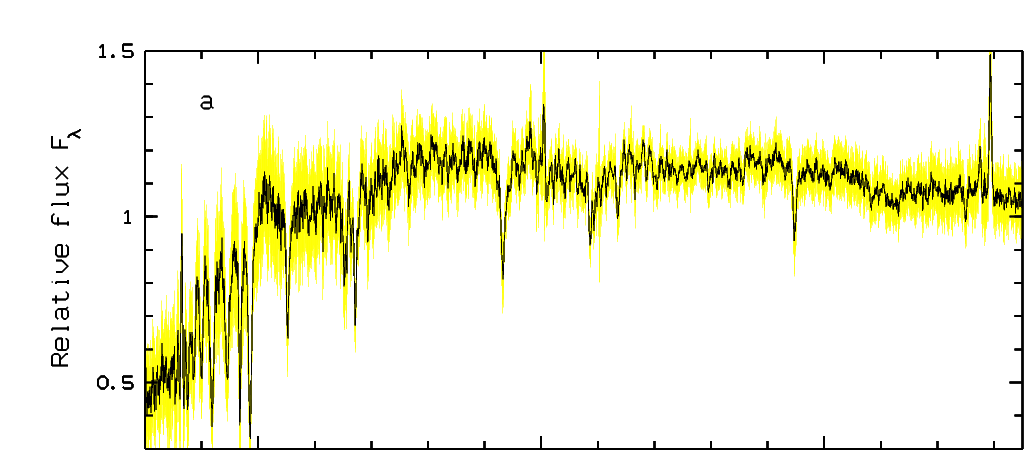} \\
\includegraphics[width=1.0\hsize]{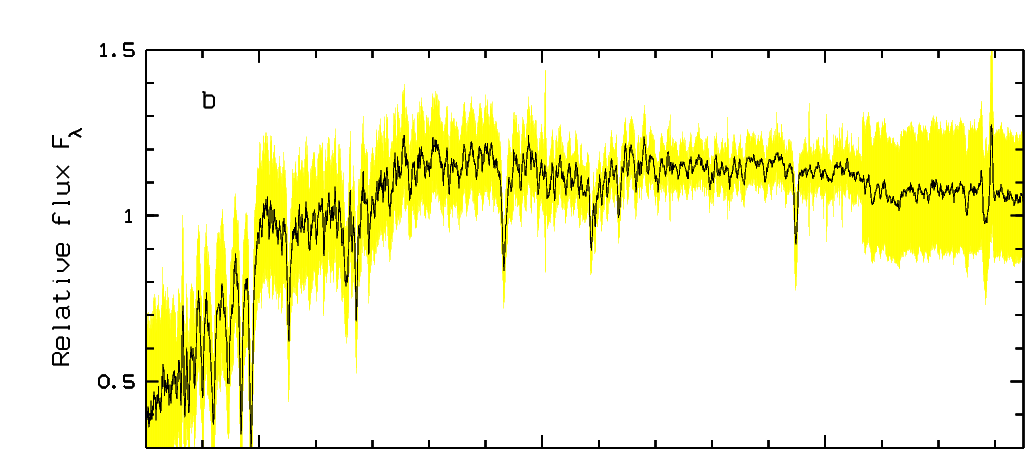} \\
\includegraphics[width=1.0\hsize]{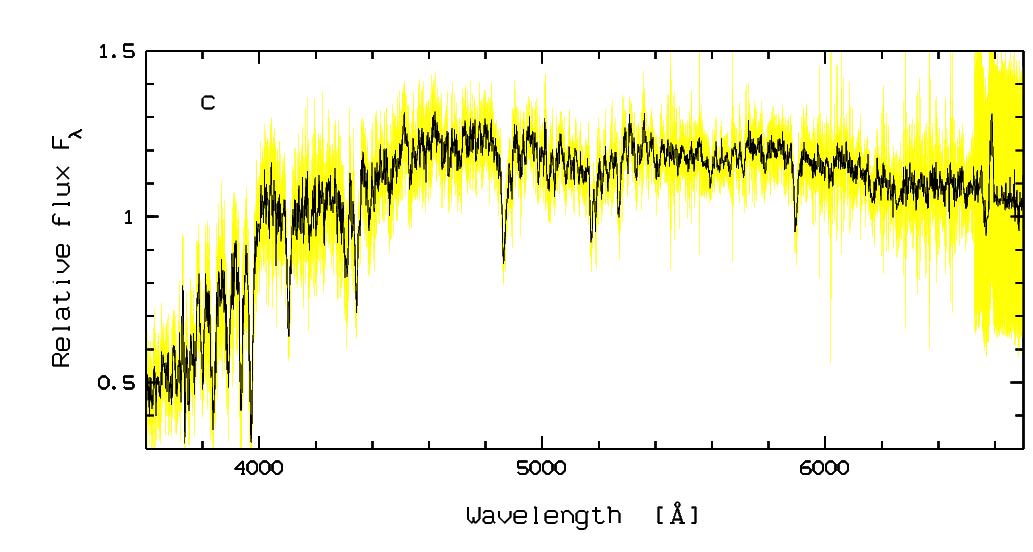}
\caption{Median composite (black) and standard deviations (yellow) for (a) 18 AGN candidates from the WISE 3-band selection, (b) 39 radio-quiet FIRST sources, (c) 8 radio-loud FIRST sources.
}
\label{fig:composites_WISE_FIRST}
\end{figure}

\subsubsection{Properties of the AGN candidates}

The combination of the WISE selection and the FIRST selection yields a sample of 62 AGN candidates (2.3\%) among 
the E+A sample: 18 objects from WISE and 47 from FIRST with an overlap of three objects in both samples.

Figure\,\ref{fig:composites_WISE_FIRST} compares  the composite spectrum of the WISE AGN candidates (top) with that of the FIRST E+A galaxies, where we distinguish between radio-quiet (middle) and radio-loud ones (bottom). The differences are only subtle. What is most conspicuous is the somewhat stronger emission from [\ion{N}{ii}]$\lambda 6584$ and [\ion{O}{iii}]$\lambda 5007$ in the WISE-selected composite (top). 

\begin{figure*}[htbp]
\begin{tabbing}
\includegraphics[viewport= 40 30 540 540,width=4.5cm,angle=270]{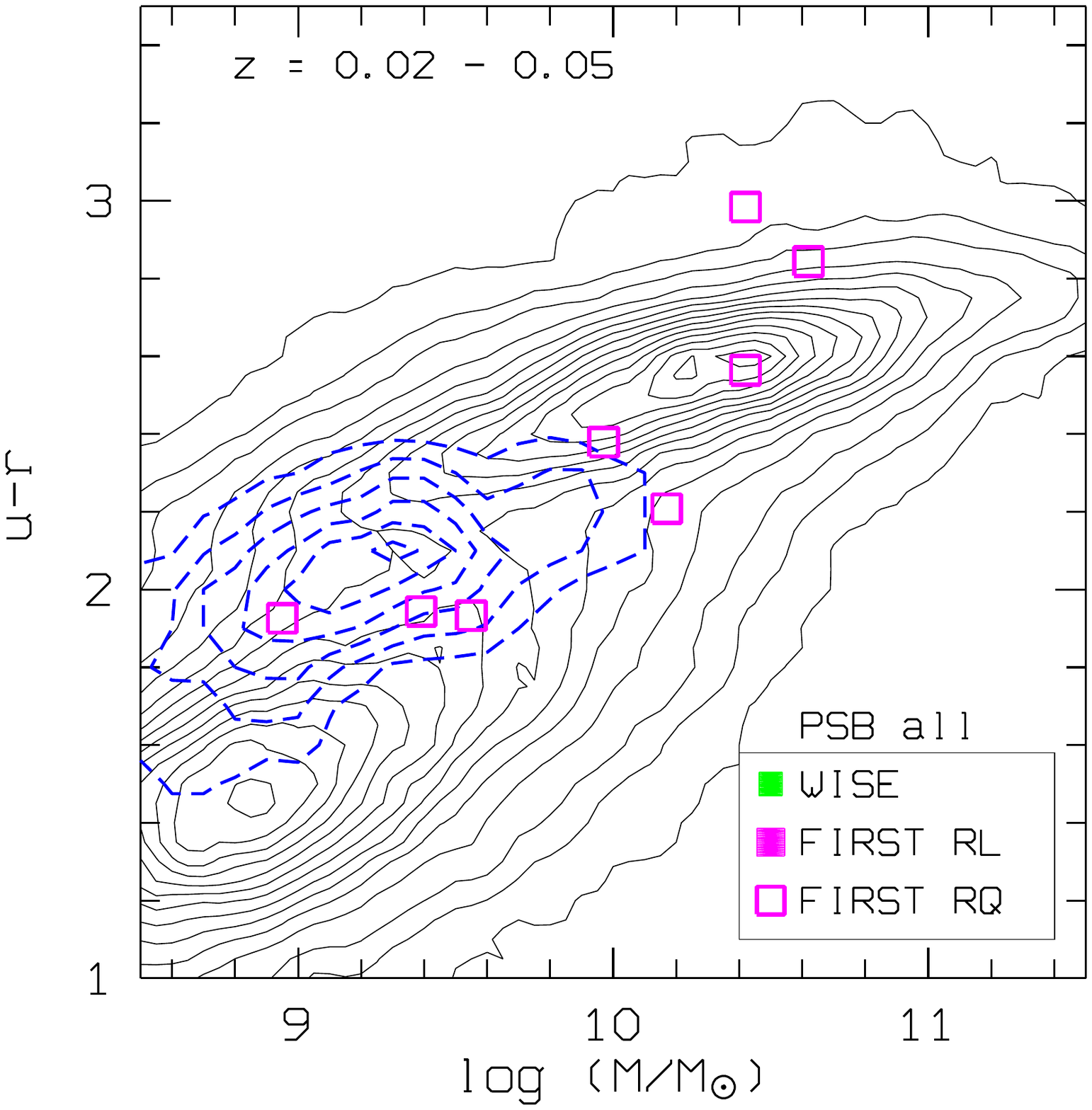}\hfill \=
\includegraphics[viewport= 40 30 540 540,width=4.5cm,angle=270]{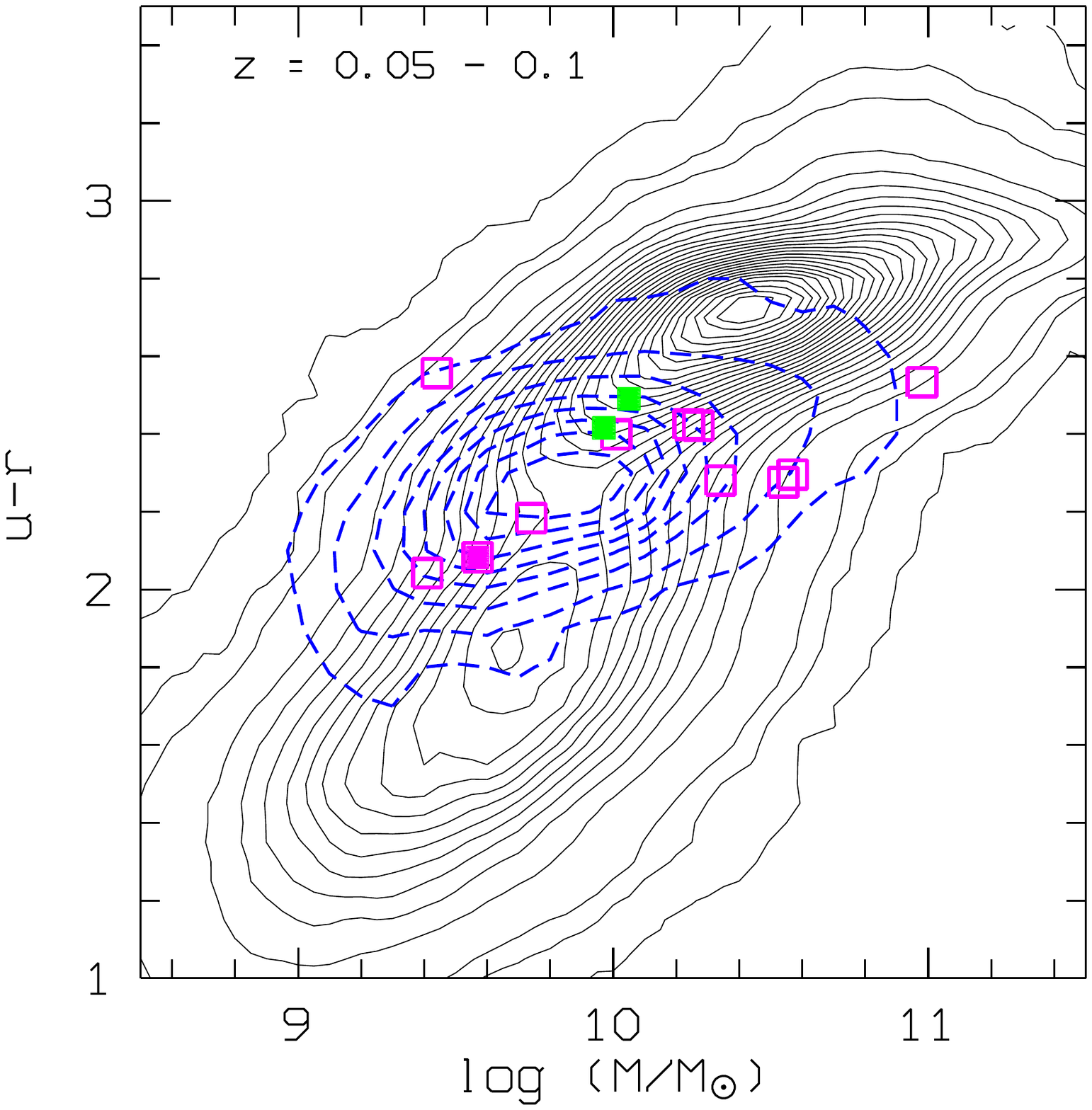}\hfill \=
\includegraphics[viewport= 40 30 540 540,width=4.5cm,angle=270]{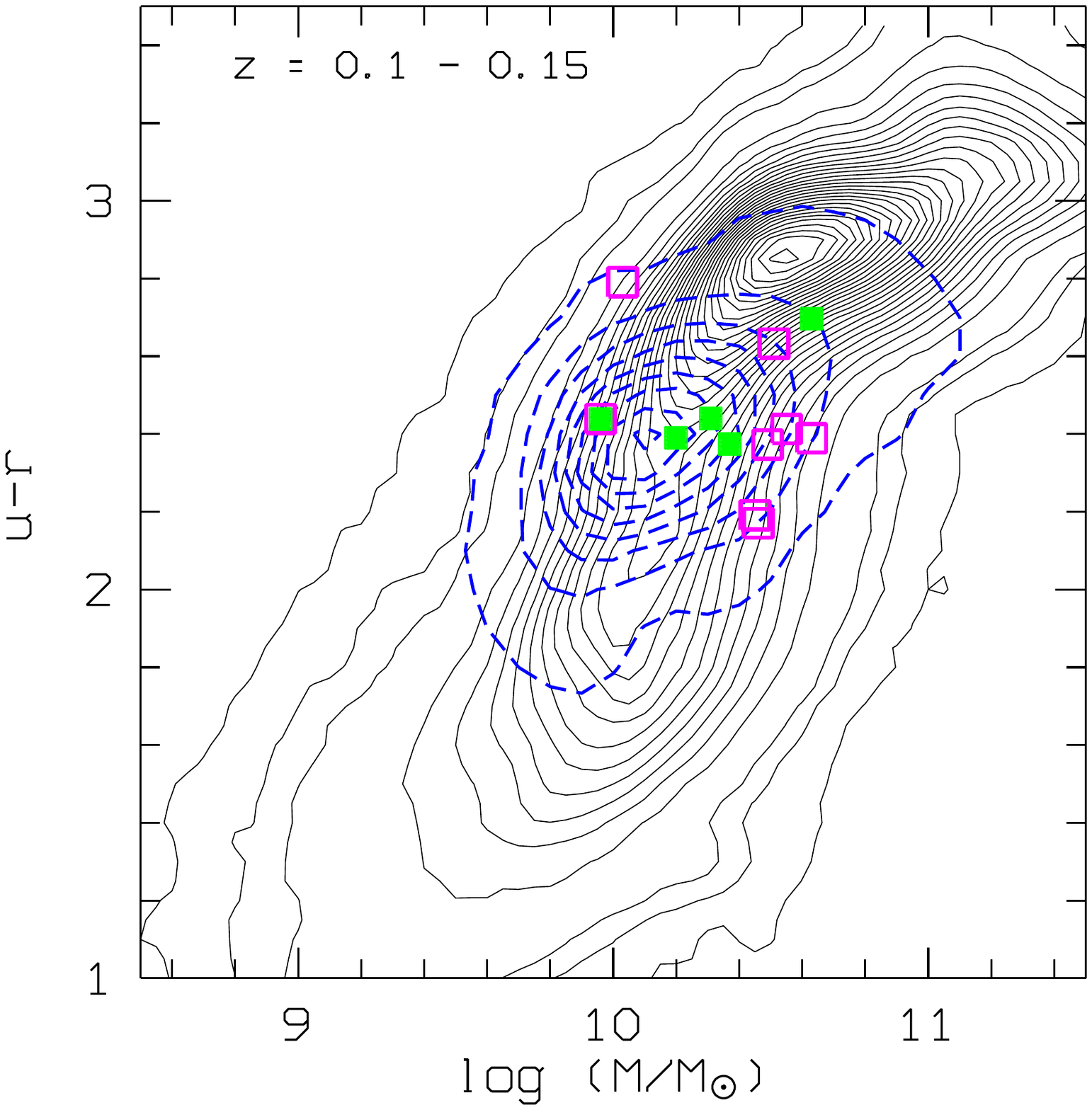}\hfill \=
\includegraphics[viewport= 40 30 540 540,width=4.5cm,angle=270]{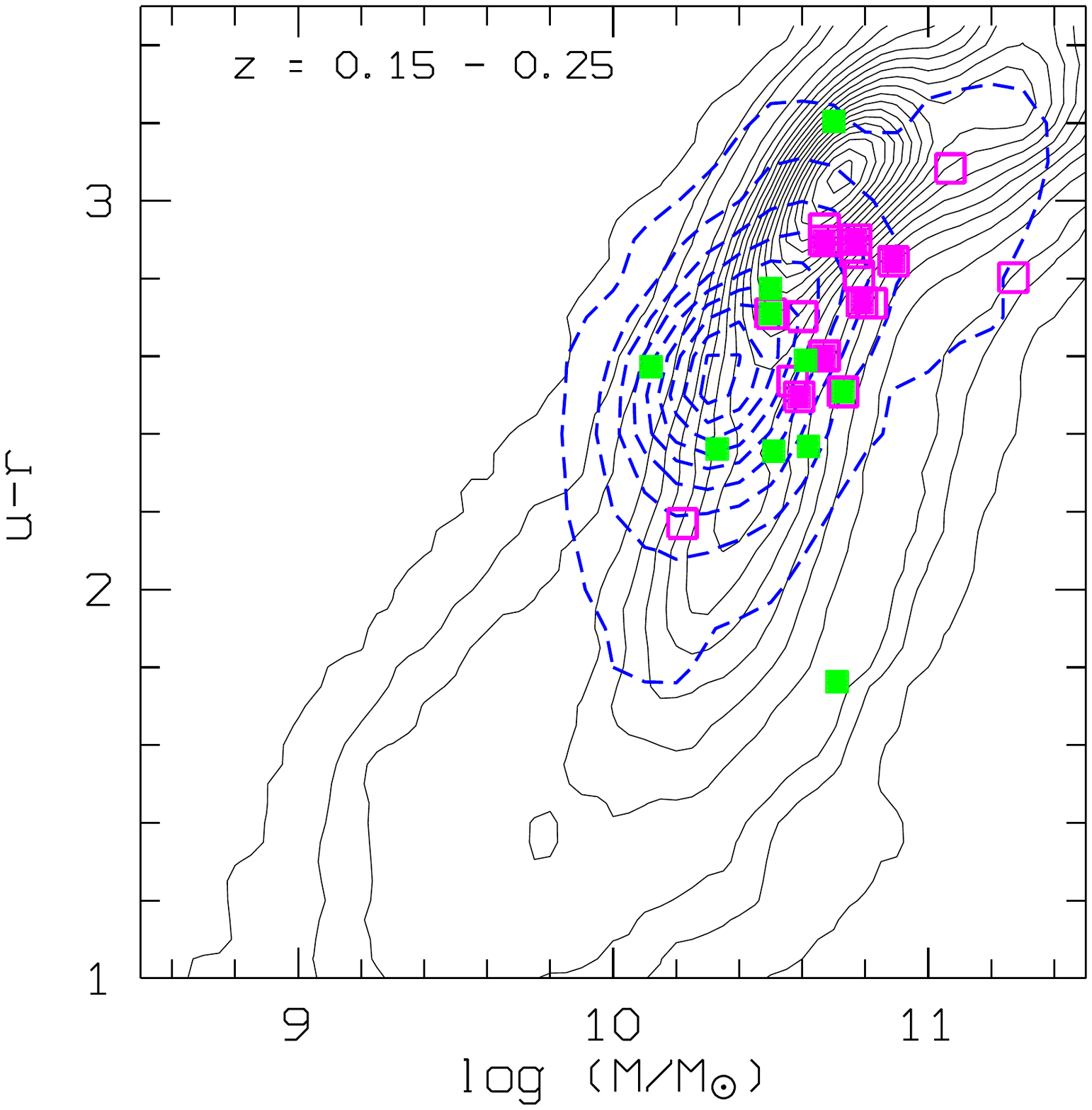}
\end{tabbing}
\caption{Colour-mass diagrams for the AGN candidates with $z<0.25$ from WISE (green) and FIRST (magenta) in four redshift bins.  The FIRST sources are subdivided in radio-loud (RL = filled squares) and radio-quiet (RQ = open squares). The contour curves show the distributions of all SDSS galaxies (black) and of our E+A galaxies (blue, dashed).
}
\label{fig:colour_mass_AGN}           
\end{figure*}

Some properties of the AGN subsamples are listed in Table\,\ref{tab:AGN}, compared with the whole E+A sample. The distribution on the colour-mass plane is shown in Fig.\,\ref{fig:colour_mass_AGN}.
The strongest difference is the higher mean redshift of the FIRST E+A galaxies. 
Both AGN subsamples have somewhat higher mean masses compared to the entire E+A sample. 
The percentage of galaxies with $\log M/M_\odot > 10.3$ is 51\% for all E+A, but 68\% for the radio-detected and 88\% for the radio-loud sources. 
It is known that radio galaxies in optical and radio flux limited samples tend to be biased
towards higher $z$ and thus towards higher optical luminosities and stellar masses \citep{Ivezic_2002}.
However, the tendency to be related to higher stellar masses is indicated for both AGN candidate samples and is seen in each panel of Fig.\,\ref{fig:colour_mass_AGN}. In each $z$ bin, the distribution of the AGN candidates is centred on a somewhat higher mass than the corresponding E+A sample.
The stellar ages show a wide scatter, the median values are 1.14, 0.90, and 1.14 Gyr for the FIRST, the WISE, and the total sample, respectively. Both AGN candidate samples have on average stronger H$\delta$ absorption and 
stronger $\ion{O}{ii}$ emission lines. No significant differences are seen in the morphological type probabilities from Galaxy Zoo.

Finally, we briefly discuss whether the E+A galaxies in our sample are massive enough for a supermassive black hole in the centre.  
The expected mean black hole mass $M_{\rm BH}$  can be estimated from the scaling relation between  
$M_{\rm BH}$ and the mass $M_{\rm sph}$ of the spheroidal component of a galaxy.  The mean stellar mass from the Portsmouth sMSP database is $\log M_\ast/M_\odot = 10.22$ (Table\,\ref{tab:AGN}). We assumed a ratio  $M_{\rm sph}/M_\ast = 0.5 \ldots 1$ because the deep S82 co-adds clearly indicate the presence of a significant spheroidal component for many E+A galaxies. 
The empirical $M_{\rm BH} - M_{\rm sp}$  relation for galaxies with $\log M_{\rm sph}/M_\odot \la  10.5$ 
derived by \citet{Scott_2013} yields $\log M_{\rm BH}/M_\odot = 7.0 \ldots 7.7$. 
Masses in this range are typical of AGNs in the local Universe.
\citet{Greene_2007} reported a characteristic mass of $\log M_{\rm BH}/M_\odot = 7.0$ 
for the local populations of broad and narrow-line AGNs . Recently, \citet{Kara_2016} listed BH masses for 11 Seyfert galaxies with reverberation detections, the mean mass is $\log M_{\rm BH}/M_\odot = 7.0 \pm 0.6$. Even values as small as $M_{\rm BH} \approx 10^6 M_\odot$ are not unreasonable for Seyfert galaxies \citep[e.g.][]{Greene_2007,Bentz_2016b,Bentz_2016a}.

%
\section{Summary and conclusions}

E+A galaxies are thought to represent PSB galaxies in the transition from the blue
cloud to the red sequence of galaxies (and perhaps back). 
We created a large sample of local E+A galaxies from the spectra database of the SDSS.
The selection is based on a Kohonen SOM for $\sim 10^6$ spectra 
from SDSS DR7. The process takes advantage of the clustering behaviour 
of Kohonen maps. We developed an interactive user interface based on modern web techniques for 
navigating through huge SOMs. Both the SDSS DR7 Kohonen map and the user interface are made available 
for the community. Using the catalogue of 837 E+A galaxies with EW(H$\delta)>5$\AA\ from Goto as 
input catalogue, we selected a large sample of similar galaxies from the Kohonen map and
compiled a new list of 2\,665 E+A galaxies with EW(H$\delta)>3$\AA\ 
in the redshift range $z = 0 - 0.4$. 

We used the galaxy data from the Portsmouth galaxy property computations to study particularly
the distribution of the E+A galaxies in the colour-mass diagram. We have shown that the E+A galaxy sample is clearly concentrated towards the green valley between the red sequence and the blue cloud, independent of the redshift.
There is a weak tendency for strong H$\delta$ galaxies to be closer to the blue cloud.
These findings are in agreement with the idea that PSB galaxies represent the transitioning phase between actively and passively evolving galaxies. 

It is commonly believed that galaxy interactions and wet mergers play a major role for 
triggering starburst activity. It was one of the main aim of this study, to create a sizable sample 
of E+A galaxies in the SDSS S82 that can be used to investigate the morphological  
peculiarities on the deep co-added images of the SDSS multi-epoch observations in S82.
We exploited the new deep co-adds from Fliri \& Trujillo (2016)
to analyse 74 local E+A galaxies found in S82 along with galaxies from a randomly selected control sample 
with the same $z$ distribution. We identified morphological peculiarities in many E+A galaxies. 
The relative frequency of distorted galaxies in the E+A sample is significantly 
higher than in the control sample and indicates a trend with the age of the stellar population. 
In the youngest age bin ($< 0.5$ Gyr), 
at least 73\% of the E+A galaxies are classified as distorted or merger. 

Our study confirms that AGNs are rare in E+A galaxies, unless they are heavily obscured or Compton thick. 
We identified 18 E+A galaxies in the AGN wedge of the WISE three-band colour-colour diagram and 47 galaxies selected as FIRST 1.4 GHz sources, with only a small overlap of these two samples. Based on the 1.4 GHz luminosities we argue that the majority of the FIRST sources are related to AGNs rather than star formation. The corresponding AGN fractions are 2.8\% for the MIR selection and 1.8\% for the radio selection. The radio AGN fraction is significantly lower than in the comparison sample, whereas the MIR AGN fraction is higher. We also confirm the MIR excess found by \citet{Melnick_2013} for the
majority of our E+A galaxies. The origin of this excess may be dust heated by young stars or low-luminosity AGNs. 
These results are not in contradiction with a scenario where stellar mass and black hole mass grow simultaneously in massive galaxies and where suppression of star formation and AGN activity rapidly follow each other.

%
\begin{acknowledgements}

The anonymous referee is thanked for helpful comments and suggestions.\\

This research has made use of data products from the Sloan
Digital Sky Survey (SDSS). Funding for the SDSS and SDSS-II has been provided by
the Alfred P. Sloan Foundation, the Participating Institutions
(see below), the National Science Foundation, the National
Aeronautics and Space Administration, the U.S. Department
of Energy, the Japanese Monbukagakusho, the Max Planck
Society, and the Higher Education Funding Council for
England. The SDSS Web site is http://www.sdss.org/.
The SDSS is managed by the Astrophysical Research
Consortium (ARC) for the Participating Institutions.
The Participating Institutions are: the American
Museum of Natural History, Astrophysical Institute
Potsdam, University of Basel, University of Cambridge
(Cambridge University), Case Western Reserve University,
the University of Chicago, the Fermi National
Accelerator Laboratory (Fermilab), the Institute
for Advanced Study, the Japan Participation Group,
the Johns Hopkins University, the Joint Institute
for Nuclear Astrophysics, the Kavli Institute for
Particle Astrophysics and Cosmology, the Korean
Scientist Group, the Los Alamos National Laboratory,
the Max-Planck-Institute for Astronomy (MPIA),
the Max-Planck-Institute for Astrophysics (MPA),
the New Mexico State University, the Ohio State
University, the University of Pittsburgh, University
of Portsmouth, Princeton University, the United
States Naval Observatory, and the University of
Washington. \\

This publication has made extensive use of the VizieR catalogue access
tool, CDS, Strasbourg, France and of data obtained from 
the NASA/IPAC Infrared Science Archive (IRSA), operated by the 
Jet Propulsion Laboratories/California Institute of Technology, founded
by the National Aeronautic and Space Administration. In particular,  
this publication makes use of data products from the Wide-field Infrared Survey Explorer, which is a joint project of the University of California, Los Angeles, and the Jet Propulsion Laboratory/California Institute of Technology, funded by the National Aeronautics and Space Administration.

\end{acknowledgements}

%

\begin{thebibliography}{105}
\expandafter\ifx\csname natexlab\endcsname\relax\def\natexlab#1{#1}\fi

\bibitem[{{Abazajian} {et~al.}(2009){Abazajian}, {Adelman-McCarthy},
  {Ag{\"u}eros}, {Allam}, {Allende Prieto}, {An}, {Anderson}, {Anderson},
  {Annis}, {Bahcall}, \& et~al.}]{Abazajian_2009}
{Abazajian}, K., {Adelman-McCarthy}, J., {Ag{\"u}eros}, M., {et~al.} 2009,
  \apjs, 182, 543

\bibitem[{{Ahn} {et~al.}(2014){Ahn}, {Alexandroff}, {Allende Prieto}, {Anders},
  {Anderson}, {Anderton}, {Andrews}, {Aubourg}, {Bailey}, {Bastien}, \&
  et~al.}]{Ahn_2014}
{Ahn}, C.~P., {Alexandroff}, R., {Allende Prieto}, C., {et~al.} 2014, \apjs,
  211, 17

\bibitem[{{Alam} {et~al.}(2015){Alam}, {Albareti}, {Allende Prieto}, {Anders},
  {Anderson}, {Anderton}, {Andrews}, {Armengaud}, {Aubourg}, {Bailey}, \&
  et~al.}]{Alam_2015}
{Alam}, S., {Albareti}, F.~D., {Allende Prieto}, C., {et~al.} 2015, \apjs, 219,
  12

\bibitem[{{Annis} {et~al.}(2014){Annis}, {Soares-Santos}, {Strauss}, {Becker},
  {Dodelson}, {Fan}, {Gunn}, {Hao}, {Ivezi{'c}}, {Jester}, {Jiang}, {Johnston},
  {Kubo}, {Lampeitl}, {Lin}, {Lupton}, {Miknaitis}, {Seo}, {Simet}, \&
  {Yanny}}]{Annis_2014}
{Annis}, J., {Soares-Santos}, M., {Strauss}, M., {et~al.} 2014, \apj, 794, 120

\bibitem[{{Assef} {et~al.}(2010){Assef}, {Kochanek}, {Brodwin}, {Cool},
  {Forman}, {Gonzalez}, {Hickox}, {Jones}, {Le Floc'h}, {Moustakas}, {Murray},
  \& {Stern}}]{Assef_2010}
{Assef}, R.~J., {Kochanek}, C.~S., {Brodwin}, M., {et~al.} 2010, \apj, 713, 970

\bibitem[{{Baldry} {et~al.}(2004){Baldry}, {Glazebrook}, {Brinkmann},
  {Ivezi{\'c}}, {Lupton}, {Nichol}, \& {Szalay}}]{Baldry_2004}
{Baldry}, I.~K., {Glazebrook}, K., {Brinkmann}, J., {et~al.} 2004, \apj, 600,
  681

\bibitem[{{Balogh} {et~al.}(2005){Balogh}, {Miller}, {Nichol}, {Zabludoff}, \&
  {Goto}}]{Balogh_2005}
{Balogh}, M.~L., {Miller}, C., {Nichol}, R., {Zabludoff}, A., \& {Goto}, T.
  2005, \mnras, 360, 587

\bibitem[{{Barton} {et~al.}(2000){Barton}, {Geller}, \& {Kenyon}}]{Barton_2000}
{Barton}, E.~J., {Geller}, M.~J., \& {Kenyon}, S.~J. 2000, \apj, 530, 660

\bibitem[{{Becker} {et~al.}(1995){Becker}, {White}, \& {Helfand}}]{Becker_1995}
{Becker}, R.~H., {White}, R.~L., \& {Helfand}, D.~J. 1995, \apj, 450, 559

\bibitem[{{Bennert} {et~al.}(2008){Bennert}, {Canalizo}, {Jungwiert},
  {Stockton}, {Schweizer}, {Peng}, \& {Lacy}}]{Bennert_2008}
{Bennert}, N., {Canalizo}, G., {Jungwiert}, B., {et~al.} 2008, \apj, 677, 846

\bibitem[{{Bentz} {et~al.}(2016{\natexlab{a}}){Bentz}, {Batiste}, {Seals},
  {Garcia}, {Kuzio de Naray}, {Peters}, {Anderson}, {Jones}, {Lester},
  {Machuca}, {Parks}, {Pope}, {Revalski}, {Roberts}, {Saylor}, {Sevrinsky}, \&
  {Turner}}]{Bentz_2016b}
{Bentz}, M.~C., {Batiste}, M., {Seals}, J., {et~al.} 2016{\natexlab{a}}, ArXiv
  e-prints [\eprint[arXiv]{1608.03893}]

\bibitem[{{Bentz} {et~al.}(2016{\natexlab{b}}){Bentz}, {Cackett}, {Crenshaw},
  {Horne}, {Street}, \& {Ou-Yang}}]{Bentz_2016a}
{Bentz}, M.~C., {Cackett}, E.~M., {Crenshaw}, D.~M., {et~al.}
  2016{\natexlab{b}}, ArXiv e-prints [\eprint[arXiv]{1608.01229}]

\bibitem[{{Bergvall} {et~al.}(2016){Bergvall}, {Marquart}, {Way}, {Blomqvist},
  {Holst}, {{\"O}stlin}, \& {Zackrisson}}]{Bergvall_2016}
{Bergvall}, N., {Marquart}, T., {Way}, M.~J., {et~al.} 2016, \aap, 587, A72

\bibitem[{{Blanton} {et~al.}(2003){Blanton}, {Hogg}, {Bahcall}, {Baldry},
  {Brinkmann}, {Csabai}, {Eisenstein}, {Fukugita}, {Gunn}, {Ivezi{\'c}},
  {Lamb}, {Lupton}, {Loveday}, {Munn}, {Nichol}, {Okamura}, {Schlegel},
  {Shimasaku}, {Strauss}, {Vogeley}, \& {Weinberg}}]{Blanton_2003}
{Blanton}, M.~R., {Hogg}, D.~W., {Bahcall}, N.~A., {et~al.} 2003, \apj, 594,
  186

\bibitem[{{Booth} \& {Schaye}(2013)}]{Booth_2013}
{Booth}, C.~M. \& {Schaye}, J. 2013, Scientific Reports, 1738

\bibitem[{{Bournaud} {et~al.}(2005){Bournaud}, {Jog}, \&
  {Combes}}]{Bournaud_2005}
{Bournaud}, F., {Jog}, C.~J., \& {Combes}, F. 2005, \aap, 437, 69

\bibitem[{{Brinchmann} {et~al.}(2004){Brinchmann}, {Charlot}, {White},
  {Tremonti}, {Kauffmann}, {Heckman}, \& {Brinkmann}}]{Brinchmann_2004}
{Brinchmann}, J., {Charlot}, S., {White}, S.~D.~M., {et~al.} 2004, \mnras, 351,
  1151

\bibitem[{{Brotherton} {et~al.}(1999){Brotherton}, {van Breugel}, {Stanford},
  {Smith}, {Boyle}, {Miller}, {Shanks}, {Croom}, \&
  {Filippenko}}]{Brotherton_1999}
{Brotherton}, M.~S., {van Breugel}, W., {Stanford}, S.~A., {et~al.} 1999,
  \apjl, 520, L87

\bibitem[{{Cales} \& {Brotherton}(2015)}]{Cales_2015}
{Cales}, S.~L. \& {Brotherton}, M.~S. 2015, \mnras, 449, 2374

\bibitem[{{Choi} {et~al.}(2009){Choi}, {Goto}, \& {Yoon}}]{Choi_2009}
{Choi}, Y., {Goto}, T., \& {Yoon}, S.-J. 2009, \mnras, 395, 637

\bibitem[{{Civano} {et~al.}(2014){Civano}, {Fabbiano}, {Pellegrini}, {Kim},
  {Paggi}, {Feder}, \& {Elvis}}]{Civano_2014}
{Civano}, F., {Fabbiano}, G., {Pellegrini}, S., {et~al.} 2014, \apj, 790, 16

\bibitem[{{Couch} \& {Sharples}(1987)}]{Couch_1987}
{Couch}, W.~J. \& {Sharples}, R.~M. 1987, \mnras, 229, 423

\bibitem[{{De Propris} \& {Melnick}(2014)}]{De_Propris_2014}
{De Propris}, R. \& {Melnick}, J. 2014, \mnras, 439, 2837

\bibitem[{{Di Matteo} {et~al.}(2007){Di Matteo}, {Combes}, {Melchior}, \&
  {Semelin}}]{DiMatteo_2007}
{Di Matteo}, P., {Combes}, F., {Melchior}, A.-L., \& {Semelin}, B. 2007, \aap,
  468, 61

\bibitem[{{Dressler} \& {Gunn}(1983)}]{Dressler_Gunn_1983}
{Dressler}, A. \& {Gunn}, J.~E. 1983, \apj, 270, 7

\bibitem[{{Dressler} \& {Gunn}(1992)}]{Dressler_Gunn_1992}
{Dressler}, A. \& {Gunn}, J.~E. 1992, \apjs, 78, 1

\bibitem[{{Duc} \& {Renaud}(2013)}]{Duc_2013}
{Duc}, P.-A. \& {Renaud}, F. 2013, in Lecture Notes in Physics, Berlin Springer
  Verlag, Vol. 861, Lecture Notes in Physics, Berlin Springer Verlag, ed.
  J.~{Souchay}, S.~{Mathis}, \& T.~{Tokieda}, 327

\bibitem[{Ester {et~al.}(1998)Ester, Kriegel, Sander, \& Xu}]{Ester_1996}
Ester, M., Kriegel, H.-P., Sander, J., \& Xu, X. 1998, Data Min. Knowl.
  Discov., 2, 169

\bibitem[{{Fabian}(2012)}]{Fabian_2012}
{Fabian}, A.~C. 2012, \araa, 50, 455

\bibitem[{{Fabian} \& {Iwasawa}(1999)}]{Fabian_1999}
{Fabian}, A.~C. \& {Iwasawa}, K. 1999, \mnras, 303, L34

\bibitem[{{Fliri} \& {Trujillo}(2016)}]{Fliri_2016}
{Fliri}, J. \& {Trujillo}, I. 2016, \mnras, 456, 1359

\bibitem[{{Gabor} {et~al.}(2011){Gabor}, {Dav{\'e}}, {Oppenheimer}, \&
  {Finlator}}]{Gabor_2011}
{Gabor}, J.~M., {Dav{\'e}}, R., {Oppenheimer}, B.~D., \& {Finlator}, K. 2011,
  \mnras, 417, 2676

\bibitem[{{Georgakakis} {et~al.}(2008){Georgakakis}, {Nandra}, {Yan},
  {Willner}, {Lotz}, {Pierce}, {Cooper}, {Laird}, {Koo}, {Barmby}, {Newman},
  {Primack}, \& {Coil}}]{Georgkakis_2008}
{Georgakakis}, A., {Nandra}, K., {Yan}, R., {et~al.} 2008, \mnras, 385, 2049

\bibitem[{{Goto}(2004)}]{Goto_2004}
{Goto}, T. 2004, \aap, 427, 125

\bibitem[{{Goto}(2005)}]{Goto_2005}
{Goto}, T. 2005, \mnras, 357, 937

\bibitem[{{Goto}(2007{\natexlab{a}})}]{Goto_2007b}
{Goto}, T. 2007{\natexlab{a}}, \mnras, 381, 187

\bibitem[{{Goto}(2007{\natexlab{b}})}]{Goto_2007a}
{Goto}, T. 2007{\natexlab{b}}, \mnras, 377, 1222

\bibitem[{{Goto} {et~al.}(2003){Goto}, {Nichol}, {Okamura}, {Sekiguchi},
  {Miller}, {Bernardi}, {Hopkins}, {Tremonti}, {Connolly}, {Castander},
  {Brinkmann}, {Fukugita}, {Harvanek}, {Ivezic}, {Kleinman}, {Krzesinski},
  {Long}, {Loveday}, {Neilsen}, {Newman}, {Nitta}, {Snedden}, \&
  {Subbarao}}]{Goto_2003}
{Goto}, T., {Nichol}, R., {Okamura}, S., {et~al.} 2003, \pasj, 55, 771

\bibitem[{{Greene} \& {Ho}(2007)}]{Greene_2007}
{Greene}, J.~E. \& {Ho}, L.~C. 2007, in Astronomical Society of the Pacific
  Conference Series, Vol. 373, The Central Engine of Active Galactic Nuclei,
  ed. L.~C. {Ho} \& J.-W. {Wang}, 33

\bibitem[{{Heckman} \& {Best}(2014)}]{Heckman_2014}
{Heckman}, T.~M. \& {Best}, P.~N. 2014, \araa, 52, 589

\bibitem[{{Helfand} {et~al.}(2015){Helfand}, {White}, \&
  {Becker}}]{Helfand_2015}
{Helfand}, D.~J., {White}, R.~L., \& {Becker}, R.~H. 2015, \apj, 801, 26

\bibitem[{{Hodge} {et~al.}(2011){Hodge}, {Becker}, {White}, {Richards}, \&
  {Zeimann}}]{Hodge_2011}
{Hodge}, J.~A., {Becker}, R.~H., {White}, R.~L., {Richards}, G.~T., \&
  {Zeimann}, G.~R. 2011, \aj, 142, 3

\bibitem[{{Holincheck} {et~al.}(2016){Holincheck}, {Wallin}, {Borne},
  {Fortson}, {Lintott}, {Smith}, {Bamford}, {Keel}, \&
  {Parrish}}]{Holincheck_2016}
{Holincheck}, A.~J., {Wallin}, J.~F., {Borne}, K., {et~al.} 2016, \mnras, 459,
  720

\bibitem[{{Hopkins} {et~al.}(2008){Hopkins}, {Cox}, {Kere{\v s}}, \&
  {Hernquist}}]{Hopkins_2008}
{Hopkins}, P.~F., {Cox}, T.~J., {Kere{\v s}}, D., \& {Hernquist}, L. 2008,
  \apjs, 175, 390

\bibitem[{{Hopkins} {et~al.}(2006){Hopkins}, {Hernquist}, {Cox}, {Di Matteo},
  {Robertson}, \& {Springel}}]{Hopkins_2006}
{Hopkins}, P.~F., {Hernquist}, L., {Cox}, T.~J., {et~al.} 2006, \apjs, 163, 1

\bibitem[{{in der}~Au {et~al.}(2012){in der}~Au, Meusinger, Shalldach, \&
  Newholm}]{inderAu_2012}
{in der}~Au, A., Meusinger, H., Shalldach, P., \& Newholm, M. 2012, \aap, 547,
  A115

\bibitem[{{Ivezi{\'c}} {et~al.}(2002){Ivezi{\'c}}, {Menou}, {Knapp}, {Strauss},
  {Lupton}, {Vanden Berk}, {Richards}, {Tremonti}, {Weinstein}, {Anderson},
  {Bahcall}, {Becker}, {Bernardi}, {Blanton}, {Eisenstein}, {Fan},
  {Finkbeiner}, {Finlator}, {Frieman}, {Gunn}, {Hall}, {Kim}, {Kinkhabwala},
  {Narayanan}, {Rockosi}, {Schlegel}, {Schneider}, {Strateva}, {SubbaRao},
  {Thakar}, {Voges}, {White}, {Yanny}, {Brinkmann}, {Doi}, {Fukugita},
  {Hennessy}, {Munn}, {Nichol}, \& {York}}]{Ivezic_2002}
{Ivezi{\'c}}, {\v Z}., {Menou}, K., {Knapp}, G.~R., {et~al.} 2002, \aj, 124,
  2364

\bibitem[{{Jiang} {et~al.}(2008){Jiang}, {Fan}, {Annis}, {Becker}, {White},
  {Chiu}, {Lin}, {Lupton}, {Richards}, {Strauss}, {Jester}, \&
  {Schneider}}]{Jiang_2008}
{Jiang}, L., {Fan}, X., {Annis}, J., {et~al.} 2008, \aj, 135, 1057

\bibitem[{{Kara} {et~al.}(2016){Kara}, {Miller}, {Reynolds}, \&
  {Dai}}]{Kara_2016}
{Kara}, E., {Miller}, J.~M., {Reynolds}, C., \& {Dai}, L. 2016, \nat, 535, 388

\bibitem[{{Kauffmann} {et~al.}(2003{\natexlab{a}}){Kauffmann}, {Heckman},
  {Tremonti}, {Brinchmann}, {Charlot}, {White}, {Ridgway}, {Brinkmann},
  {Fukugita}, {Hall}, {Ivezi{\'c}}, {Richards}, \&
  {Schneider}}]{2003MNRAS.346.1055K}
{Kauffmann}, G., {Heckman}, T.~M., {Tremonti}, C., {et~al.} 2003{\natexlab{a}},
  \mnras, 346, 1055

\bibitem[{{Kauffmann} {et~al.}(2003{\natexlab{b}}){Kauffmann}, {Heckman},
  {White}, {Charlot}, {Tremonti}, {Peng}, {Seibert}, {Brinkmann}, {Nichol},
  {SubbaRao}, \& {York}}]{2003MNRAS.341...33K}
{Kauffmann}, G., {Heckman}, T.~M., {White}, S.~D.~M., {et~al.}
  2003{\natexlab{b}}, \mnras, 341, 54

\bibitem[{{Kaviraj} {et~al.}(2007){Kaviraj}, {Kirkby}, {Silk}, \&
  {Sarzi}}]{Kaviraj_2007}
{Kaviraj}, S., {Kirkby}, L., {Silk}, J., \& {Sarzi}, M. 2007, \mnras, 382, 960

\bibitem[{{Knobel} {et~al.}(2015){Knobel}, {Lilly}, {Woo}, \& {Kova{\v
  c}}}]{Knobel_2015}
{Knobel}, C., {Lilly}, S.~J., {Woo}, J., \& {Kova{\v c}}, K. 2015, \apj, 800,
  24

\bibitem[{Kohonen(2001)}]{Kohonen_2001}
Kohonen, T. 2001, Self-organizing maps

\bibitem[{{Koss} {et~al.}(2011){Koss}, {Mushotzky}, {Veilleux}, {Winter},
  {Baumgartner}, {Tueller}, {Gehrels}, \& {Valencic}}]{Koss_2011}
{Koss}, M., {Mushotzky}, R., {Veilleux}, S., {et~al.} 2011, \apj, 739, 57

\bibitem[{{LaMassa} {et~al.}(2016){LaMassa}, {Urry}, {Cappelluti},
  {B{\"o}hringer}, {Comastri}, {Glikman}, {Richards}, {Ananna}, {Brusa},
  {Cardamone}, {Chon}, {Civano}, {Farrah}, {Gilfanov}, {Green}, {Komossa},
  {Lira}, {Makler}, {Marchesi}, {Pecoraro}, {Ranalli}, {Salvato}, {Schawinski},
  {Stern}, {Treister}, \& {Viero}}]{LaMassa_2016}
{LaMassa}, S.~M., {Urry}, C.~M., {Cappelluti}, N., {et~al.} 2016, \apj, 817,
  172

\bibitem[{{LaMassa} {et~al.}(2013){LaMassa}, {Urry}, {Cappelluti}, {Civano},
  {Ranalli}, {Glikman}, {Treister}, {Richards}, {Ballantyne}, {Stern},
  {Comastri}, {Cardamone}, {Schawinski}, {B{\"o}hringer}, {Chon}, {Murray},
  {Green}, \& {Nandra}}]{LaMassa_2013}
{LaMassa}, S.~M., {Urry}, C.~M., {Cappelluti}, N., {et~al.} 2013, \mnras, 436,
  3581

\bibitem[{{Larson} \& {Tinsley}(1978)}]{Larson_1978}
{Larson}, R.~B. \& {Tinsley}, B.~M. 1978, \apj, 219, 46

\bibitem[{{Li} {et~al.}(2008){Li}, {Kauffmann}, {Heckman}, {White}, \&
  {Jing}}]{Li_2008}
{Li}, C., {Kauffmann}, G., {Heckman}, T.~M., {White}, S.~D.~M., \& {Jing},
  Y.~P. 2008, \mnras, 385, 1915

\bibitem[{{Lintott} {et~al.}(2011){Lintott}, {Schawinski}, {Bamford}, {Slosar},
  {Land}, {Thomas}, {Edmondson}, {Masters}, {Nichol}, {Raddick}, {Szalay},
  {Andreescu}, {Murray}, \& {Vandenberg}}]{Lintott_2011}
{Lintott}, C., {Schawinski}, K., {Bamford}, S., {et~al.} 2011, \mnras, 410, 166

\bibitem[{{Liu} {et~al.}(2007){Liu}, {Hooper}, {O'Neil}, {Thompson}, {Wolf}, \&
  {Lisker}}]{Liu_2007}
{Liu}, C.~T., {Hooper}, E.~J., {O'Neil}, K., {et~al.} 2007, \apj, 658, 249

\bibitem[{{Maraston} {et~al.}(2006){Maraston}, {Daddi}, {Renzini}, {Cimatti},
  {Dickinson}, {Papovich}, {Pasquali}, \& {Pirzkal}}]{Maraston_2006}
{Maraston}, C., {Daddi}, E., {Renzini}, A., {et~al.} 2006, \apj, 652, 85

\bibitem[{{Maraston} {et~al.}(2013){Maraston}, {Pforr}, {Henriques}, {Thomas},
  {Wake}, {Brownstein}, {Capozzi}, {Tinker}, {Bundy}, {Skibba}, {Beifiori},
  {Nichol}, {Edmondson}, {Schneider}, {Chen}, {Masters}, {Steele}, {Bolton},
  {York}, {Weaver}, {Higgs}, {Bizyaev}, {Brewington}, {Malanushenko},
  {Malanushenko}, {Snedden}, {Oravetz}, {Pan}, {Shelden}, \&
  {Simmons}}]{Maraston_2013}
{Maraston}, C., {Pforr}, J., {Henriques}, B.~M., {et~al.} 2013, \mnras, 435,
  2764

\bibitem[{{Maraston} {et~al.}(2009){Maraston}, {Str{\"o}mb{\"a}ck}, {Thomas},
  {Wake}, \& {Nichol}}]{Maraston_2009}
{Maraston}, C., {Str{\"o}mb{\"a}ck}, G., {Thomas}, D., {Wake}, D.~A., \&
  {Nichol}, R.~C. 2009, \mnras, 394, L107

\bibitem[{{Mateos} {et~al.}(2012){Mateos}, {Alonso-Herrero}, {Carrera},
  {Blain}, {Watson}, {Barcons}, {Braito}, {Severgnini}, {Donley}, \&
  {Stern}}]{Mateos_2012}
{Mateos}, S., {Alonso-Herrero}, A., {Carrera}, F.~J., {et~al.} 2012, \mnras,
  426, 3271

\bibitem[{{Melnick} \& {De Propris}(2013)}]{Melnick_2013}
{Melnick}, J. \& {De Propris}, R. 2013, \mnras, 431, 2034

\bibitem[{{Melnick} {et~al.}(2015){Melnick}, {Telles}, {De Propris}, \&
  {Chu}}]{Melnick_2015}
{Melnick}, J., {Telles}, E., {De Propris}, R., \& {Chu}, Z.-H. 2015, \aap, 582,
  A37

\bibitem[{{Meusinger} \& {Balafkan}(2014)}]{Meusinger_2014}
{Meusinger}, H. \& {Balafkan}, N. 2014, \aap, 568, A114

\bibitem[{{Meusinger} {et~al.}(2016){Meusinger}, {Schalldach}, {Mirhosseini},
  \& {Pertermann}}]{Meusinger_2016}
{Meusinger}, H., {Schalldach}, P., {Mirhosseini}, A., \& {Pertermann}, F. 2016,
  \aap, 587, A83

\bibitem[{{Meusinger} {et~al.}(2012){Meusinger}, {Schalldach}, {Scholz}, {in
  der Au}, {Newholm}, {de Hoon}, \& {Kaminsky}}]{Meusinger_2012}
{Meusinger}, H., {Schalldach}, P., {Scholz}, R.-D., {et~al.} 2012, \aap, 541,
  A77

\bibitem[{{Mihos} \& {Hernquist}(1996)}]{Mihos_1996}
{Mihos}, J.~C. \& {Hernquist}, L. 1996, \apj, 464, 641

\bibitem[{{Morrison} {et~al.}(2003){Morrison}, {Owen}, {Ledlow}, {Keel},
  {Hill}, {Voges}, \& {Herter}}]{Morrison_2003}
{Morrison}, G.~E., {Owen}, F.~N., {Ledlow}, M.~J., {et~al.} 2003, \apjs, 146,
  267

\bibitem[{{Nielsen} {et~al.}(2012){Nielsen}, {Ridgway}, {De Propris}, \&
  {Goto}}]{Nielsen_2012}
{Nielsen}, D.~M., {Ridgway}, S.~E., {De Propris}, R., \& {Goto}, T. 2012,
  \apjl, 761, L16

\bibitem[{{Paggi} {et~al.}(2016){Paggi}, {Fabbiano}, {Civano}, {Pellegrini},
  {Elvis}, \& {Kim}}]{Paggi_2016}
{Paggi}, A., {Fabbiano}, G., {Civano}, F., {et~al.} 2016, ArXiv e-prints
  [\eprint[arXiv]{1507.03170v2}]

\bibitem[{{Poggianti} {et~al.}(1999){Poggianti}, {Smail}, {Dressler}, {Couch},
  {Barger}, {Butcher}, {Ellis}, \& {Oemler}}]{Poggianti_1999}
{Poggianti}, B.~M., {Smail}, I., {Dressler}, A., {et~al.} 1999, \apj, 518, 576

\bibitem[{{Poggianti} \& {Wu}(2000)}]{Poggianti_2000}
{Poggianti}, B.~M. \& {Wu}, H. 2000, \apj, 529, 157

\bibitem[{{Quintero} {et~al.}(2004){Quintero}, {Hogg}, {Blanton}, {Schlegel},
  {Eisenstein}, {Gunn}, {Brinkmann}, {Fukugita}, {Glazebrook}, \&
  {Goto}}]{Quintero_2004}
{Quintero}, A.~D., {Hogg}, D.~W., {Blanton}, M.~R., {et~al.} 2004, \apj, 602,
  190

\bibitem[{{Reichard} {et~al.}(2009){Reichard}, {Heckman}, {Rudnick},
  {Brinchmann}, {Kauffmann}, \& {Wild}}]{Reichard_2009}
{Reichard}, T.~A., {Heckman}, T.~M., {Rudnick}, G., {et~al.} 2009, \apj, 691,
  1005

\bibitem[{{Rodr{\'{\i}}guez Del Pino} {et~al.}(2014){Rodr{\'{\i}}guez Del
  Pino}, {Bamford}, {Arag{\'o}n-Salamanca}, {Milvang-Jensen}, {Merrifield}, \&
  {Balcells}}]{Rodriguez_2014}
{Rodr{\'{\i}}guez Del Pino}, B., {Bamford}, S.~P., {Arag{\'o}n-Salamanca}, A.,
  {et~al.} 2014, \mnras, 438, 1038

\bibitem[{Sachs(1982)}]{Sachs_1982}
Sachs, L. 1982, Applied Statistics. A Handbook of Techniques, Vol. Springer,
  New York

\bibitem[{{Sadler} {et~al.}(2002){Sadler}, {Jackson}, {Cannon}, {McIntyre},
  {Murphy}, {Bland-Hawthorn}, {Bridges}, {Cole}, {Colless}, {Collins}, {Couch},
  {Dalton}, {De Propris}, {Driver}, {Efstathiou}, {Ellis}, {Frenk},
  {Glazebrook}, {Lahav}, {Lewis}, {Lumsden}, {Maddox}, {Madgwick}, {Norberg},
  {Peacock}, {Peterson}, {Sutherland}, \& {Taylor}}]{Sadler_2002}
{Sadler}, E.~M., {Jackson}, C.~A., {Cannon}, R.~D., {et~al.} 2002, \mnras, 329,
  227

\bibitem[{{Sanders} {et~al.}(1988){Sanders}, {Soifer}, {Elias}, {Madore},
  {Matthews}, {Neugebauer}, \& {Scoville}}]{Sanders_1988}
{Sanders}, D.~B., {Soifer}, B.~T., {Elias}, J.~H., {et~al.} 1988, \apj, 325, 74

\bibitem[{{Schneider} {et~al.}(2010){Schneider}, {Richards}, {Hall}, {Strauss},
  {Anderson}, {Boroson}, {Ross}, {Shen}, {Brandt}, {Fan}, {Inada}, {Jester},
  {Knapp}, {Krawczyk}, {Thakar}, {Vanden Berk}, {Voges}, {Yanny}, {York},
  {Bahcall}, {Bizyaev}, {Blanton}, {Brewington}, {Brinkmann}, {Eisenstein},
  {Frieman}, {Fukugita}, {Gray}, {Gunn}, {Hibon}, {Ivezi{\'c}}, {Kent}, {Kron},
  {Lee}, {Lupton}, {Malanushenko}, {Malanushenko}, {Oravetz}, {Pan}, {Pier},
  {Price}, {Saxe}, {Schlegel}, {Simmons}, {Snedden}, {SubbaRao}, {Szalay}, \&
  {Weinberg}}]{Schneider_2010}
{Schneider}, D.~P., {Richards}, G.~T., {Hall}, P.~B., {et~al.} 2010, \aj, 139,
  2360

\bibitem[{{Scott} {et~al.}(2013){Scott}, {Graham}, \& {Schombert}}]{Scott_2013}
{Scott}, N., {Graham}, A.~W., \& {Schombert}, J. 2013, \apj, 768, 76

\bibitem[{{Sell} {et~al.}(2014){Sell}, {Tremonti}, {Hickox}, {Diamond-Stanic},
  {Moustakas}, {Coil}, {Williams}, {Rudnick}, {Robaina}, {Geach}, {Heinz}, \&
  {Wilcots}}]{Sell_2014}
{Sell}, P.~H., {Tremonti}, C.~A., {Hickox}, R.~C., {et~al.} 2014, \mnras, 441,
  3417

\bibitem[{{Shen} {et~al.}(2011){Shen}, {Richards}, {Strauss}, {Hall},
  {Schneider}, {Snedden}, {Bizyaev}, {Brewington}, {Malanushenko},
  {Malanushenko}, {Oravetz}, {Pan}, \& {Simmons}}]{Shen_2011}
{Shen}, Y., {Richards}, G.~T., {Strauss}, M.~A., {et~al.} 2011, \apjs, 194, 45

\bibitem[{{Snyder} {et~al.}(2011){Snyder}, {Cox}, {Hayward}, {Hernquist}, \&
  {Jonsson}}]{Snyder_2011}
{Snyder}, G., {Cox}, T., {Hayward}, C., {Hernquist}, L., \& {Jonsson}, P. 2011,
  \apj, 741, 77

\bibitem[{{Springel} {et~al.}(2005){Springel}, {Di Matteo}, \&
  {Hernquist}}]{Springel_2005}
{Springel}, V., {Di Matteo}, T., \& {Hernquist}, L. 2005, \apjl, 620, L79

\bibitem[{{Stern} {et~al.}(2012){Stern}, {Assef}, {Benford}, {Blain}, {Cutri},
  {Dey}, {Eisenhardt}, {Griffith}, {Jarrett}, {Lake}, {Masci}, {Petty},
  {Stanford}, {Tsai}, {Wright}, {Yan}, {Harrison}, \& {Madsen}}]{Stern_2012}
{Stern}, D., {Assef}, R.~J., {Benford}, D.~J., {et~al.} 2012, \apj, 753, 30

\bibitem[{{Stoughton} {et~al.}(2002){Stoughton}, {Lupton}, {Bernardi},
  {Blanton}, {Burles}, {Castander}, {Connolly}, {Eisenstein}, {Frieman},
  {Hennessy}, {Hindsley}, {Ivezi{\'c}}, {Kent}, {Kunszt}, {Lee}, {Meiksin},
  {Munn}, {Newberg}, {Nichol}, {Nicinski}, {Pier}, {Richards}, {Richmond},
  {Schlegel}, {Smith}, {Strauss}, {SubbaRao}, {Szalay}, {Thakar}, {Tucker},
  {Vanden Berk}, {Yanny}, {Adelman}, {Anderson}, {Anderson}, {Annis},
  {Bahcall}, {Bakken}, {Bartelmann}, {Bastian}, {Bauer}, {Berman},
  {B{\"o}hringer}, {Boroski}, {Bracker}, {Briegel}, {Briggs}, {Brinkmann},
  {Brunner}, {Carey}, {Carr}, {Chen}, {Christian}, {Colestock}, {Crocker},
  {Csabai}, {Czarapata}, {Dalcanton}, {Davidsen}, {Davis}, {Dehnen},
  {Dodelson}, {Doi}, {Dombeck}, {Donahue}, {Ellman}, {Elms}, {Evans}, {Eyer},
  {Fan}, {Federwitz}, {Friedman}, {Fukugita}, {Gal}, {Gillespie}, {Glazebrook},
  {Gray}, {Grebel}, {Greenawalt}, {Greene}, {Gunn}, {de Haas}, {Haiman},
  {Haldeman}, {Hall}, {Hamabe}, {Hansen}, {Harris}, {Harris}, {Harvanek},
  {Hawley}, {Hayes}, {Heckman}, {Helmi}, {Henden}, {Hogan}, {Hogg}, {Holmgren},
  {Holtzman}, {Huang}, {Hull}, {Ichikawa}, {Ichikawa}, {Johnston}, {Kauffmann},
  {Kim}, {Kimball}, {Kinney}, {Klaene}, {Kleinman}, {Klypin}, {Knapp},
  {Korienek}, {Krolik}, {Kron}, {Krzesi{\'n}ski}, {Lamb}, {Leger},
  {Limmongkol}, {Lindenmeyer}, {Long}, {Loomis}, {Loveday}, {MacKinnon},
  {Mannery}, {Mantsch}, {Margon}, {McGehee}, {McKay}, {McLean}, {Menou},
  {Merelli}, {Mo}, {Monet}, {Nakamura}, {Narayanan}, {Nash}, {Neilsen},
  {Newman}, {Nitta}, {Odenkirchen}, {Okada}, {Okamura}, {Ostriker}, {Owen},
  {Pauls}, {Peoples}, {Peterson}, {Petravick}, {Pope}, {Pordes}, {Postman},
  {Prosapio}, {Quinn}, {Rechenmacher}, {Rivetta}, {Rix}, {Rockosi}, {Rosner},
  {Ruthmansdorfer}, {Sandford}, {Schneider}, {Scranton}, {Sekiguchi}, {Sergey},
  {Sheth}, {Shimasaku}, {Smee}, {Snedden}, {Stebbins}, {Stubbs}, {Szapudi},
  {Szkody}, {Szokoly}, {Tabachnik}, {Tsvetanov}, {Uomoto}, {Vogeley}, {Voges},
  {Waddell}, {Walterbos}, {Wang}, {Watanabe}, {Weinberg}, {White}, {White},
  {Wilhite}, {Wolfe}, {Yasuda}, {York}, {Zehavi}, \& {Zheng}}]{Stoughton_2002}
{Stoughton}, C., {Lupton}, R.~H., {Bernardi}, M., {et~al.} 2002, \aj, 123, 485

\bibitem[{{Strateva} {et~al.}(2001){Strateva}, {Ivezi{\'c}}, {Knapp},
  {Narayanan}, {Strauss}, {Gunn}, {Lupton}, {Schlegel}, {Bahcall}, {Brinkmann},
  {Brunner}, {Budav{\'a}ri}, {Csabai}, {Castander}, {Doi}, {Fukugita}, {Gy{\H
  o}ry}, {Hamabe}, {Hennessy}, {Ichikawa}, {Kunszt}, {Lamb}, {McKay},
  {Okamura}, {Racusin}, {Sekiguchi}, {Schneider}, {Shimasaku}, \&
  {York}}]{Strateva_2001}
{Strateva}, I., {Ivezi{\'c}}, {\v Z}., {Knapp}, G.~R., {et~al.} 2001, \aj, 122,
  1861

\bibitem[{{Swinbank} {et~al.}(2012){Swinbank}, {Balogh}, {Bower}, {Zabludoff},
  {Lucey}, {McGee}, {Miller}, \& {Nichol}}]{Swinbank_2012}
{Swinbank}, A.~M., {Balogh}, M.~L., {Bower}, R.~G., {et~al.} 2012, \mnras, 420,
  672

\bibitem[{{Symeonidis} {et~al.}(2013){Symeonidis}, {Kartaltepe}, {Salvato},
  {Bongiorno}, {Brusa}, {Page}, {Ilbert}, {Sanders}, \& {van der
  Wel}}]{Symeonidis_2013}
{Symeonidis}, M., {Kartaltepe}, J., {Salvato}, M., {et~al.} 2013, \mnras, 433,
  1015

\bibitem[{{Toomre} \& {Toomre}(1972)}]{Toomre_1972}
{Toomre}, A. \& {Toomre}, J. 1972, \apj, 178, 623

\bibitem[{{Urrutia} {et~al.}(2008){Urrutia}, {Lacy}, \&
  {Becker}}]{Urrutia_2008}
{Urrutia}, T., {Lacy}, M., \& {Becker}, R.~H. 2008, \apj, 674, 80

\bibitem[{{Willett} {et~al.}(2013){Willett}, {Lintott}, {Bamford}, {Masters},
  {Simmons}, {Casteels}, {Edmondson}, {Fortson}, {Kaviraj}, {Keel}, {Melvin},
  {Nichol}, {Raddick}, {Schawinski}, {Simpson}, {Skibba}, {Smith}, \&
  {Thomas}}]{Willett_2013}
{Willett}, K.~W., {Lintott}, C.~J., {Bamford}, S.~P., {et~al.} 2013, \mnras,
  435, 2835

\bibitem[{{Williams}(1983)}]{Williams_1983}
{Williams}, L. 1983, SIGGRAPH '83 Proceedings of the 10th annual conference on
  Computer graphics and interactive techniques, 1

\bibitem[{{Wong} {et~al.}(2012){Wong}, {Schawinski}, {Kaviraj}, {Masters},
  {Nichol}, {Lintott}, {Keel}, {Darg}, {Bamford}, {Andreescu}, {Murray},
  {Raddick}, {Szalay}, {Thomas}, \& {Vandenberg}}]{Wong_2012}
{Wong}, O.~I., {Schawinski}, K., {Kaviraj}, S., {et~al.} 2012, \mnras, 420,
  1684

\bibitem[{{Wright} {et~al.}(2010){Wright}, {Eisenhardt}, {Mainzer}, {Ressler},
  {Cutri}, {Jarrett}, {Kirkpatrick}, {Padgett}, {McMillan}, {Skrutskie},
  {Stanford}, {Cohen}, {Walker}, {Mather}, {Leisawitz}, {Gautier}, {McLean},
  {Benford}, {Lonsdale}, {Blain}, {Mendez}, {Irace}, {Duval}, {Liu}, {Royer},
  {Heinrichsen}, {Howard}, {Shannon}, {Kendall}, {Walsh}, {Larsen}, {Cardon},
  {Schick}, {Schwalm}, {Abid}, {Fabinsky}, {Naes}, \& {Tsai}}]{Wright_2010}
{Wright}, E.~L., {Eisenhardt}, P.~R.~M., {Mainzer}, A.~K., {et~al.} 2010, \aj,
  140, 1868

\bibitem[{{Wu} {et~al.}(2014){Wu}, {Gal}, {Lemaux}, {Kocevski}, {Lubin},
  {Rumbaugh}, \& {Squires}}]{Wu_2014}
{Wu}, P.-F., {Gal}, R., {Lemaux}, B., {et~al.} 2014, \apj, 792, 16

\bibitem[{{Yamauchi} {et~al.}(2008){Yamauchi}, {Yagi}, \&
  {Goto}}]{Yamauchi_2008}
{Yamauchi}, C., {Yagi}, M., \& {Goto}, T. 2008, \mnras, 390, 383

\bibitem[{{Yan} {et~al.}(2006){Yan}, {Newman}, {Faber}, {Konidaris}, {Koo}, \&
  {Davis}}]{Yan_2006}
{Yan}, R., {Newman}, J.~A., {Faber}, S.~M., {et~al.} 2006, \apj, 648, 281

\bibitem[{{Yang} {et~al.}(2008){Yang}, {Zabludoff}, {Zaritsky}, \&
  {Mihos}}]{Yang_2008}
{Yang}, Y., {Zabludoff}, A.~I., {Zaritsky}, D., \& {Mihos}, J.~C. 2008, \apj,
  688, 945

\bibitem[{{York} {et~al.}(2000){York}, {Adelman}, {Anderson}, {Anderson},
  {Annis}, {Bahcall}, {Bakken}, {Barkhouser}, {Bastian}, {Berman}, {Boroski},
  {Bracker}, {Briegel}, {Briggs}, {Brinkmann}, {Brunner}, {Burles}, {Carey},
  {Carr}, {Castander}, {Chen}, {Colestock}, {Connolly}, {Crocker}, {Csabai},
  {Czarapata}, {Davis}, {Doi}, {Dombeck}, {Eisenstein}, {Ellman}, {Elms},
  {Evans}, {Fan}, {Federwitz}, {Fiscelli}, {Friedman}, {Frieman}, {Fukugita},
  {Gillespie}, {Gunn}, {Gurbani}, {de Haas}, {Haldeman}, {Harris}, {Hayes},
  {Heckman}, {Hennessy}, {Hindsley}, {Holm}, {Holmgren}, {Huang}, {Hull},
  {Husby}, {Ichikawa}, {Ichikawa}, {Ivezi{\'c}}, {Kent}, {Kim}, {Kinney},
  {Klaene}, {Kleinman}, {Kleinman}, {Knapp}, {Korienek}, {Kron}, {Kunszt},
  {Lamb}, {Lee}, {Leger}, {Limmongkol}, {Lindenmeyer}, {Long}, {Loomis},
  {Loveday}, {Lucinio}, {Lupton}, {MacKinnon}, {Mannery}, {Mantsch}, {Margon},
  {McGehee}, {McKay}, {Meiksin}, {Merelli}, {Monet}, {Munn}, {Narayanan},
  {Nash}, {Neilsen}, {Neswold}, {Newberg}, {Nichol}, {Nicinski}, {Nonino},
  {Okada}, {Okamura}, {Ostriker}, {Owen}, {Pauls}, {Peoples}, {Peterson},
  {Petravick}, {Pier}, {Pope}, {Pordes}, {Prosapio}, {Rechenmacher}, {Quinn},
  {Richards}, {Richmond}, {Rivetta}, {Rockosi}, {Ruthmansdorfer}, {Sandford},
  {Schlegel}, {Schneider}, {Sekiguchi}, {Sergey}, {Shimasaku}, {Siegmund},
  {Smee}, {Smith}, {Snedden}, {Stone}, {Stoughton}, {Strauss}, {Stubbs},
  {SubbaRao}, {Szalay}, {Szapudi}, {Szokoly}, {Thakar}, {Tremonti}, {Tucker},
  {Uomoto}, {Vanden Berk}, {Vogeley}, {Waddell}, {Wang}, {Watanabe},
  {Weinberg}, {Yanny}, {Yasuda}, \& {SDSS Collaboration}}]{York_2000}
{York}, D.~G., {Adelman}, J., {Anderson}, Jr., J.~E., {et~al.} 2000, \aj, 120,
  1579

\bibitem[{{Zabludoff} {et~al.}(1996){Zabludoff}, {Zaritsky}, {Lin}, {Tucker},
  {Hashimoto}, {Shectman}, {Oemler}, \& {Kirshner}}]{Zabludoff_1996}
{Zabludoff}, A.~I., {Zaritsky}, D., {Lin}, H., {et~al.} 1996, \apj, 466, 104

\end{thebibliography}
%
%

%

\appendix

%
\section{List of E+A galaxies in S82}\label{sect:list_S82}

\begin{table*}[bhpt]
\caption{E+A galaxies in SDSS S82.}
\begin{tabular}{rccrrrcccccrrc}
\hline\hline
&&&&&&&&&&&&&\\
  & Name & $z \ \ $ & \multicolumn{3}{c}{EWs} & \multicolumn{5}{c}{GZ2 S82 co-add2} & $t_{\mathrm m}$ & \multicolumn{2}{c}{sMSP} \\
\cline{4-6}  \cline{7-11} \cline{13-14}
  &  
  &  
  & $[\ion{O}{ii}]$
  & H$\delta$
  & H$\alpha$
  & $a14$
  & $a21$  
  & $a22$
  & $a23$  
  & $a24$  
  & 
  & log $M/M_\odot$
  & age
 \\
   &                       &        &(\AA) &(\AA) &(\AA)  &      &      &      &      &      &    &              & (Gyr)\\ 
\hline
\multicolumn{14}{l}{Sample A} \\
 1 & J$010858.19-001740.2$ & 0.0951 & -0.8 &  5.4 &  0.7  &  -   &  -   &  -   &  -   &  -    &  0 &  9.8 $\quad$ & 1.0  \\
 2 & J$011942.23+010751.5$ & 0.0900 & -0.8 &  7.1 &  2.8  & 0.20 & 0.00 & 0.00 & 0.67 & 0.33  &  0 &  9.9 $\quad$ & 0.6  \\
 3 & J$014447.13+003215.1$ & 0.1789 & -0.6 &  5.2 &  2.1  & 0.29 & 0.00 & 0.00 & 0.44 & 0.15  &  0 & 10.4 $\quad$ & 0.9  \\
 4 & J$015107.01-005636.7$ & 0.1981 & -0.2 &  6.4 &  2.9  &  -   &  -   &  -   &  -   &  -    &  3 & 10.5 $\quad$ & 0.5  \\
 5 & J$020505.99-004345.1$ & 0.1134 & -0.6 &  5.1 & -2.6  & 0.58 & 0.36 & 0.46 & 0.18 & 0.00  &  3 & 10.0 $\quad$ & 1.4  \\
 6 & J$022743.21-001523.0$ & 0.2192 & -1.5 &  7.5 & -0.3  & 0.45 & 0.33 & 0.33 & 0.00 & 0.22  &  3 & 10.6 $\quad$ & 1.1  \\
 7 & J$022957.37-005412.8$ & 0.0859 & -1.7 &  6.1 & -2.8  & 0.42 & 0.25 & 0.25 & 0.38 & 0.13  & -1 & 10.3 $\quad$ & 3.0  \\
 8 & J$023346.93-010128.3$ & 0.0491 & -3.4 &  6.9 & -1.1  &  -   &  -   &  -   &  -   &  -    &  0 &  9.0 $\quad$ & 0.5  \\
 9 & J$023446.37+003035.8$ & 0.1405 & -1.9 &  7.4 &  3.0  & 0.66 & 0.00 & 0.00 & 1.00 & 0.00  &  2 & 10.1 $\quad$ & 0.7  \\
10 & J$025758.99+005144.8$ & 0.0746 & -4.2 &  6.6 & -4.8  & 0.47 & 0.00 & 0.66 & 0.22 & 0.11  &  2 &  9.1 $\quad$ & 0.2  \\
11 & J$025850.52+003458.7$ & 0.1941 & -2.3 &  5.9 & -4.6  & 0.44 & 0.12 & 0.00 & 0.25 & 0.50  &  3 & 10.5 $\quad$ & 0.9  \\
12 & J$030228.42-005618.1$ & 0.1684 & -4.0 &  6.8 &  0.1  & 0.60 & 0.08 & 0.67 & 0.25 & 0.00  &  3 & 10.5 $\quad$ & 0.5  \\
13 & J$030404.25-004913.0$ & 0.2166 & -1.7 &  5.0 & -2.8  & 0.50 & 0.00 & 0.11 & 0.11 & 0.00  &  3 & 10.8 $\quad$ & 0.9  \\
14 & J$032033.65-002021.2$ & 0.0379 & -2.9 &  5.9 & -3.0  &  -   &  -   &  -   &  -   &  -    &  2 &  9.0 $\quad$ & 3.2  \\
15 & J$032802.61+004502.3$ & 0.2020 & -0.8 &  6.5 & -0.1  & 0.10 & 0.00 & 1.00 & 0.00 & 0.00  & -1 & 10.5 $\quad$ & 0.5  \\
16 & J$204133.67-002116.4$ & 0.2160 &  0.1 &  5.5 &  3.2  & 0.26 & 0.20 & 0.00 & 0.60 & 0.00  &  3 &  8.8 $\quad$ & 0.1  \\
17 & J$204457.86-001010.8$ & 0.0802 & -0.8 &  5.7 &  3.0  & 0.21 & 0.15 & 0.28 & 0.28 & 0.28  &  0 &  9.8 $\quad$ & 0.8  \\
18 & J$204834.46-003300.9$ & 0.1513 & -0.5 &  5.0 &  2.7  &  -   &  -   &  -   &  -   &  -    &  0 & 10.1 $\quad$ & 1.0  \\
19 & J$205535.61-001712.5$ & 0.1962 & -2.6 &  7.7 & -0.4  & 0.09 & 0.00 & 0.00 & 0.00 & 0.00  &  1 & 10.3 $\quad$ & 0.5  \\
20 & J$211230.60-005022.4$ & 0.2146 & -2.3 &  9.2 & -0.8  & 0.53 & 0.11 & 0.22 & 0.33 & 0.22  &  3 & 10.4 $\quad$ & 0.4  \\
21 & J$211400.54+003206.3$ & 0.0271 & -0.7 &  7.7 &  0.8  & 0.52 & 0.00 & 0.00 & 0.54 & 0.09  &  3 &   -  $\quad$ &  -   \\
22 & J$213556.91+000328.7$ & 0.0856 & -1.2 &  6.5 &  3.9  &  -   &  -   &  -   &  -   &  -    &  1 &  9.6 $\quad$ & 0.6  \\
23 & J$214413.54+005751.4$ & 0.0793 & -1.4 &  6.1 & -1.5  & 0.76 & 0.00 & 0.00 & 0.44 & 0.25  &  0 &  9.8 $\quad$ & 0.9  \\
24 & J$215735.44-005021.3$ & 0.2186 &  0.7 &  5.7 &  1.8  & 0.77 & 0.06 & 0.00 & 0.34 & 0.54  &  3 & 10.5 $\quad$ & 0.5  \\
25 & J$220122.89+002516.3$ & 0.2230 & -1.6 &  6.9 &  0.2  &  -   &  -   &  -   &  -   &  -    &  3 & 10.6 $\quad$ & 1.4  \\
26 & J$223006.83-004031.3$ & 0.1926 & -3.6 &  5.7 & -1.2  & 0.82 & 0.00 & 0.00 & 0.17 & 0.56  &  3 & 10.3 $\quad$ & 0.5  \\
27 & J$225506.79+005840.0$ & 0.0534 & -0.9 &  5.0 &  0.9  & 0.81 & 0.23 & 0.08 & 0.31 & 0.23  &  3 & 10.4 $\quad$ & 1.6  \\
28 & J$231947.69+004316.0$ & 0.1186 & -1.9 &  6.0 &  1.3  & 0.50 & 0.50 & 0.00 & 0.00 & 0.30  & -1 &  9.9 $\quad$ & 0.8  \\
\hline
\multicolumn{14}{l}{Sample B} \\
 1 & J$003220.86+004429.1$ & 0.0945 & -0.6 &  4.3 &  4.1  & 0.11 & 0.50 & 0.00 & 0.00 & 0.00 & -1 &  9.9 $\quad$ & 0.8  \\
 2 & J$005715.51-004930.5$ & 0.0477 &  0.1 &  4.1 &  2.7  & 0.07 & 0.00 & 0.00 & 0.00 & 0.00 &  0 &  9.1 $\quad$ & 1.6  \\
 3 & J$005954.43-005105.7$ & 0.1535 & -0.1 &  4.0 &  2.6  & 0.22 & 0.50 & 0.00 & 0.50 & 0.00 &  2 & 10.6 $\quad$ & 1.7  \\
 4 & J$011119.58+010706.7$ & 0.1480 & -1.4 &  4.1 &  1.1  & 0.16 & 0.00 & 0.67 & 0.00 & 0.00 &  3 & 10.1 $\quad$ & 1.1  \\
 5 & J$011553.74+002237.3$ & 0.0484 &  1.2 &  4.1 &  2.7  & 0.28 & 0.00 & 1.00 & 0.00 & 0.00 &  1 &  9.0 $\quad$ & 0.6  \\
 6 & J$013250.17-005617.5$ & 0.0445 &  4.1 &  4.4 &  2.3  & 0.04 & 0.00 & 0.00 & 0.00 & 0.00 &  0 &  9.2 $\quad$ & 0.9  \\
 7 & J$014203.82-003542.8$ & 0.1006 & -4.7 &  4.8 &  0.4  & 0.00 & 0.00 & 0.00 & 0.00 & 0.00 &  0 & 10.0 $\quad$ & 1.4  \\
 8 & J$015012.99+000504.8$ & 0.1274 & -2.5 &  4.5 &  1.3  & 0.58 & 0.36 & 0.27 & 0.00 & 0.09 &  3 & 10.7 $\quad$ & 2.7  \\
 9 & J$213213.21+000924.5$ & 0.1376 & -0.3 &  4.9 &  2.8  & 0.74 & 0.07 & 0.00 & 0.50 & 0.36 &  3 & 10.3 $\quad$ & 0.6  \\
10 & J$213547.32+000436.3$ & 0.0854 & -1.3 &  4.9 &  3.9  & 0.32 & 0.33 & 0.50 & 0.00 & 0.00 &  3 & 10.3 $\quad$ & 1.0  \\
11 & J$214532.49+004726.6$ & 0.2022 & -0.0 &  4.0 &  2.8  & 0.22 & 0.00 & 0.00 & 0.25 & 0.50 &  0 & 10.4 $\quad$ & 0.7  \\
12 & J$215738.85+000416.9$ & 0.1445 & -4.6 &  4.6 & -0.2  & 0.73 & 0.00 & 0.00 & 0.37 & 0.63 &  3 & 10.3 $\quad$ & 1.8  \\
13 & J$220003.56-000313.7$ & 0.1807 & -0.5 &  4.0 &  2.4  & 0.05 & 0.00 & 0.00 & 0.00 & 0.00 & -1 & 10.3 $\quad$ & 0.6  \\
14 & J$220806.10-005424.9$ & 0.0380 & -0.5 &  4.2 &  2.5  & 0.53 & 0.00 & 0.00 & 0.44 & 0.11 &  3 &  9.9 $\quad$ & 1.6  \\
15 & J$224603.63-000918.7$ & 0.2053 & -0.3 &  4.6 &  2.9  & 0.00 & 0.00 & 0.00 & 0.00 & 0.00 &  2 & 10.6 $\quad$ & 0.5  \\
16 & J$232703.21-004417.8$ & 0.0929 & -0.7 &  4.2 &  2.7  & 0.28 & 0.00 & 0.00 & 0.80 & 0.00 &  0 &  9.7 $\quad$ & 1.1  \\
17 & J$233823.04+003831.8$ & 0.1375 & -2.0 &  4.5 & -0.4  & 0.05 & 0.00 & 0.00 & 0.00 & 0.00 &  3 & 10.4 $\quad$ & 0.7  \\
\hline
\end{tabular}
\end{table*}


\addtocounter{table}{-1}
\begin{table*}[hbpt]
\caption{(continued.)}
\begin{tabular}{rccrrrcccccrrc}
\hline\hline
&&&&&&&&&&&&&\\
  & Name & $z \ \ $ & \multicolumn{3}{c}{EWs} & \multicolumn{5}{c}{Galaxy Zoo 2} & $t_{\mathrm m}$ & \multicolumn{2}{c}{sMSP} \\
\cline{4-6}  \cline{7-11} \cline{13-14}
  &  
  &  
  & $[\ion{O}{ii}]$
  & H$\delta$
  & H$\alpha$
  & $a14$
  & $a21$  
  & $a22$
  & $a23$  
  & $a24$  
  & 
  & log $M/M_\odot$
  & age
 \\
   &                       &        &(\AA) &(\AA) &(\AA)  &      &      &      &      &      &    &              & (Gyr)\\ 
\hline
\multicolumn{11}{l}{Sample C} \\
 1 & J$000130.48-003030.5$ & 0.1083 & -3.0 &  3.9 & -0.0 & 0.90 & 0.06 & 0.06 & 0.50 & 0.28 &  3 & 10.1 $\quad$ & 1.1 \\
 2 & J$000328.51+002730.1$ & 0.1003 &  2.0 &  3.1 &  2.1 & 0.12 & 0.00 & 0.00 & 0.00 & 0.37 &  2 & 10.5 $\quad$ & 1.6 \\
 3 & J$002119.36-011330.5$ & 0.1066 & -1.3 &  3.2 & -0.8 & 0.14 & 0.00 & 0.67 & 0.33 & 0.00 &  1 & 10.0 $\quad$ & 1.4 \\
 4 & J$003350.41+004233.3$ & 0.1635 & -0.3 &  3.4 &  2.0 & 0.12 & 0.00 & 0.00 & 0.37 & 0.00 &  0 & 10.6 $\quad$ & 1.6 \\
 5 & J$005141.08+011121.8$ & 0.2064 & -1.9 &  3.9 &  3.4 &  -   &  -   &  -   &  -   &  -   & -1 &  9.8 $\quad$ & 3.2 \\
 6 & J$005455.23-005223.6$ & 0.1349 & -0.7 &  3.7 &  1.3 &  -   &  -   &  -   &  -   &  -   &  0 & 10.1 $\quad$ & 1.1 \\
 7 & J$010828.85+002734.6$ & 0.0456 & -0.4 &  3.0 &  2.2 & 0.00 & 0.00 & 0.00 & 0.00 & 0.00 &  0 &  9.3 $\quad$ & 1.4 \\
 8 & J$010833.88+010630.5$ & 0.0877 &  2.6 &  3.9 &  1.6 &  -   &  -   &  -   &  -   &  -   &  0 &  9.8 $\quad$ & 1.7 \\
 9 & J$011447.22+003755.5$ & 0.0473 & -0.4 &  3.2 &  1.8 & 0.00 & 0.00 & 0.00 & 0.00 & 0.00 & -1 &  9.6 $\quad$ & 3.0 \\
10 & J$014147.01-005732.4$ & 0.1545 & -0.3 &  3.8 &  1.4 & 0.15 & 0.33 & 0.00 & 0.00 & 0.00 &  0 & 10.4 $\quad$ & 1.4 \\
11 & J$014325.38-001742.6$ & 0.0735 & -0.7 &  3.2 &  1.6 &  -   &  -   &  -   &  -   &  -   &  0 &  9.6 $\quad$ & 1.0 \\
12 & J$014709.81+005950.5$ & 0.1124 & -0.2 &  3.4 &  1.9 &  -   &  -   &  -   &  -   &  -   &  0 & 10.0 $\quad$ & 1.4 \\
13 & J$015656.17+004325.1$ & 0.0418 &  2.0 &  3.8 &  1.8 & 0.35 & 0.00 & 0.00 & 0.43 & 0.00 &  1 &  9.4 $\quad$ & 0.9 \\
14 & J$020627.01+003805.0$ & 0.0413 & -1.5 &  3.5 & -0.5 & 0.32 & 0.00 & 0.00 & 0.00 & 0.00 &  1 &  9.4 $\quad$ & 1.0 \\
15 & J$030305.75+002940.2$ & 0.2183 &  0.0 &  3.5 &  1.4 & 0.41 & 0.25 & 0.08 & 0.08 & 0.25 &  3 & 10.8 $\quad$ & 1.2 \\
16 & J$032333.26-002618.8$ & 0.0239 &  8.1 &  3.8 &  2.1 & 0.02 & 0.00 & 0.00 & 0.00 & 0.00 &  0 &  9.7 $\quad$ & 2.3 \\
17 & J$032411.43-002343.2$ & 0.1521 & -0.5 &  3.6 &  1.6 & 0.05 & 1.00 & 0.00 & 0.00 & 0.00 & -1 & 10.2 $\quad$ & 1.1 \\
18 & J$033241.31+003142.1$ & 0.2378 &  1.2 &  3.7 &  3.1 & 0.33 & 0.43 & 0.00 & 0.29 & 0.00 &  3 & 10.8 $\quad$ & 0.7 \\
19 & J$205449.29-001639.2$ & 0.1591 & -0.3 &  3.1 &  0.9 &  -   &  -   &  -   &  -   &  -   & -1 & 10.3 $\quad$ & 1.1 \\
20 & J$205636.04+001542.6$ & 0.1597 & -3.0 &  3.0 &  0.3 & 0.16 & 0.00 & 0.00 & 0.66 & 0.33 &  0 & 10.2 $\quad$ & 1.0 \\
21 & J$205916.89+001723.7$ & 0.1063 & -0.0 &  3.3 &  2.3 & 0.47 & 0.25 & 0.25 & 0.00 & 0.00 &  2 & 10.7 $\quad$ & 1.7 \\
22 & J$211948.17+004021.7$ & 0.0344 & -1.0 &  3.7 &  2.4 & 0.00 & 0.00 & 0.00 & 0.00 & 0.00 &  2 &  9.1 $\quad$ & 1.7 \\
23 & J$212515.31+001241.5$ & 0.1133 & -1.0 &  3.7 &  5.0 &  -   &  -   &  -   &  -   &  -   &  0 & 10.3 $\quad$ & 1.1 \\
24 & J$212644.11+001835.4$ & 0.1157 & -0.7 &  3.1 &  2.5 & 0.04 & 0.00 & 0.00 & 0.00 & 0.00 &  0 & 10.0 $\quad$ & 1.2 \\
25 & J$220014.80-005038.6$ & 0.2200 &  0.1 &  3.6 &  3.1 & 0.24 & 0.00 & 0.20 & 0.00 & 0.60 &  3 & 10.6 $\quad$ & 0.8 \\
26 & J$221711.08-001528.7$ & 0.1112 & -0.4 &  3.4 &  2.1 & 0.17 & 0.00 & 0.00 & 0.67 & 0.33 &  0 & 10.3 $\quad$ & 1.8 \\
27 & J$230648.18-000155.3$ & 0.1158 & -1.1 &  3.4 &  1.8 &  -   &  -   &  -   &  -   &  -   &  0 & 10.5 $\quad$ & 2.1 \\
28 & J$231648.19-003425.3$ & 0.1516 & -0.2 &  3.1 &  2.1 &  -   &  -   &  -   &  -   &  -   &  0 & 10.3 $\quad$ & 1.1 \\
29 & J$231858.45-010459.2$ & 0.0299 & -1.9 &  3.5 & -0.6 & 0.17 & 0.00 & 0.00 & 0.00 & 0.00 &  0 &  9.1 $\quad$ & 1.6 \\
\hline
\end{tabular}
\label{table:S82}
\end{table*}

%
\section{Images of morphologically peculiar E+A galaxies in S82}

\begin{figure*}[h]
\begin{tabbing}
\fbox{\includegraphics[width=5.5cm,height=5.5cm]{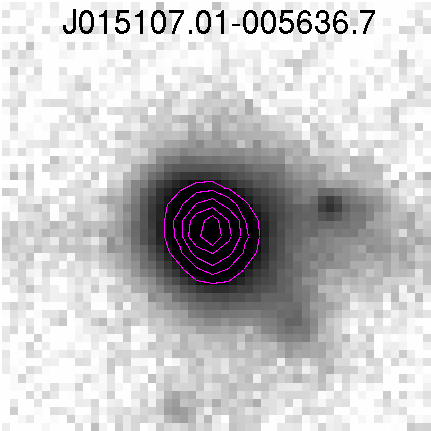}}\hfill \=
\fbox{\includegraphics[width=5.5cm,height=5.5cm]{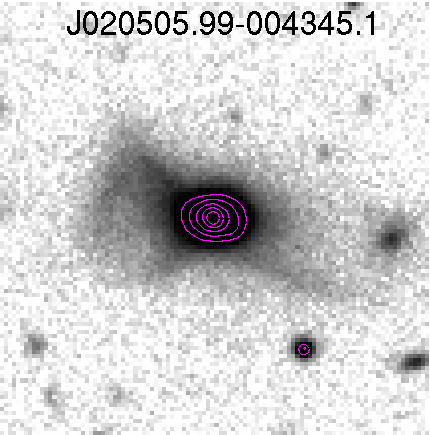}}\hfill \=
\fbox{\includegraphics[width=5.5cm,height=5.5cm]{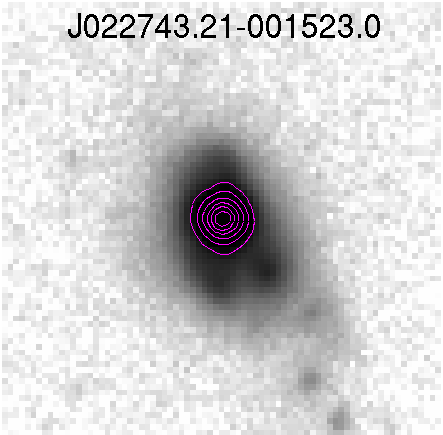}}\\
\fbox{\includegraphics[width=5.5cm,height=5.5cm]{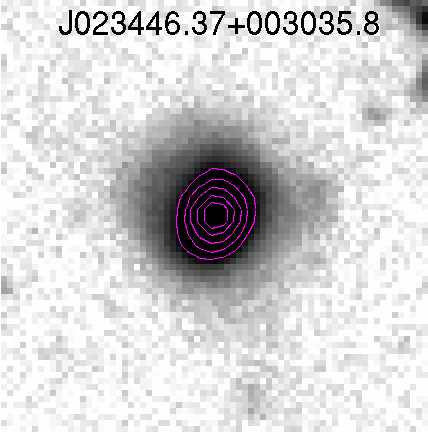}}\hfill \=
\fbox{\includegraphics[width=5.5cm,height=5.5cm]{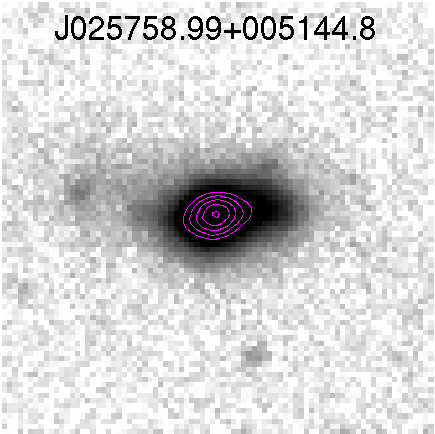}}\hfill \=
\fbox{\includegraphics[width=5.5cm,height=5.5cm]{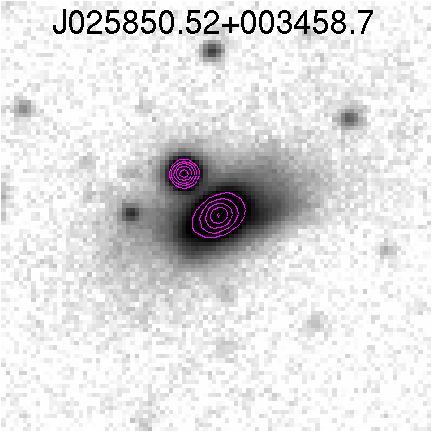}}\\
\fbox{\includegraphics[width=5.5cm,height=5.5cm]{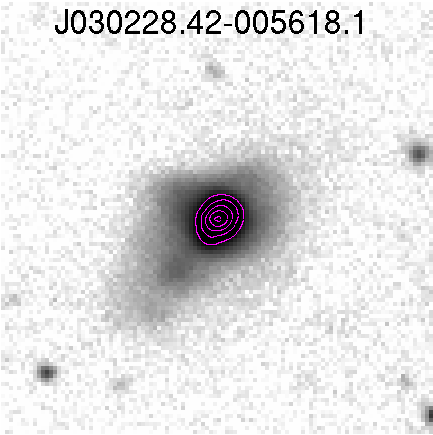}}\hfill \=
\fbox{\includegraphics[width=5.5cm,height=5.5cm]{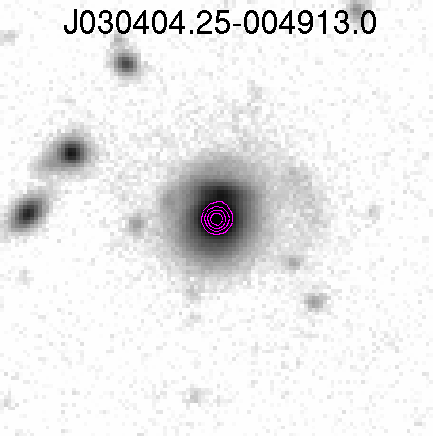}}\hfill \=
\fbox{\includegraphics[width=5.5cm,height=5.5cm]{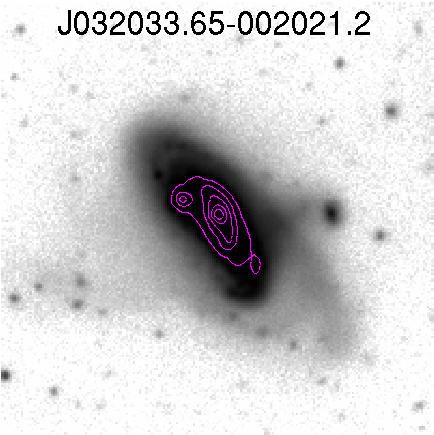}}\\
\fbox{\includegraphics[width=5.5cm,height=5.5cm]{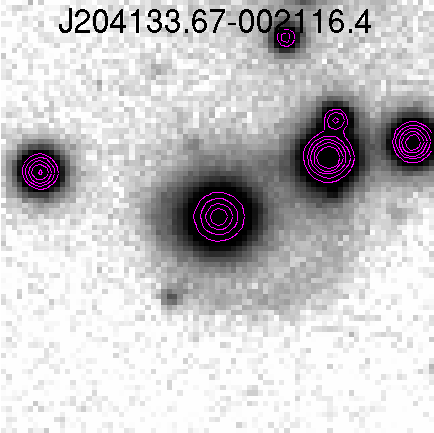}}\hfill \=
\fbox{\includegraphics[width=5.5cm,height=5.5cm]{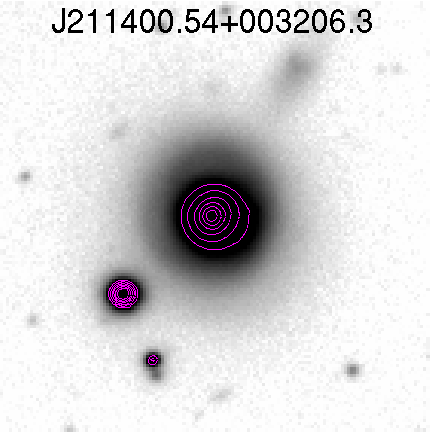}}\hfill \=
\fbox{\includegraphics[width=5.5cm,height=5.5cm]{Fig_B1_l.png}}\\
\end{tabbing}
\caption{Distorted galaxies with $t_m \ge 2$ from sample A 
(left to right, than top to bottom in the same order as in Table\,\ref{table:S82}).}
\label{fig:images_merger_A}
\end{figure*}

\clearpage
\newpage

\setcounter{figure}{0}
\begin{figure*}[h]
\begin{tabbing}
\fbox{\includegraphics[width=5.5cm,height=5.5cm]{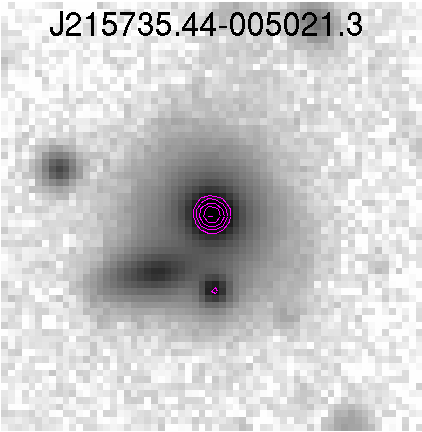}}\hfill \=
\fbox{\includegraphics[width=5.5cm,height=5.5cm]{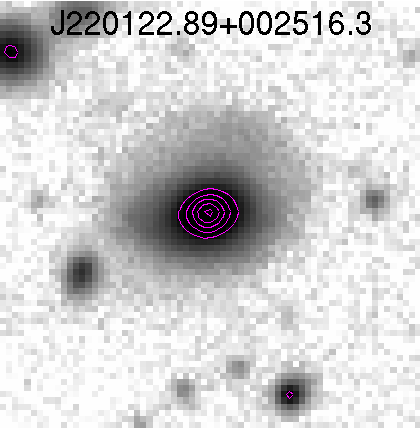}}\hfill \=
\fbox{\includegraphics[width=5.5cm,height=5.5cm]{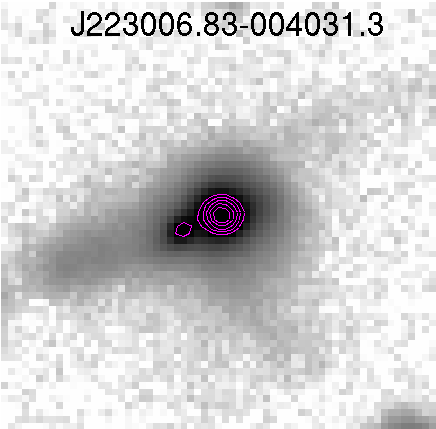}}\\
\fbox{\includegraphics[width=5.5cm,height=5.5cm]{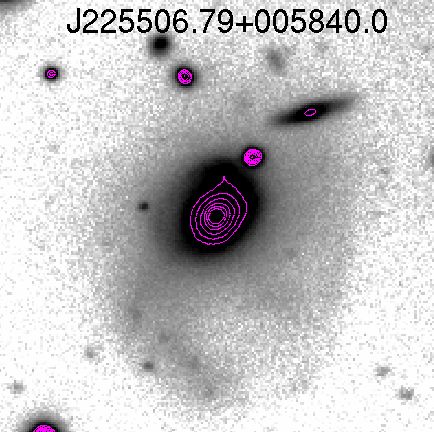}}\hfill \=
\end{tabbing}
\caption{continued.}
\label{fig:images_merger_A2}
\end{figure*}

\clearpage
\newpage

\begin{figure*}[htbp]
\begin{tabbing}
\fbox{\includegraphics[width=5.5cm,height=5.5cm]{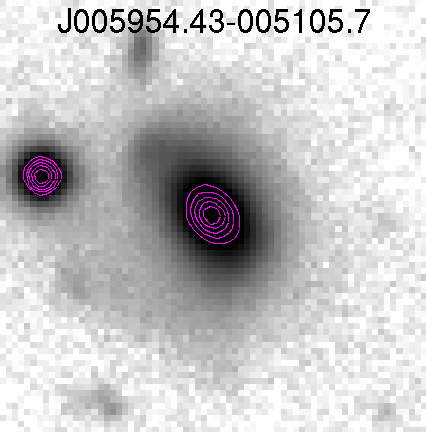}}\hfill \=
\fbox{\includegraphics[width=5.5cm,height=5.5cm]{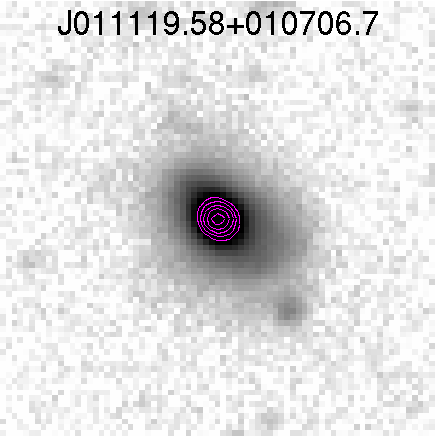}}\hfill \=
\fbox{\includegraphics[width=5.5cm,height=5.5cm]{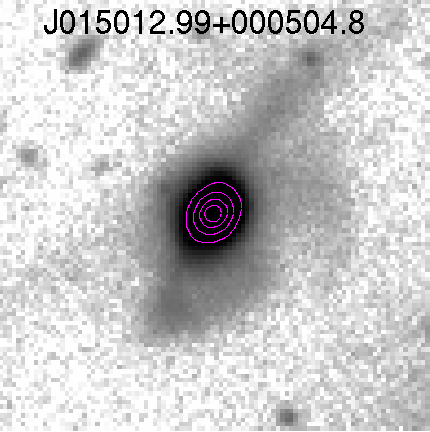}}\\
\fbox{\includegraphics[width=5.5cm,height=5.5cm]{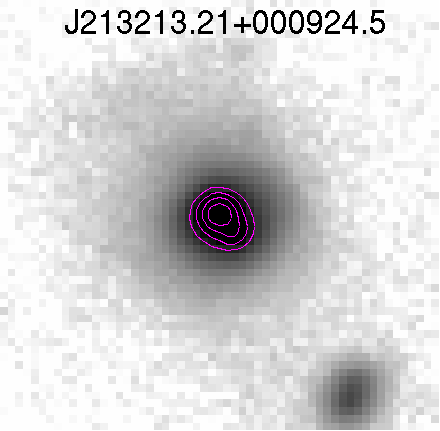}}\hfill \=
\fbox{\includegraphics[width=5.5cm,height=5.5cm]{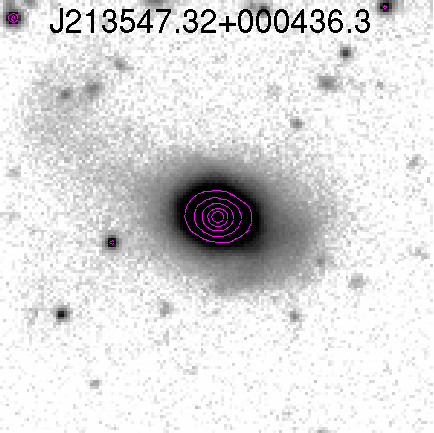}}\hfill \=
\fbox{\includegraphics[width=5.5cm,height=5.5cm]{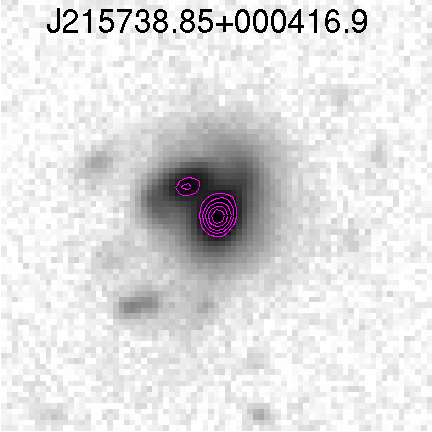}}\\
\fbox{\includegraphics[width=5.5cm,height=5.5cm]{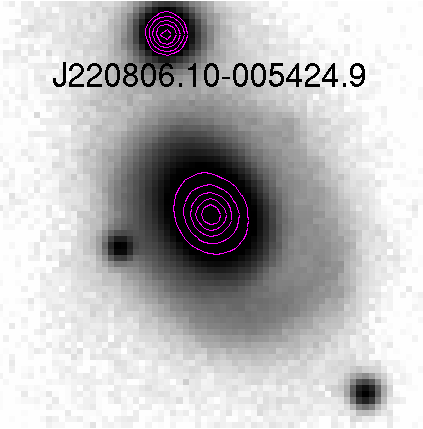}}\hfill \=
\fbox{\includegraphics[width=5.5cm,height=5.5cm]{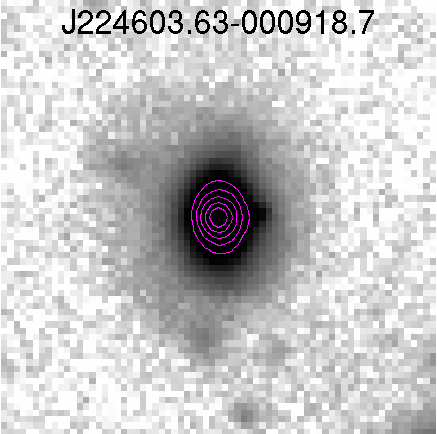}}\hfill \=
\fbox{\includegraphics[width=5.5cm,height=5.5cm]{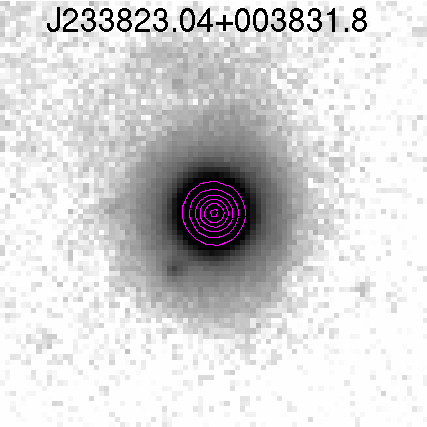}}
\end{tabbing}
\caption{As Fig.\,\ref{fig:images_merger_A} but for sample B.}
\label{fig:images_merger_B}
\end{figure*}

\clearpage
\newpage

\begin{figure*}[htbp]
\begin{tabbing}
\fbox{\includegraphics[width=5.5cm,height=5.5cm]{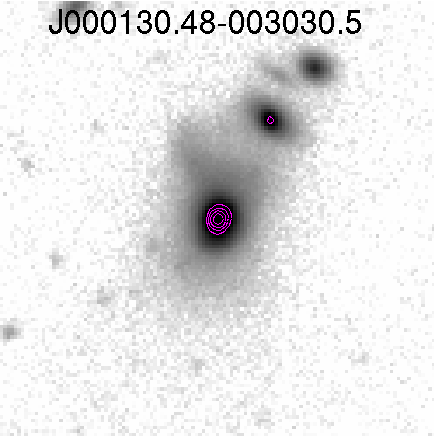}}\hfill \=
\fbox{\includegraphics[width=5.5cm,height=5.5cm]{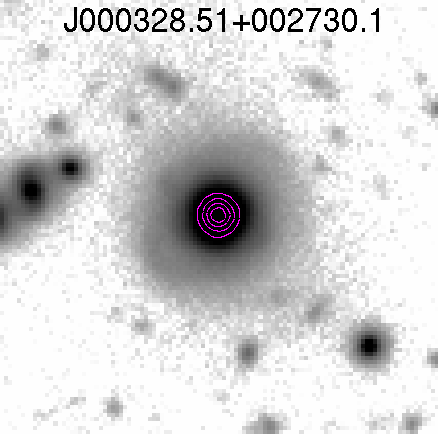}}\hfill \=
\fbox{\includegraphics[width=5.5cm,height=5.5cm]{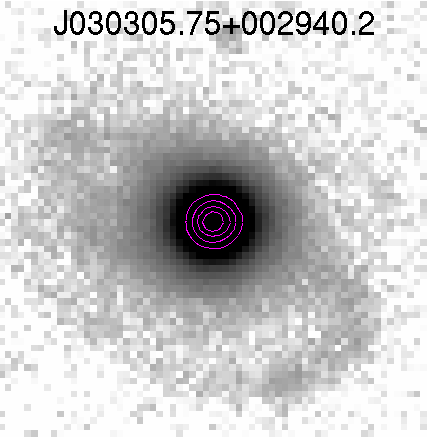}}\\
\fbox{\includegraphics[width=5.5cm,height=5.5cm]{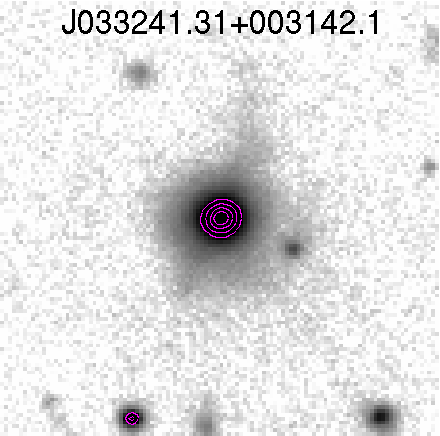}}\hfill \=
\fbox{\includegraphics[width=5.5cm,height=5.5cm]{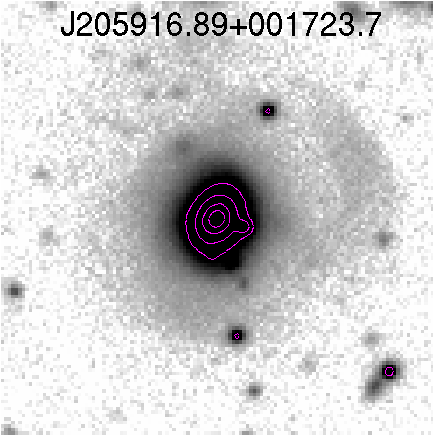}}\hfill \=
\fbox{\includegraphics[width=5.5cm,height=5.5cm]{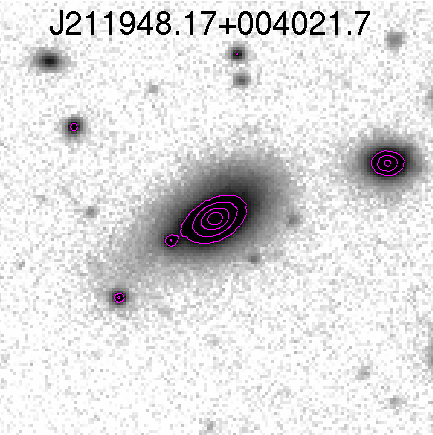}}\\
\fbox{\includegraphics[width=5.5cm,height=5.5cm]{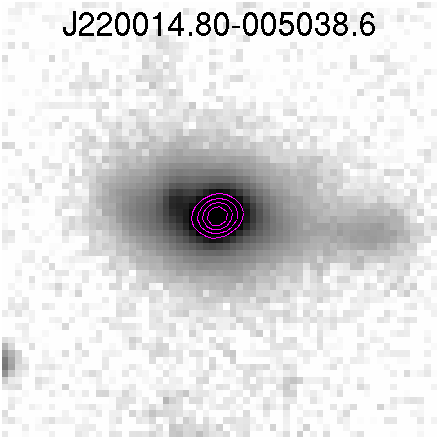}} 
\end{tabbing}
\caption{As Fig.\,\ref{fig:images_merger_A} but for sample C.}
\label{fig:images_merger_C}
\end{figure*}

\end{document}